# Micromagnetorotation effects in micropolar magnetohydrodynamic blood flow through stenosis

Kyriaki-Evangelia Aslani[1] *, Ioannis E. Sarris[2] and Efstratios Tzirtzilakis[3]

[1]*Department of Mechanical Engineering, University of the Peloponnese, 26334 Patras, Greece*
[2]*Department of Mechanical Engineering, University of West Attica, 12244 Athens, Greece*
[3]*Department of Civil Engineering, University of the Peloponnese, 26334 Patras, Greece*

* Corresponding author: k.aslani@go.uop.gr, kaslani@uniwa.gr

**Abstract**: This study presents a numerical investigation of a three-dimensional (3D) micropolar magnetohydrodynamic (MHD) blood flow through stenosis, with and without the effects of micromagnetorotation (MMR). MMR refers to the magnetic torque caused by the misalignment of the magnetization of magnetic particles in the fluid with the magnetic field, which affects the internal rotation (microrotation) of these particles. Blood can be modeled as a micropolar fluid with magnetic particles due to the magnetization of erythrocytes. In this manner, this study analyzes important flow features, i.e., streamlines, vorticity, velocity, microrotation, wall shear stress, and pressure drop—under varying stenosis, hematocrit levels, and magnetic fields, using two newly developed transient OpenFOAM solvers: epotMicropolarFoam and epotMMRFoam. Results indicate that micropolar effects become more pronounced at severe stenosis due to the significant reduction in artery size, resulting also in higher wall shear stress and pressure drop. Furthermore, when MMR is disregarded, the magnetic field does not significantly alter blood flow, regardless of its intensity, due to the minimal impact of the Lorentz force on blood. Conversely, MMR substantially affects blood flow, particularly at higher hematocrit levels and severe stenoses, leading to reductions of up to 30% in velocity and vorticity and up to 99.9% in microrotation and higher wall shear stress and pressure drop. Simultaneously, any vortices or disturbances are dampened. These findings underscore the critical role of MMR (which was ignored so far) in altering flow behavior in stenosed arteries, suggesting that it should be considered in future MHD micropolar blood flow studies.

## I. INTRODUCTION

A micropolar fluid is a type of fluid that contains small, rigid particles capable of independent motion within the mother liquid carrier. The theory of micropolar fluids, introduced by Eringen in 1966 [1, 2], is an extension of the Navier-Stokes equations, which are traditionally used to describe Newtonian fluids. Unlike the Navier-Stokes equations, the micropolar fluid theory accounts for the antisymmetric part of the stress tensor, which is zero in Newtonian fluids. Moreover, the antisymmetric part of the stress tensor plays an important role in the conservation of angular momentum, which is not inherently satisfied as in the Navier-Stokes equations. As a result, an additional equation must be solved, namely, the equation for the change in internal angular momentum. For this reason, in micropolar fluids, the fluid's behavior is described not only by its axial velocity $\boldsymbol{v}$ but also by its microrotation, denoted as $\boldsymbol{\omega}$. Microrotation refers to the averaged angular velocity of the fluid's internal microstructure [3]. It differs from the vorticity of the fluid given by $\boldsymbol{\Omega} = \frac{1}{2} \nabla \times \boldsymbol{v}$, because if the two were identical, the fluid would still behave as a Newtonian fluid, even when microrotation is nonzero. However, due to the introduction of an additional stress tensor associated with the microrotation's diffusion—known as the couple stress tensor—microrotation and vorticity generally do not coincide, except in specific cases. As a consequence of the introduction of the antisymmetric part of the stress tensor and the couple stress tensor in the micropolar fluid theory, two new viscosities are also introduced, i.e. the rotational viscosity $\mu_r$ related to the microrotation-vorticity difference [4, 5] and the spin viscosity $\gamma$ related to the flux of the internal angular momentum. Micropolar fluid theory has been applied in modeling various complex fluids, including exotic lubricants [6], colloidal suspensions (such as ferrofluids [7]), liquid crystals [8], and blood [9].

Blood is a fluid of the human circulatory system that delivers essential substances, such as oxygen and nutrients, to cells while also removing metabolic waste. It is primarily composed of plasma (~55% by volume), which consists mainly of water (~92% by volume), along with proteins, glucose, and mineral ions, as well as blood cells. There are three main types of blood cells: red blood cells (erythrocytes), white blood cells (leukocytes), and platelets (thrombocytes) [10]. Due to these blood cells in the plasma, blood is widely modeled as a micropolar fluid. Ariman, in 1971, was among the first researchers who used Eringen's micropolar fluid theory accounting for stretch to examine blood flow in small arteries of a diameter of ~100 $\mu m$ [11]. The study's findings were quite promising. In his words: "It is possible that micropolar fluid theory can yield more realistic results for blood velocity compared to the Navier-Stokes equations or the power-law model for non-Newtonian fluids." Later, in 1974, Ariman et al. conducted an experimental study on steady and pulsatile blood flow using classical micropolar fluid theory without considering stretch [12]. Their findings aligned well with the predictions of the micropolar fluid theory for these flows despite not accounting for erythrocyte deformability and nonlinear viscous effects. The above results motivated numerous researchers to study blood flows using micropolar fluid theory. Chaturani and Mahajan conducted an analytical study on Poiseuille blood flow using micropolar fluid theory [13]. Their findings were compared with experimental data on Poiseuille blood flow, and the results showed strong agreement between the analytical and experimental data, supporting the description of blood as a micropolar fluid. Karvelas et al. [14] examined blood flow within a human carotid model by modeling blood as a micropolar fluid. The study highlighted the differences in blood flow when its microstructure is considered. Findings indicated that micropolarity influences velocity through rotational viscosity by 4%. An increase in rotational viscosity resulted in higher velocities at the vessel's center and lower velocities near the boundaries. Moreover, they found a significant decrease in the shear stress at the walls when the rotational viscosity and the microrotation increased. Aslani et al. investigated the 2D Poiseuille and Couette flows of blood and ferrofluid by reviewing the cases of vorticity $\boldsymbol{\Omega}$ and microrotation $\boldsymbol{\omega}$ being equal [15]. The influence of all dimensionless parameters relevant to the mathematical model on the difference between $\boldsymbol{\omega}$ and $\boldsymbol{\Omega}$ was analyzed. These parameters were specifically determined based on the physical properties of ferrofluid and blood to accurately define the conditions under which these fluids can be modeled as either micropolar or Newtonian. For blood, it was found that when the size effect parameter $\lambda$ is small (for the definition of this dimensionless parameter see Equation 25), the difference between $\boldsymbol{\omega}$ and $\boldsymbol{\Omega}$ is intensified, leading to higher micropolarity. This occurs when the channel's height matches the diameter of human arterioles, a scenario that does not allow blood to be modeled as a Newtonian fluid.

An important subject regarding hemodynamics is the study of blood flows in pathogenic artery geometries. Pathogenic artery geometries refer to abnormal shapes and structures of arteries that contribute to or result from vascular diseases. These geometrical variations can significantly impact blood flow, shear stress, and the likelihood of developing atherosclerosis, aneurysms, stenoses, or other cardiovascular conditions. Micropolar fluid theory has been widely used for examining blood flows through pathogenic artery geometries. Hogan and Henriksen [16] conducted a numerical study on micropolar blood flow through an idealized stenosis. They employed the finite element method to obtain the solution to the problem. Their findings revealed a notable increase in wall shear stress (approximately 25%) due to the microrotation of erythrocytes. However, they observed no significant differences in the streamlines between the Newtonian and micropolar models. Mekheimer and Kot [17] analyzed a blood flow through an axially nonsymmetric but radially symmetric mild stenosis using micropolar fluid theory. Their results were obtained analytically. It was found that the resistance of the micropolar fluid increases with a higher coupling parameter, which is linked to particle size. Additionally, they demonstrated that as micropolarity increases, the velocity of the flow both inside and outside the stenotic region decreases, whereas it increases with a higher coupling parameter. Asadi et al. [18] studied a micropolar blood flow through a simple artery, represented as Hagen–Poiseuille flow within a tube, and an idealized stenosis. Using the finite element method, they numerically obtained the flow solution. Their findings indicated that axial velocity decreases with an increasing micropolar effect parameter, while both velocity and microrotation increase with the stenosis size. Furthermore, they confirmed that micropolar effects play a significant role in small vessels.

Magnetohydrodynamics (MHD) has been applied to conductive micropolar fluids, as introduced by Eringen [19, 20]. This model is derived by combining Maxwell's equations and the micropolar balance laws. In this framework, a force, namely the Lorentz force, occurs when electric currents are induced in the direction of the applied magnetic field, affecting the fluid flow. Over the years, MHD micropolar fluid theory has been widely applied in biomedical engineering, where magnetic fields influence blood flow, especially through pathogenic artery geometries. Abdullah et al. [21] investigated the MHD micropolar blood flow through an irregular stenosis. They numerically solved the problem using the finite difference method. Their findings indicated that increasing the micropolar effect parameter and the Hartmann number led to a reduction in axial velocity and flow rate. Conversely, an increase in the Hartmann number resulted in a decrease in wall shear stress. Jaiswal and Yadav [22] studied a two-phase MHD blood flow through a porous, layered artery. In their model, blood was treated as Newtonian near the vessel walls, while the core region was represented as micropolar. They analytically solved the problem using a modified Bessel function. Their findings revealed that increasing the Hartmann number suppressed both velocity and microrotation. Additionally, higher effective viscosity ratios led to reductions in flow rate and microrotation. Finally, an increase in the Hartmann number significantly lowered the wall shear stress. Bourantas [23] investigated a micropolar MHD blood flow through asymmetrical single stenosis and irregular multiple stenoses. The flow solutions were obtained numerically using a meshless point collocation method. His findings indicated that as the Hartmann number increased, the vortex formed outside the stenotic region was slightly suppressed, likely due to the decrease in flow velocity caused by the application of the magnetic field.

A popular approach to modeling blood flows under the influence of an applied magnetic field is ferrohydrodynamics (FHD) [24]. FHD deals with the modeling of ferrofluids' motion. Ferrofluids are colloidal magnetic fluids made by dispersing ferromagnetic particles (such as magnetite) into a mother-liquid carrier (such as water or kerosene). In this framework, the particles within a ferrofluid experience resistance due to magnetic polarization (magnetization $\boldsymbol{M}$) when exposed to an external magnetic field $\boldsymbol{H}$ [25]. As anticipated, ferrofluids display rotational degrees of freedom linked to the rotational motion of the suspended ferromagnetic particles, which is caused by magnetization. This rotational motion is treated in the same way as the microrotation in micropolar fluid. Specifically, the magnetization tends to "relax" and align with the external magnetic field. This alignment generates a magnetic moment, $\boldsymbol{M} \times \boldsymbol{H}$, which contributes to the equation of change of the internal angular momentum, significantly impacting the microrotation of ferromagnetic nanoparticles within the fluid [15]. At this point, it is important to mention that FHD should be distinguished from MHD, which refers to the behavior of conductive fluids influenced by the Lorentz force. FHD and MHD can be combined when dealing with a conductive ferrofluid. Ferrohydrodynamics was extensively studied by Shliomis [4] in 1972 and Rosensweig [26] in 1985. Later, in 1986, Shizawa & Tanahashi [27] developed a comprehensive mathematical model for ferrofluids based on micropolar fluid theory and Maxwell's equations. Their model introduced a new equation for magnetization, accounting for the existing microrotation of ferromagnetic nanoparticles—a concept known as micromagnetorotation (MMR). The flow equations were derived using the principles of irreversible thermodynamics, ensuring that the dissipation function remains positive at all times.

As mentioned in the previous paragraph, blood can be considered a ferrofluid when exposed to an external magnetic field, such as in a magnetic resonance imaging (MRI) scanner. This is due to the existence of the hemoglobin molecule in the erythrocytes, which is an iron oxide and behaves like a magnetic particle, while blood plasma is the mother-liquid carrier. As a result, the applied magnetic field may affect the microrotation of erythrocytes due to hemoglobin magnetization, influencing blood viscosity and velocity [28, 29]. Experimental studies have shown statistically significant symptoms—including vertigo, nausea, and a metallic taste—at magnetic field intensities of $1.5T$ and $4T$, which are linked to reduced blood velocity [30-33]. However, these effects cannot be explained by the Lorentz force, as blood's low electrical conductivity prevents it from being significantly influenced by this force. To the authors' knowledge, despite this evidence, the classification of blood as a ferrofluid acknowledging micromagnetorotation remains debated, and it is more commonly studied as a classical magnetohydrodynamic (MHD) Newtonian or micropolar fluid when it is under the influence of an external magnetic field (see the discussion above considering micropolar MHD blood flows).

Recently, Aslani et al. [15, 34-36] conducted several analytical studies on the influence of MMR in micropolar MHD flows, such as blood, using the Shizawa–Tanahashi mathematical model.

Their research compared the effects of including and neglecting the MMR term while always considering the Lorentz force. The findings revealed that MMR has a significant braking effect on velocity and microrotation. In plane MHD micropolar Poiseuille flow, velocity decreased by up to 16%, while microrotation dropped by up to 99%. Stability analysis further showed that MMR strongly stabilizes the flow, similar to the Lorentz force [35]. In the study of Aslani & Sarris, it was shown that MMR suppresses convection, reducing heat transfer and causing up to an 8.5% temperature drop due to velocity reduction [36]. In the paper of Aslani et al., it was demonstrated that an externally applied magnetic field influences erythrocyte microrotation via MMR, which in turn affects blood velocity through vorticity—a phenomenon observed in many experiments that cannot be explained solely by the Lorentz force [15].

Motivated by the above literature review, this paper concerns the numerical study of a three-dimensional (3D) micropolar MHD blood flow through stenosis by acknowledging and ignoring the effects of MMR. It examines two stenotic regions commonly associated with cardiovascular diseases, one with 50% and the other with 80% narrowing. Key flow variables are analyzed, including streamlines, vorticity, and microrotation contours, as well as velocity and microrotation profiles, both within and downstream of the stenotic regions. Wall shear stress (WSS) and pressure drop are also examined. For comparison, these variables are evaluated in four scenarios: Newtonian blood flow, micropolar blood flow, MHD micropolar blood flow without the MMR effect, and MHD micropolar blood flow with the MMR effect. Two hematocrit levels are considered: $\varphi = 25\%$ and $\varphi = 45\%$. The applied magnetic field is varied across three values: 1 $T$, 3 $T$, and 8 $T$. The chosen hematocrit and applied magnetic field values are sourced from previous numerical and experimental studies relevant to various biomedical applications. Simulations are conducted using two newly developed OpenFOAM solvers - epotMicropolarFoam and epotMMRFoam - which were validated against the analytical results of an MHD micropolar Poiseuille blood flow from the paper by Aslani et al., using different values for the hematocrit and the intensity of the applied magnetic field. The findings of this study are anticipated to be beneficial for bioengineering applications that involve high-intensity applied magnetic fields on blood flow, such as magnetic hyperthermia and magnetic drug delivery.

## II. MATHEMATICAL PRELIMINARIES

### A. Problem setup

As mentioned above, in this paper, a 3D micropolar MHD blood is numerically examined by acknowledging and ignoring MMR. The cylindrical coordinate system $(r, \theta, z)$ is used, where, $r$ is the radial coordinate, $\theta$ is the azimuthal angle, and $z$ is the axial coordinate. As can be seen from Figure 1, an idealized stenosis is considered with an external magnetic field $H_0$, which is applied transverse to the flow. The velocity components are given as $\boldsymbol{\upsilon} = \upsilon_r, 0, \upsilon_z$, while the microrotation components are given as $\boldsymbol{\omega} = 0, \omega_\theta, 0$. No-slip boundary conditions are imposed for both velocity and microrotation. The blood flow is driven by a uniform pressure gradient in the $z$ direction. The geometry of the flow is a cylinder of radius $R = 0.0015\ m$ (diameter $D = 2R = 0.003\ m$) and length $l = 0.03\ m$.

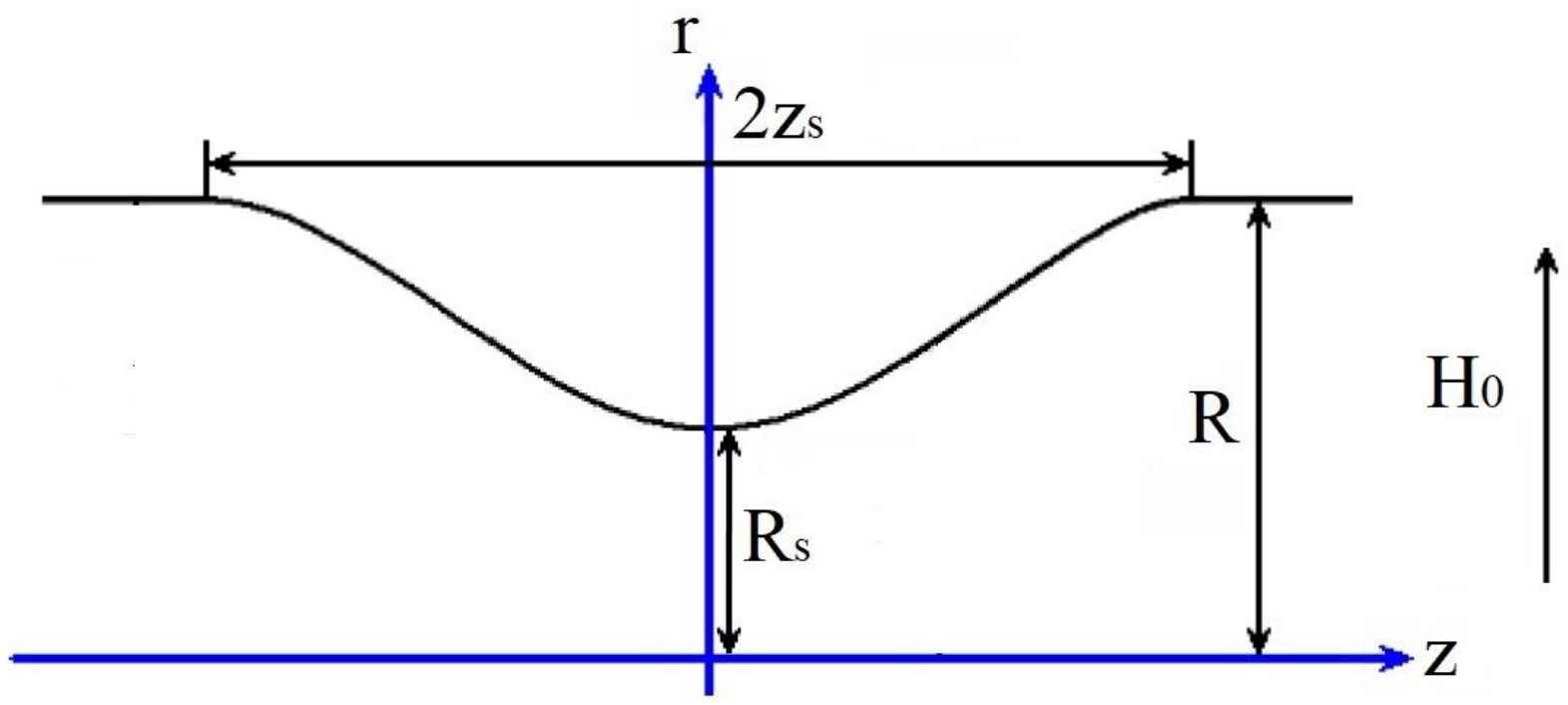

FIG. 1. A schematic diagram of the artery stenosis.

The center of the stenosis is located at $z = 0.009\ m$, considering that $z = 0$ at the inlet. The idealized stenosis is drawn according to Equation 1 as follows [16, 18, 24]:

$$l_s(z) = R\left[1 - \frac{R - R_s}{2R}\left(1 + \cos\frac{\pi z}{z_s}\right)\right], \tag{1}$$

where $R_s$ is the radius of the stenosis (stenosis' diameter $D_s = 2R_s$) and $2z_s$ is the total length of the stenosis. According to these, two stenotic areas are considered in this study, one at 50% of the nominal artery radius $R$ (i.e., $R_s = 0.00075\ m$) and one at 80% of the nominal radius $R$ (i.e., $R_s = 0.0003\ m$). The corresponding total length for the two stenotic regions is $2z_s = 0.002\ m$ for the 50% stenosis and $2z_s = 0.0024\ m$ for the 80% stenosis.

### B. Governing equations

According to Shizawa and Tanahashi [15, 27], the governing equations for modelling an MHD micropolar flow with ferromagnetic particles, such as blood, are:

$$\rho\frac{D\boldsymbol{v}}{Dt} = -\boldsymbol{\nabla}p + \mu\boldsymbol{\nabla}^2\boldsymbol{v} + 2\mu_r\boldsymbol{\nabla}\times(\boldsymbol{\omega} - \boldsymbol{\Omega}) + (\boldsymbol{M}\cdot\boldsymbol{\nabla})\boldsymbol{H} + \mu_0(\boldsymbol{j}\times\boldsymbol{H}), \tag{2}$$

$$\rho j\frac{D\boldsymbol{\omega}}{Dt} = 4\mu_r(\boldsymbol{\Omega} - \boldsymbol{\omega}) + \gamma\boldsymbol{\nabla}^2\boldsymbol{\omega} + \boldsymbol{M}\times\boldsymbol{H}, \tag{3}$$

$$\boldsymbol{\nabla}\cdot\boldsymbol{B} = 0, \tag{4}$$

$$\boldsymbol{\nabla}\times\boldsymbol{H} = \boldsymbol{j}, \tag{5}$$

$$\boldsymbol{j} = \sigma(\boldsymbol{v}\times\boldsymbol{B}), \tag{6}$$

$$\boldsymbol{M} = \frac{M_0}{H}[\boldsymbol{H} - \tau(\boldsymbol{H}\times\boldsymbol{\omega})], \tag{7}$$

where $\rho$ is fluid's density, $t$ is time, $p$ is pressure, $\mu$ is the dynamic viscosity, $\mu_r$ is the rotational viscosity, $\boldsymbol{M}$ is the magnetisation vector, $\boldsymbol{H}$ is the applied magnetic field from the magnetic flux density $\boldsymbol{B} = \mu_0\boldsymbol{H} + \boldsymbol{M}$ with $\mu_0$ being the magnetic permeability of free space ($4\pi\cdot 10^{-7}\ ^{H}/_{m}$), $\boldsymbol{j}$ is the current density vector, $j$ is the microinertia coefficient, $\gamma$ is spin viscosity, $\sigma$ is the electrical conductivity, $M_0$ is the equilibrium magnetisation and $\tau$ is the magnetization relaxation time. The first relation represents the conservation of linear momentum, while the second relation represents the equation of change of the internal angular momentum, which appears in micropolar fluids and is derived from the antisymmetric part of the stress tensor that contributes to the conservation of the total angular momentum. In the first equation, the term $2\mu_r\boldsymbol{\nabla}\times(\boldsymbol{\omega} - \boldsymbol{\Omega})$ represents force due to microrotation-vorticity difference that characterises micropolar fluids, the term $(\boldsymbol{M}\cdot\boldsymbol{\nabla})\boldsymbol{H}$ is the magnetic body force, and the term $\mu_0(\boldsymbol{j}\times\boldsymbol{H})$ is the Lorentz force. In the second equation, the term $4\mu_r(\boldsymbol{\Omega} - \boldsymbol{\omega})$ represents the internal torque of the suspended particles due to the microrotation-vorticity difference, the term $\gamma\boldsymbol{\nabla}^2\boldsymbol{\omega}$ is the microrotation diffusion and the last term $\boldsymbol{M} \times \boldsymbol{H}$ is the magnetic torque generated due to the magnetisation of the particles (micromagnetorotation). Equations (4), (5) and (6) represent Gauss law for magnetic monopoles, Ampere's law and Ohm's law, respectively (Maxwell's equations). Equation (7) is the constitutive magnetization equation from the mathematical model of Shizawa and Tanahashi [27]. The vorticity of the fluid remains $\boldsymbol{\Omega} = \frac{1}{2}\,\boldsymbol{\nabla} \times \boldsymbol{v}$ as in Newtonian fluids. It is evident from Equations (2)-(3) that when $\mu_r = 0$ or when $\boldsymbol{\omega} = \boldsymbol{\Omega}$, the classical Newtonian hydrodynamic equations are retrieved. For this reason, it is mandatory that $\boldsymbol{\omega} \neq \boldsymbol{\Omega}$ for a fluid to be classified as micropolar, even when $\boldsymbol{\omega} \neq 0$ [15]. Moreover, from Equation (7) it can be seen that when no microrotation exists, i.e., when $\boldsymbol{\omega} = \boldsymbol{0}$, the magnetization attains its equilibrium value $M_0$ parallel to the applied magnetic field $\boldsymbol{H}$. Provided that the blood flow is viscous and incompressible, the mass conservation law should be included:

$$\boldsymbol{\nabla} \cdot \boldsymbol{v} = \boldsymbol{0}, \tag{8}$$

$$\boldsymbol{\nabla} \cdot \boldsymbol{\Omega} = \boldsymbol{0}. \tag{9}$$

As mentioned above, no-slip boundary conditions are imposed for velocity and microrotation as follows:

$$v_r(R) = v_z(R) = 0, \tag{10}$$

$$\omega_\theta(R) = 0, \tag{11}$$

$$\left.\frac{\partial v_z}{\partial r}\right|_{r=0} = 0, \tag{12}$$

$$v_z, \omega_\theta \text{ are finite at } r = 0. \tag{13}$$

It should be mentioned that $\rho j$ at the left side of Equation (3) can be replaced by $I = \rho j$, where $I$ is the sum of the particles' moment of inertia per unit volume. According to the mathematical model of Shizawa & Tanahashi [27], the dynamic viscosity $\mu$, the vortex viscosity coefficient $\mu_r$, the spin viscosity $\gamma$ and the moment of inertia $I$ are correlated as:

$$\mu_r = \frac{I}{4\tau_s}, \tag{14}$$

$$\gamma = \mu j \;\; or \;\; \gamma = \mu\frac{I}{\rho}, \tag{15}$$

where $\tau_s = \frac{\alpha^2 \rho_\alpha}{15\mu}$ is the spin relaxation time, $\alpha$ being the radius of the suspended particles and $\rho_\alpha$ their density.

The governing equations (Equations 2-9) can be non-dimensionalized using the following dimensionless parameters:

$$\begin{gathered} \bar{\boldsymbol{\nabla}} = R\boldsymbol{\nabla}, \quad \bar{t} = t\frac{u_0}{R}, \quad \bar{\boldsymbol{v}} = \frac{\boldsymbol{v}}{u_0}, \quad \bar{\boldsymbol{\Omega}} = \boldsymbol{\Omega}\frac{R}{u_0}, \quad \bar{\boldsymbol{\omega}} = \boldsymbol{\omega}\frac{R}{u_0}, \\ \bar{p} = \frac{p}{\rho {u_0}^2}, \quad \bar{\boldsymbol{H}} = \frac{\boldsymbol{H}}{H_0}, \quad \bar{\boldsymbol{M}} = \frac{\boldsymbol{M}}{M_0}, \quad \bar{\boldsymbol{j}} = \frac{\boldsymbol{j}}{\sigma\mu_0 H_0 u_0}. \end{gathered} \tag{16}$$

Using Equation (16), the dimensionless governing equations are written as:

$$\frac{\partial \bar{\boldsymbol{v}}}{\partial \bar{t}} + (\bar{\boldsymbol{v}} \cdot \bar{\boldsymbol{\nabla}})\bar{\boldsymbol{v}} = -\bar{\boldsymbol{\nabla}}\bar{p} + \frac{1}{Re}\bar{\boldsymbol{\nabla}}^2\bar{\boldsymbol{v}} + \frac{2\varepsilon}{Re}\bar{\boldsymbol{\nabla}} \times (\bar{\boldsymbol{\omega}} - \bar{\boldsymbol{\Omega}}) + \frac{Mn_f\varepsilon}{Re}(\bar{\boldsymbol{M}} \cdot \bar{\boldsymbol{\nabla}})\bar{\boldsymbol{H}} + \frac{Ha^2}{Re}(\bar{\boldsymbol{J}} \times \bar{\boldsymbol{H}}), \tag{17}$$

$$\frac{\partial \bar{\boldsymbol{\omega}}}{\partial \bar{t}} + (\bar{\boldsymbol{\omega}} \cdot \bar{\boldsymbol{\nabla}})\bar{\boldsymbol{\omega}} = \frac{4\varepsilon\lambda^2}{Re}(\bar{\boldsymbol{\Omega}} - \bar{\boldsymbol{\omega}}) + \frac{1}{Re}\bar{\boldsymbol{\nabla}}^2\bar{\boldsymbol{\omega}} + \frac{Mn_f\varepsilon\lambda^2}{Re}(\bar{\boldsymbol{M}} \times \bar{\boldsymbol{H}}), \tag{18}$$

$$\bar{\boldsymbol{\nabla}} \cdot \bar{\boldsymbol{B}} = 0, \tag{19}$$

$$\bar{\boldsymbol{\nabla}} \times \bar{\boldsymbol{H}} = Re_m\bar{\boldsymbol{J}}, \tag{20}$$

$$\bar{\boldsymbol{J}} = \bar{\boldsymbol{v}} \times \bar{\boldsymbol{H}} + \chi_m(\bar{\boldsymbol{v}} \times \bar{\boldsymbol{M}}), \tag{21}$$

$$\bar{\boldsymbol{M}} = \frac{\bar{\boldsymbol{H}}}{\bar{\bar{H}}} - Per\frac{1}{\bar{\bar{H}}}(\bar{\boldsymbol{H}} \times \bar{\boldsymbol{\omega}}) \tag{22}$$

$$\bar{\boldsymbol{\nabla}} \cdot \bar{\boldsymbol{v}} = \mathbf{0}, \tag{23}$$

$$\bar{\boldsymbol{\nabla}} \cdot \bar{\boldsymbol{\Omega}} = \mathbf{0}. \tag{24}$$

The dimensionless parameters that are seen in Equations (17)-(24) are defined as:

$$Re = \frac{\rho u_0 R}{\mu}, \quad \varepsilon = \frac{\mu_r}{\mu}, \quad \lambda = \frac{R}{\iota}, \quad Mn_f = \frac{M_0 H_0 R}{\mu_r u_0},$$
$$Ha = \mu_0 H_0 R\sqrt{\sigma/\mu}\,, \quad Re_m = \sigma\mu_0 H_0 R, \quad \chi_m = \frac{M_0}{\mu_0 H_0}, \quad Per = \frac{u_0\tau}{R}. \tag{25}$$

where $\iota = \sqrt{j}$. In Equation (25), $Re$ is the Reynolds number, $\varepsilon$ is the micropolar effect parameter, $\lambda$ is the size effect parameter, $Ha$ is the Hartmann number, $Re_m$ is the magnetic Reynolds number, $\chi_m$ is the magnetic susceptibility, $Mn_f$ is the magnetic number and $Per$ is the rotational Peclet number. In Equations (17)-(22), the magnetization effect parameter $\sigma_m$ from the previous papers of Aslani et al. [15, 34-36] is not explicitly shown. Still, it can be easily calculated from the relation $Mn_f = \frac{4\sigma_m}{Per}$. All these parameters are extensively discussed in the next subsection.

The boundary conditions are also non-dimensionalized as follows:

$$\bar{v}_r(1) = \bar{v}_z(1) = 0, \tag{26}$$

$$\bar{\omega}_\theta(1) = 0, \tag{27}$$

$$\left.\frac{\partial \bar{v}_z}{\partial \bar{r}}\right|_{\bar{r}=0} = 0, \tag{28}$$

$$\bar{v}_z, \bar{\omega}_\theta \text{ are finite at } \bar{r} = 0. \tag{29}$$

It should be noted, that in this study, as can be seen from Equations (11) and (27), the standard no-spin microrotation boundary condition $\omega_\theta = 0$ was imposed, which corresponds to the commonly used choice $\delta = 0$, where $\delta$ is the microrotation boundary condition parameter $0 \leq \delta \leq 1$ [4, 16, 27]. When $\delta = 1$, microrotation equals the fluid's vorticity at the walls. This condition is widely adopted in micropolar hemodynamic studies due to the lack of conclusive experimental evidence supporting a finite spin–vorticity coupling at the walls of the vessel [12, 14, 16, 18, 22, 33, 37], and because nonzero values of $\delta$ progressively suppress micropolar effects, driving the solution toward the Newtonian limit. Our previous analysis of microrotation–vorticity coupling in micropolar MHD flows[15] demonstrated that increasing $\delta$ systematically reduces microrotation gradients and ultimately recovers Newtonian behavior, indicating that the qualitative trends reported here would remain unchanged. A more detailed investigation of alternative microrotation boundary conditions in fully three-dimensional, anatomically realistic blood arteries is an interesting direction for future research.

Since near-wall hemodynamics are central in stenosed arteries, the wall shear stress (WSS) $\tau_w$ is computed, defined as the magnitude of the tangential viscous traction acting on the vessel wall. In cylindrical coordinates, this corresponds to the $r - z$ component of the stress tensor at $r = R$. For a Newtonian fluid, the shear stress is:

$$T_{rz_{Newt}} = \mu\left(\frac{\partial v_z}{\partial r} + \frac{\partial v_r}{\partial z}\right), \tag{30}$$

and, because the flow is predominantly axial near the wall, we approximate:

$$\tau_{w_{Newt}} = \mu \frac{\partial v_z}{\partial r}\bigg|_{r=R}. \tag{31}$$

For the micropolar fluid (even with or without the effect of an applied magnetic field), the corresponding shear stress includes the contribution of the antisymmetric part, i.e, the part that contains the microrotation. Using the constitutive relations of the micropolar theory [1, 27, 38], the $r - z$ component can be written as:

$$T_{rz_{micr}} = \left(\mu + \mu_r\right)\frac{\partial v_r}{\partial z} + \mu_r \omega_\theta. \tag{32}$$

Together with the no-spin microrotation boundary condition $\omega_\theta = 0$ at the walls, Equation (32) yields:

$$\tau_{w_{micr}} = (\mu + \mu_r)\frac{\partial v_z}{\partial r}\bigg|_{r=R}. \tag{33}$$

### C. Calculation and analysis of the dimensionless parameters

As can be seen from Equation (25), the problem depends on various dimensionless parameters. The first parameter is the Reynolds number $Re$, which is perhaps the most popular dimensionless parameter in fluid dynamics. The Reynolds number represents the ratio of inertial forces to viscous forces in a fluid experiencing internal motion caused by a velocity gradient. This motion creates fluid friction, which contributes to the development of turbulence. Viscosity, on the other hand, acts to resist this turbulence by dampening the flow irregularities. In this manner, the Reynolds number is used to predict the flow regime of a fluid — whether it will be laminar, transitional, or turbulent.

One of the key dimensionless parameters in this study and in general in micropolar fluids is the micropolar effect parameter, $\varepsilon$. This parameter quantifies the degree of micropolarity in a fluid. As shown in Equation (25), $\varepsilon$ is directly proportional to the rotational viscosity, $\mu_r$— meaning that an increase in $\mu_r$ leads to a corresponding increase in $\varepsilon$. According to Equation (14), this increase in $\mu_r$ can result from either a rise in the particles' moment of inertia per unit volume, $I$, or a decrease in the spin relaxation time, $\tau_s$. In general, a great increase of $\varepsilon$ leads to equal vorticity and microrotation, which results in a Newtonian behavior for the fluid, even when microrotation is nonzero [15]. Finally, it should be noted that $\varepsilon$ is directly proportional to the volume fraction $\varphi$ by the equation $\varepsilon = \frac{3}{2}\varphi$, indicating that an increase in $\varphi$ also leads to an increase in $\varepsilon$. For blood, the volume fraction $\varphi$ is coincides with the hematocrit [11, 12, 15].

Another important dimensionless parameter for this study and micropolar flows in general is the size effect parameter $\lambda$. According to Equation (25), $\lambda$ is directly proportional to the radius of the vessel $R$ and inversely proportional to the square root of the microinertia coefficient $j$. Since $j$ is inversely related to the spin viscosity $\gamma$ (see Equation 15), $\lambda$ is also inversely proportional to $\gamma$. This implies that as microrotation diffusion increases, the size effect parameter $\lambda$ decreases. Additionally, it has been shown that larger values of $\lambda$ reduce the difference between microrotation and vorticity. As a result, micropolar fluids with high $\lambda$ values can be approximated as Newtonian fluids. However, in the case of blood flow, especially in human arterioles, $\lambda$ tends to be small. This means that blood flow in small vessels cannot be accurately modeled as Newtonian [15].

Two parameters that are commonly encountered in magnetohydrodynamics and are particularly important for this study are the Hartmann number $Ha$ and the magnetic Reynolds number $Re_m$. $Ha$ is influenced by the strength of the applied magnetic field $H_0$ and the electrical conductivity $\sigma$ of the fluid. As shown in Equation (25), the Hartmann number becomes zero when no external magnetic field is applied ($H_0 = 0$) and/or when the fluid is non-conductive ($\sigma = 0$). $Re_m$ is associated with the ratio of the advection (transport) of the magnetic field to the diffusion of the latter due to the fluid's finite electrical conductivity. Many studies use the assumption of the low-magnetic-Reynolds number, where $Re_m \ll 1$ [27, 39-42]. This approximation omits the induction equation, due to the assumption that the induced magnetic field is smaller than the applied magnetic field and its impact on the flow can be ignored. Thereby, the number of equations that must be addressed is reduced. As in the previous studies of Aslani et al. [15, 34-36], here, the low-$Re_m$ approximation is also utilized and Equation (20) is not solved.

Finally, three dimensionless parameters from Equation (25), namely the magnetic number $Mn_f$, the magnetic susceptibility $\chi_m$ and the rotational Peclet number $Per$ are associated with the effect of micromagnetorotation (MMR) on the blood flow, due to their relation with the magnetization effect parameter $\sigma_m$ (see Aslani et al. [15, 34-36]). $Mn_f$ is associated with the ratio of the magnetic force to the inertial force and is frequently used in magnetic fluid studies [43-46]. The magnetic susceptibility $\chi_m$ is a popular quantity in magnetic materials that measures how much a material will become magnetized in an applied magnetic field. In this manner, a material can be diamagnetic with $\chi_m < 0$ (such as copper and water), paramagnetic with $\chi_m > 0$ (such as aluminum and oxygen), and ferromagnetic with $\chi_m \gg 0$ (such as iron and cobalt). The magnetic susceptibility of blood depends on the magnetic properties of hemoglobin and can result in paramagnetic properties with $\chi_{m_{blood}} \sim 10^{-4}$. The rotational Peclet number $Per$ quantifies how quickly the magnetization will reach its equilibrium value. For blood, the magnetization relaxation time is small, i.e., $\tau \sim 10^{-3}\ sec$ and $Per \ll 1$. Using the relation $Mn_f = \frac{4\sigma_m}{Per}$, the magnetization effect parameter $\sigma_m = \frac{\tau\tau_s H_0 M_0}{I}$ can be calculated, which is analogous to the effect of MMR on micropolar magnetic fluids, such as blood (see Aslani et al. [15, 34-36]). One can notice that when no external applied magnetic field is applied on the flow ($H_0 = 0$) and/or when the fluid does not contain magnetic particles ($M_0 = 0$ and $\tau = 0$), then $\sigma_m = 0$ and, thereby, $Mn_f = 0$, $Per = 0$ and $\chi_m = 0$.

All dimensionless parameters are calculated using blood's physical properties, elements of the geometry of the stenosis, and the intensity of the applied magnetic field. Table I shows the physical properties of blood as they were derived from the papers of [12, 24, 28, 33]. For the sake of comparison, results for hematocrit $\varphi = 25\%$ and $\varphi = 45\%$ are presented. Both hematocrit values are realistic; the value $\varphi = 25\%$ is below the normal range of 40-54% for males and 36-48% for females, but it is a common value for patients with diseases, such as cancer or sickle cell anemia [47]. The saturation magnetization of erythrocytes is taken as $M_s = 100\ A/m$, consistent with measurements of hemoglobin magnetization [33].

TABLE I. Blood's physical properties.

| Physical properties | Value | |
|---|---|---|
| volume fraction (hematocrit) $\varphi$ (%) | 25 | 45 |
| micropolar effect parameter $\varepsilon$ (-) | 0.375 | 0.675 |
| dynamic viscosity $\mu$ ($Pa \cdot s$) | $4 \cdot 10^{-3}$ | $4 \cdot 10^{-3}$ |
| rotational viscosity $\mu_r$ ($Pa \cdot s$) | $1.5 \cdot 10^{-3}$ | $2.7 \cdot 10^{-3}$ |
| microinertia coefficient $j$ ($m^2$) | $1.75 \cdot 10^{-10}$ | $7.5 \cdot 10^{-10}$ |
| spin viscosity $\gamma$ ($\frac{kg \cdot m}{sec}$) | $7 \cdot 10^{-13}$ | $3 \cdot 10^{-12}$ |
| fluid density $\rho_f$ ($kg \cdot m^{-3}$) | 1050 | 1050 |
| saturation magnetization $M_s$ ($A \cdot m^{-1}$) | 100 | 100 |
| electrical conductivity $\sigma$ ($S \cdot m^{-1}$) | 0.7 | 0.7 |
| magnetization relaxation time $\tau$ ($sec$) | 0.001 | 0.001 |
| rotational Peclet number $Per$ (-) | 0.0667 | 0.0667 |

Using Equation (25), the size effect parameter $\lambda$, the Hartmann number $Ha$, the magnetic number $Mn_f$, the magnetic susceptibility $\chi_m$, and the magnetization effect parameter $\sigma_m$ are calculated for different hematocrit values and presented in Tables II, III and IV. As in the analytical study of Aslani et al. [15], three values of the applied magnetic field 's intensity are used, which are frequently found in experimental and numerical studies associated with biomedical applications, i.e. $H_0 = 795{,}774.72\ \frac{A}{m}$, $H_0 = 2{,}387{,}324.15\ \frac{A}{m}$ and $H_0 = 6{,}366{,}197.72\frac{A}{m}$ [28, 30, 31]. Considering that $H_0 = \frac{B_0}{\mu_0}$, it is derived that $H_0 = 795{,}774.72\ \frac{A}{m}$ corresponding to $1\ T$, $H_0 = 2{,}387{,}324.15\ \frac{A}{m}$ corresponding to $3\ T$ and $H_0 = 6{,}366{,}197.72\frac{A}{m}$ corresponding to $8\ T$.

TABLE II. Blood's size effect parameter $\lambda$ for different values of hematocrit.

| $D$ ($m$) | $\varphi$ ($-$) | $\lambda$ ($-$) |
|---|---|---|
| 0.003 | 25 % | 133.39 |
| | 45 % | 54.77 |

TABLE III. Blood's Hartmann number $Ha$ and magnetic susceptibility $\chi_m$ for different values of magnetic field intensity.

| $D\ (m)$ | $H_0\ (\frac{A}{m})$ | $Ha\ (-)$ | $\chi_m\ (-)$ |
|---|---|---|---|
| | 795,774.72 (1 $T$) | 0.02 | 0.000126 |
| 0.003 | 2,387,324.15 (3 $T$) | 0.06 | 0.000042 |
| | 6,366,197.72 (8 $T$) | 0.16 | 0.000016 |

TABLE IV. Blood's magnetization effect parameter $\sigma_m$ and magnetic number $Mn_f$ for different values of magnetic field intensity and hematocrit.

| $\varphi\ (-)$ | $H_0\ (\frac{A}{m})$ | $\sigma_m\ (-)$ | $Mn_f\ (-)$ |
|---|---|---|---|
| | 795,774.72 (1 $T$) | 16.66 | 1000.00 |
| 25% | 2,387,324.15 (3 $T$) | 49.99 | 3000.00 |
| | 6,366,197.72 (8 $T$) | 133.33 | 8000.00 |
| | 795,774.72 (1 $T$) | 9.26 | 555.56 |
| 45% | 2,387,324.15 (3 $T$) | 27.78 | 1666.67 |
| | 6,366,197.72 (8 $T$) | 74.10 | 4444.44 |

For the simulation of a blood flow, a suitable pressure gradient $G$ or an inlet velocity must be specified. Here, in order for the numerical simulations to resemble a realistic human blood flow, no value for the inlet velocity was applied. According to Poiseuille law, the artery is modeled as a pipe flow and a maximum velocity value was selected from the bibliography [48, 49]. In this manner, the corresponding pressure gradient was calculated using the relation $v_{z_{max}} = \frac{GR^2}{4\mu}$. For an artery with diameter $D = 0.003\ m$, $v_{z_{max}} = 0.1\ {}^{m}/_{sec}$ was selected. Consequently, it was derived that $G = 711.11\ \frac{Pa}{m}$. The Reynolds number was calculated as $Re = 39.375$. Considering that the length of the artery is $l = 0.03\ m$ and using the relation $\frac{P_2 - P_1}{l} = G$, where $P_1$ is the inlet pressure and $P_2$ is the outlet pressure, it was derived that $P_1 = 21.33\ Pa$ (setting $P_2 = 0\ Pa$). Finally, the inlet pressure $P_1$ values was divided by blood's density $\rho = 1050\ \frac{kg}{m^3}$, because OpenFOAM uses kinematic pressure $\frac{P}{\rho}$ as input.

### D. Solution algorithm and code validation

Transient solvers for the numerical solution of Equations (17)–(24) were developed using the open-source OpenFOAM library. For simulating Newtonian blood flow through a stenosis, the widely used transient solver icoFoam was employed. This solver implements the PISO (Pressure Implicit with Splitting of Operators) algorithm. To model the simple micropolar blood flow through stenosis without the influence of an external magnetic field, the solver micropolarFoam was used. This solver was originally developed by Manolis and Koutsoukos [50] by modifying icoFoam to incorporate the force term arising from the microrotation–vorticity difference and to solve the internal angular momentum equation.

For modeling the MHD micropolar blood flow through stenosis without considering the effect of micromagnetorotation, a new solver named epotMicropolarFoam was developed. This solver is a modified version of epotFoam [51], which is designed for incompressible, laminar flows of conducting fluids under the influence of external magnetic fields.

EpotMicropolarFoam applies the low-magnetic-Reynolds number approximation ($Re_m \ll 1$), where the magnetic induction is neglected and the electromagnetic field is computed using an electric-potential (epot) formulation employed in epotFoam-type solvers. In this framework, the Ohm's law (Equation 6) is written as:

$$\boldsymbol{j} = \sigma(-\boldsymbol{\nabla}\varphi + \boldsymbol{v} \times \boldsymbol{B}), \tag{34}$$

where $\boldsymbol{\nabla}\varphi$ is the electric potential derived from the electric field $\boldsymbol{E} = -\boldsymbol{\nabla}\varphi$. This formulation is consistent with the Gauss law for magnetic monopoles and the Ampere's law (Equations 4-5). The electric potential $\varphi$ is obtained from the Poisson equation:

$$\boldsymbol{\nabla} \cdot (\sigma\boldsymbol{\nabla}\varphi) = \boldsymbol{\nabla} \cdot (\sigma\, \boldsymbol{v} \times \boldsymbol{B}). \tag{35}$$

The Lorentz force $\mu_0(\boldsymbol{j} \times \boldsymbol{H})$ appearing in Equation (2) is therefore computed directly from the potential-based current density. The initial and boundary conditions for the electric potential are described separately in Section II.F.

Finally, to simulate MHD micropolar blood flow through stenosis while accounting for micromagnetorotation (MMR) effects, a new solver called epotMMRFoam was created. This solver extends epotMicropolarFoam by including the magnetic torque term $\boldsymbol{M} \times \boldsymbol{H}$ in the internal angular momentum equation as a source term, as well as the constitutive equation for magnetization (Equation 7). Magnetization does not contribute an additional induced-current term in the low-Rm approximation; instead, it enters only through the constitutive magnetization relation and the microrotation equation. A detailed description of the discretization, implementation, and verification of the epotMicropolarFoam and epotMMRFoam solvers can be found in our companion manuscript [52].

All equations in the solvers were discretized using the finite volume method (FVM). Specifically, the transient terms were discretized with a second-order upwind Euler scheme, while for the diffusion terms, the Gauss linear corrected scheme was utilized. For the convection terms, a second-order unbounded scheme was employed, and a Taylor-like series expansion was applied for the source terms. To solve the resulting algebraic equations, an explicit under-relaxation method was combined with the preconditioned conjugate gradient method.

The new transient solvers epotMicropolarFoam and epotMMRFoam were validated using the plane MHD micropolar Poiseuille blood flow from the study of Aslani et al. [15]. The results were drawn for different values of the hematocrit and the intensity of the applied magnetic field. These values were the same as those of this study, i.e., $\varphi = 25\%$ and $\varphi = 45\%$ and $B_0 = 1\,T, 3\,T$ and $8\,T$. For the sake of comparison, the numerical results of the simple Newtonian blood flow and the simple micropolar blood without an applied magnetic field were drawn, using the solvers icoFoam and micropolarFoam. These results are presented in Figure 2, where the dimensionless velocity $v_x$ and the dimensionless microrotation $\omega_x$ are shown. The error between the analytical and numerical results does not exceed the value of $0.5\%$ for the velocity and $2\%$ for the microrotation. It is evident that the OpenFOAM solvers icoFoam and micropolarFoam are suitable for the simulation of Newtonian and micropolar blood flows, giving accurate results for velocity and microrotation. Finally, it should be noted that the difference between the Newtonian velocity profile and the micropolar ones for hematocrit values $\varphi = 25\%$ and $45\%$ is barely noticeable. The same applies to the microrotation profiles for the same hematocrit values. This is due to the relatively high values of $\lambda$, due to the artery's size. In smaller arteries (such as arterioles), where $\lambda$ would take smaller values for the same hematocrit, the differences in both velocity and microrotation would be more pronounced (see the paper of Aslani et al. [15]).

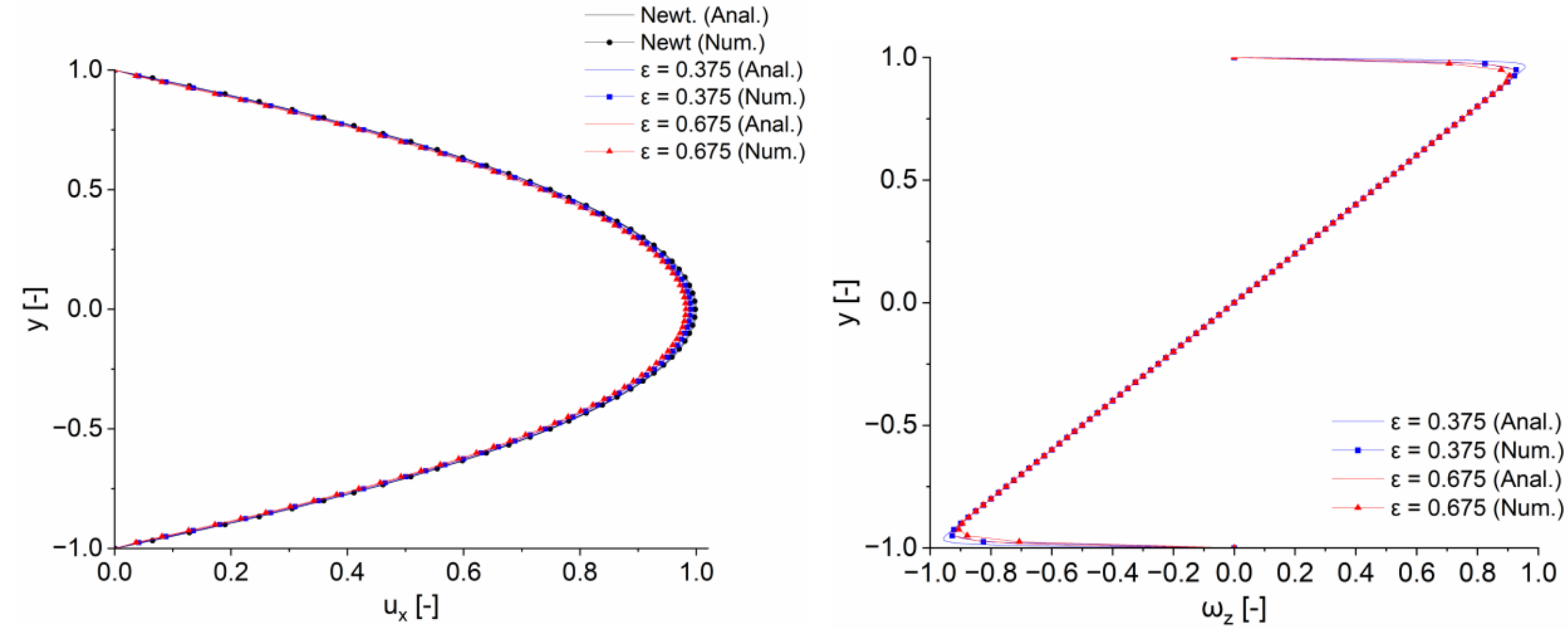

*(a)* *(b)*

FIG. 2. Analytical and numerical results for the (a) dimensionless velocity $\upsilon_x$ and (b) dimensionless microrotation $\omega_x$ of the Newtonian and micropolar Poiseuille blood flows with hematocrit $\varphi = 25\%$ and $\varphi = 45\%$ ($\varepsilon = 0.375$ and $\varepsilon = 0.675$, respectively).

In Figure 3, a comparison between the analytical and the numerical results for the dimensionless velocity $\upsilon_x$ and the dimensionless microrotation $\omega_x$. for the MHD (without acknowledging micromagnetorotation) and the MMR micropolar Poiseuille blood flows is presented. The hematocrit is held constant at $\varphi = 25\%$, while the applied magnetic field is varied at 1 $T$, 3 $T$, and 8 $T$. The error does not exceed the value of 0.6% for velocity and 1.24% for microrotation. It is evident that the two new solvers epotMicropolarFoam and epotMMRFoam give accurate results for different values of the applied magnetic field. Finally, the difference observed in both velocity and microrotation when micromagnetorotation is taken into account is noteworthy. The reduction in velocity reaches 25% for all values of the magnetic field, while for microrotation, it approaches 98% at a magnetic field of 8 $T$. One can notice that the MMR effect on the blood flow is significant, whereas when it is not taken into account, the influence of the magnetic field on the blood flow is negligible.

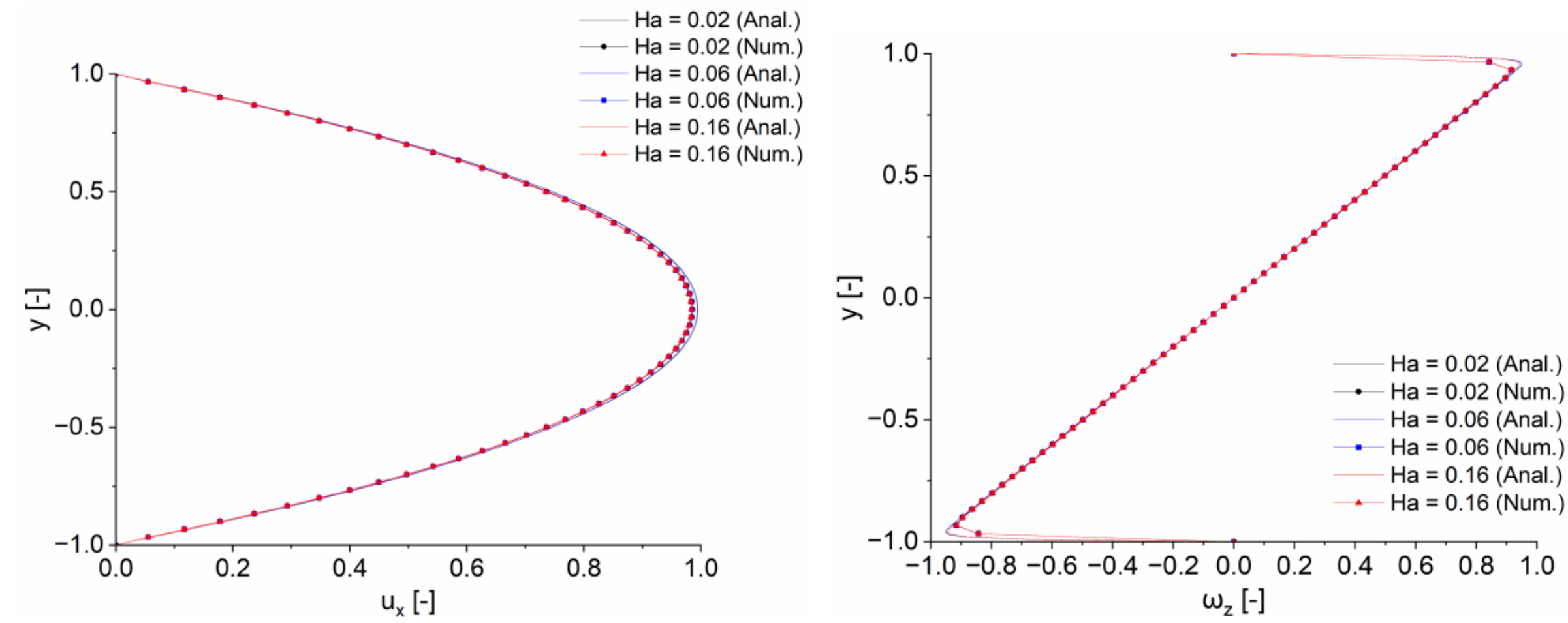

*(a)*

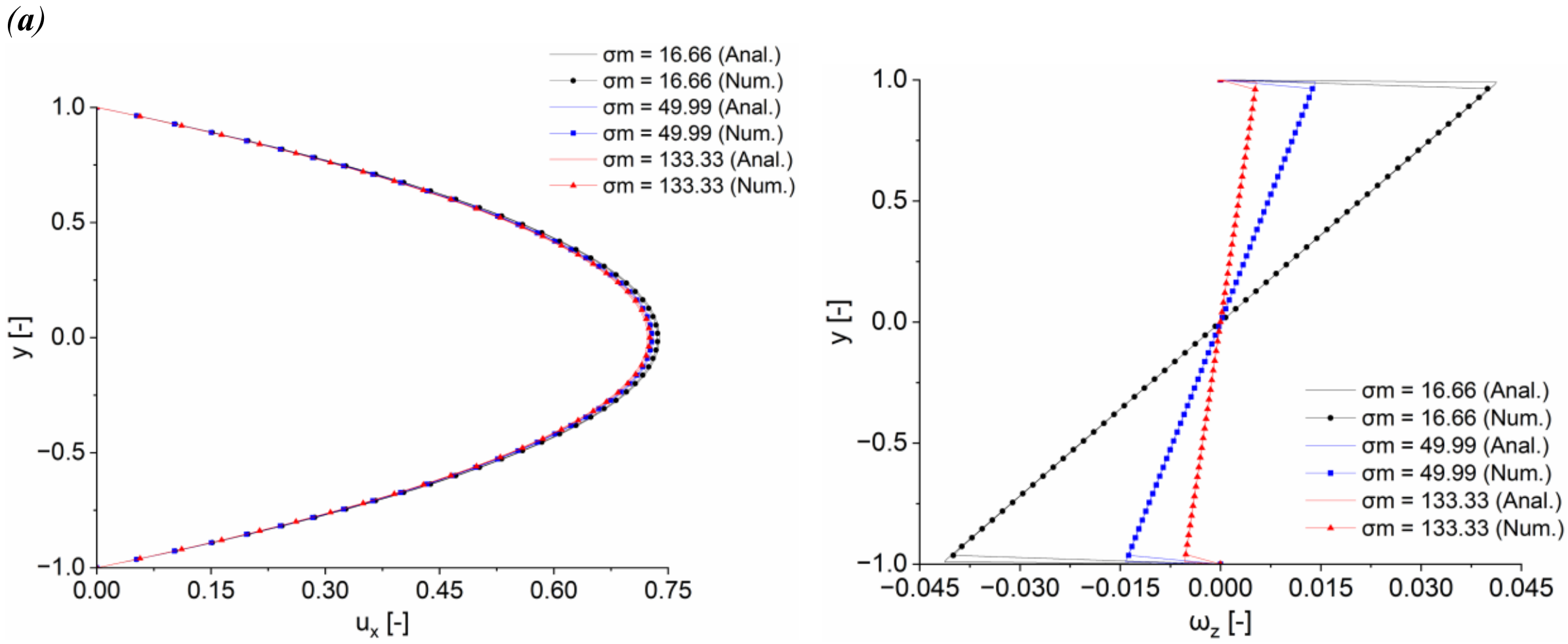

*(b)*

FIG. 3. Analytical and numerical results for the dimensionless velocity $v_x$ (left) and dimensionless microrotation $\omega_x$ (right) of the (a) MHD and (b) MMR micropolar Poiseuille blood flows with hematocrit $\varphi = 25\%$ and applied magnetic field of 1 $T$ ($Ha = 0.02, \sigma_m = 16.66$), 3 $T$ ($Ha = 0.06, \sigma_m = 49.99$) and 8 $T$ ($Ha = 0.16, \sigma_m = 133.33$).

Same as before, in Figure 4, a comparison between the analytical and the numerical results for the dimensionless velocity $v_x$ and the dimensionless microrotation $\omega_x$. for the MHD and the MMR micropolar Poiseuille blood flows is shown. In this case, the hematocrit is held constant at $\varphi = 45\%$, while the applied magnetic field is varied at 1 $T$, 3 $T$, and 8 $T$. Here, the error does not exceed the value of 1.54% for velocity and 1.55% for microrotation. It is evident that the solvers epotMicropolarFoam and epotMMRFoam give accurate results even for higher hematocrit values. Again, one can see that the difference observed in velocity and microrotation when MMR is taken into account is significant and larger than the case of $\varphi = 25\%$. The reduction in velocity reaches 37% for all values of the magnetic field, while for microrotation, it approaches 99% at a magnetic field of 8 $T$. The MMR effect on a blood flow is notable, especially for higher hematocrit values.

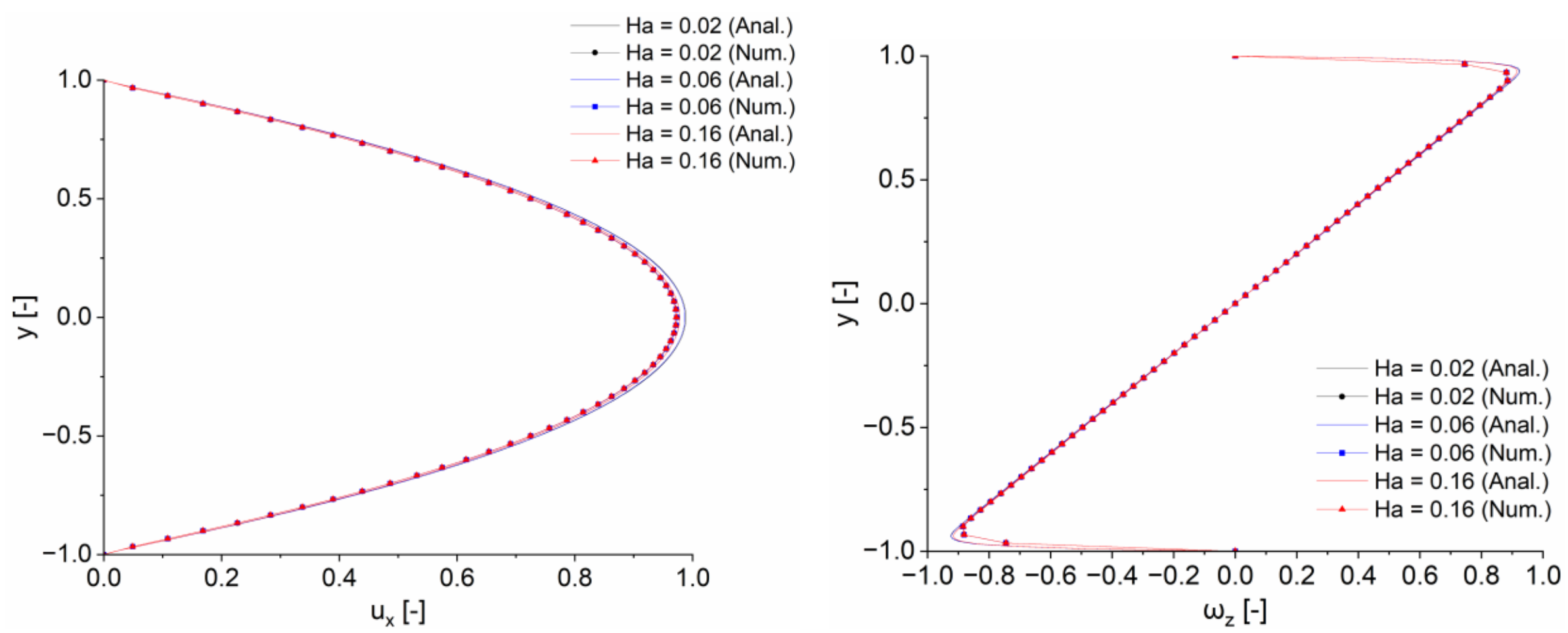

*(a)*

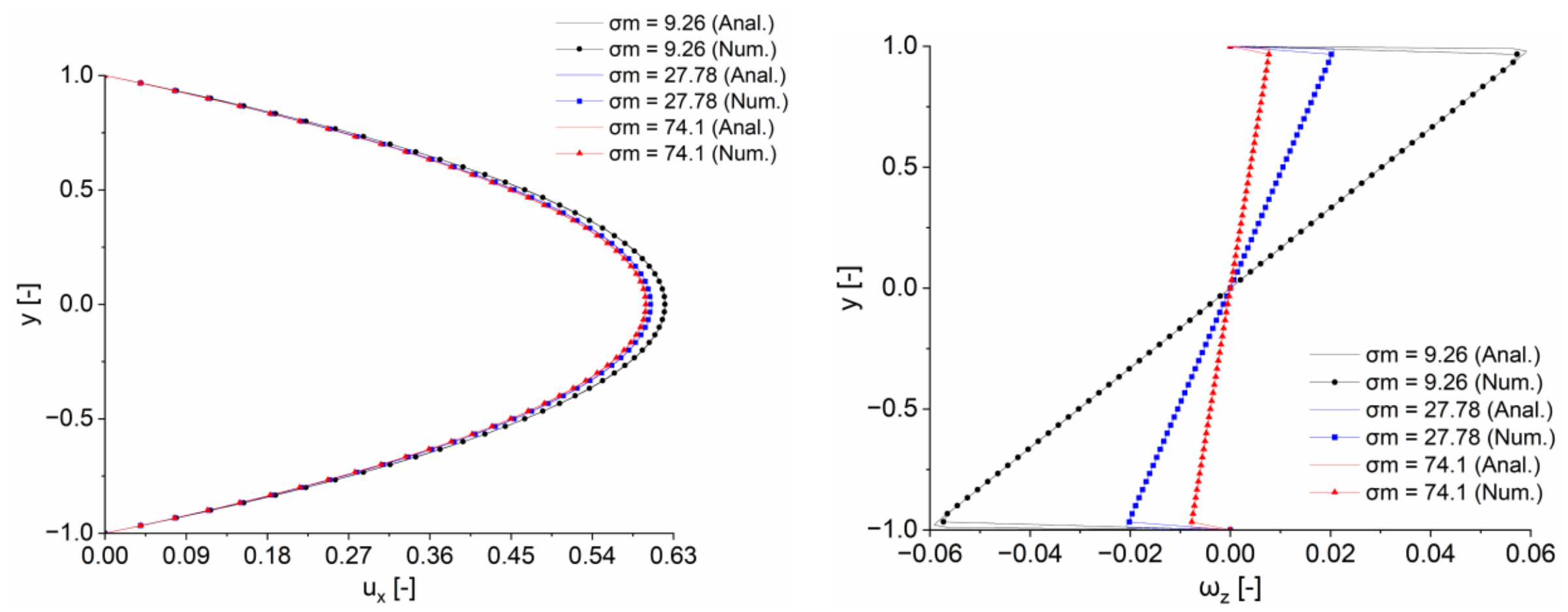

***(b)***

FIG. 4. Analytical and numerical results for the dimensionless velocity $v_x$ (left) and dimensionless microrotation $\omega_x$ (right) of the (a) MHD and (b) MMR micropolar Poiseuille blood flows with hematocrit $\varphi = 45\%$ and applied magnetic field of 1 $T$ $(Ha = 0.02, \sigma_m = 9.26)$, 3 $T$ $(Ha = 0.06, \sigma_m = 27.78)$ and 8 $T$ $(Ha = 0.16, \sigma_m = 74.1)$.

Since the results of this study concern 3D configurations of stenosed arteries, an additional validation was performed using the classical benchmark work of Hogan and Henriksen [16], who studied Newtonian and micropolar flows through an idealized 50% stenosis. As expected, the same microrotation boundary condition was imposed, i.e., the microrotation being zero at the walls of the stenosis. The computed maximum wall shear stress and maximum stenosis velocity, reported in the Table 1 of their paper, did not exceed the value of 2% for both the Newtonian and micropolar cases. Furthermore, the computed centerline pressure drop and recirculation length fall within the ranges reported the paper. Finally, the streamline and vorticity contours exhibit the well-known acceleration in the throat, followed by symmetric vortex formation at the end of the stenosis and downstream recovery. These results provide validation specifically for stenotic regions and give additional confidence that the present solvers accurately capture the essential hemodynamic features of idealized stenotic geometries.

### E. Geometry and mesh refinement

As mentioned above, the geometry of the problem is an artery represented by a 3D pipe with an idealized stenosis according to Equation (1). For this study, two values were considered for the stenosis diameter $D_s$ yielding to two stenoses, i.e., 50% and 80%. The discretization of the computational domain was achieved using a structured hexahedral O-grid mesh, which was created with the open-source platform Salome. This mesh naturally provides increased resolution near the vessel wall and in regions of geometric curvature. In Figure 5 the computational mesh for the 50% stenosis is illustrated as seen from Paraview.

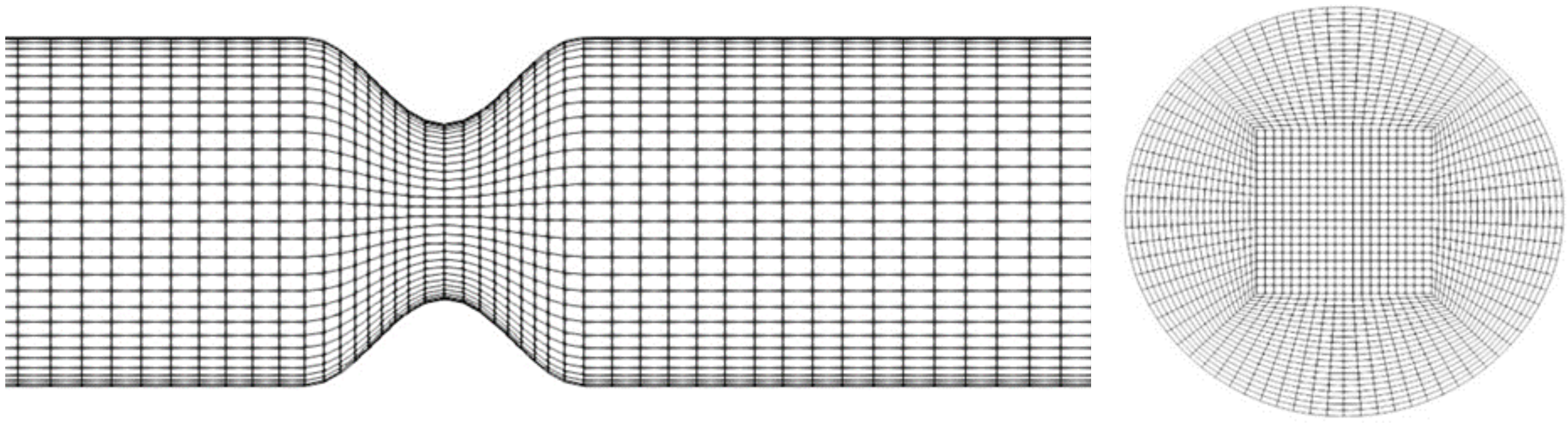

FIG. 5. Computational mesh for the 50% stenosis as depicted in Paraview.

Three meshes of increasing density were examined to assess grid independence. These corresponded to $36 \times 65$, $60 \times 95$, and $75 \times 110$ cells across the vessel diameter and along the stenosis length, respectively, with the medium grid containing 135,000 hexahedral cells, as reported by OpenFOAM checkMesh. Although no explicit local refinement was imposed at the throat or near the wall, the O-grid topology yields sufficiently fine resolution to capture wall shear stress and near-wall velocity gradients in laminar stenotic flows, consistent with common practice in such studies.

Comparisons among the three grids show that maximum velocity, vorticity, and microrotation differ by 2.1%, 4.1%, and 4.5%, respectively, between the coarse and medium grids, and by only 0.7%, 2.2%, and 2.6% between the medium and fine grids. Additional quantities—including centerline velocity profiles, recirculation length, and peak wall shear stress—change by less than 2% between the medium and fine meshes. These results indicate that the medium grid lies within the asymptotic convergence range and adequately resolves both core-flow and near-wall dynamics for the present 3D stenosis configuration. The Courant number was kept below unity throughout the simulations, and residuals for velocity, pressure, and microrotation were reduced below $10^{-8}$, ensuring robust convergence.

### F. Initial and Boundary Conditions

Since the present problem is pressure-driven (as discussed in Section II.C), the inlet pressure was prescribed as a fixed value $P_1$, for the outlet was assigned a common reference pressure $P_2 = 0$, whereas a zeroGradient condition was applied on all vessel walls. No-slip condition was imposed for both velocity and microrotation ($\boldsymbol{v} = \boldsymbol{\omega} = 0$) on all vessel walls (see Equations 10-13), whereas zeroGradient condition were used at the inlet and outlet. For the scalar electric potential $\varphi$, zeroGradient condition were imposed at the walls, inlet and outlet. For more information, complete OpenFOAM dictionaries for velocity, pressure, microrotation, and electric potential are documented can be found in our companion manuscript [52], where the solvers are described.

## III. RESULTS

This section presents numerical results for a 3D micropolar MHD blood by acknowledging and ignoring the effects of MMR. Two stenotic regions are considered, which are frequently found in cardiovascular diseases, one of 50% and 80%. The analysis is done on key flow variables, including streamlines, vorticity, and microrotation contours, as well as velocity and microrotation profiles both within and outside the stenotic region. Results for the wall shear stress and pressure drop are also presented. For comparison, these variables are analyzed across four cases: Newtonian blood flow, micropolar blood flow, MHD blood flow without the MMR effect, and MHD blood flow with the MMR effect included. Two hematocrit values are used, one at $\varphi = 25\%$ and one at $\varphi = 45\%$, while the applied magnetic field is varied using three values, 1 $T$, 3 $T$, and 8 $T$. The streamlines and the contours are plotted at the stenosis section, in the y-z plane.

### A. Results for 50% stenosis

Figure 6 presents the streamlines for the Newtonian and the micropolar blood flow through a 50% stenosis using two hematocrit values, $\varphi = 25\%$ and $\varphi = 45\%$. No external magnetic field is applied to the flow. One can immediately notice that the velocity of the flow increases at the stenosis, reaching its maximum value at the position where the stenosis is most tight; then it returns to its initial inlet value downstream of the stenosis. At the end of the stenosis, two vortices are formed, a phenomenon that is frequently observed in blood stenoses [16, 24]. It is noteworthy to mention that the differences observed in streamlines between the Newtonian and the micropolar modeling are negligible, even for high hematocrit values. This was expected due to the relatively high $\lambda$ value corresponding to the size of the artery, which minimizes any micropolar effects on velocity.

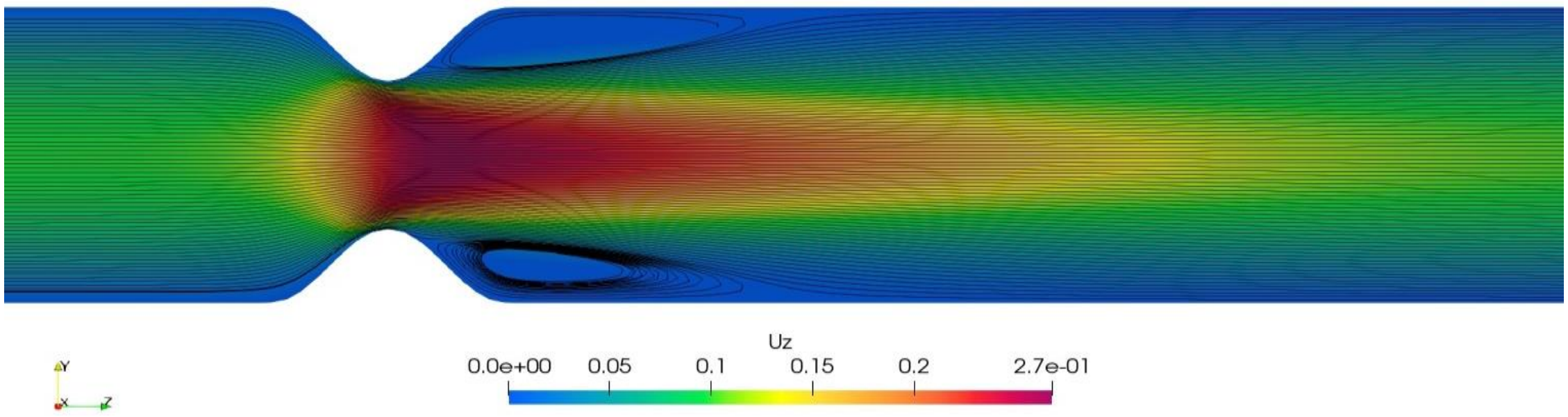

*(a)*

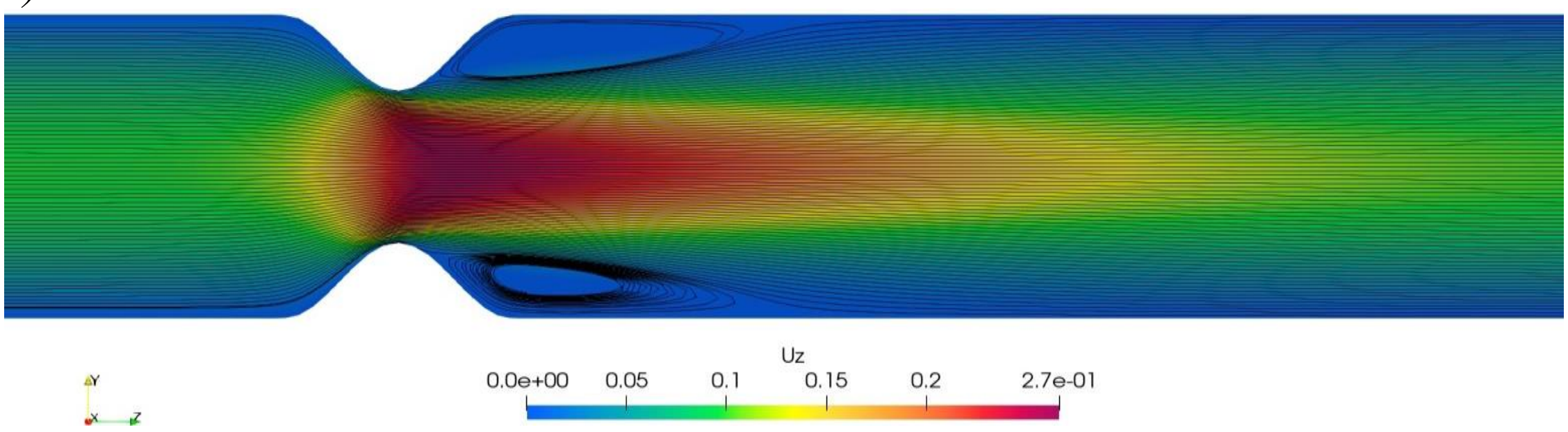

*(b)*

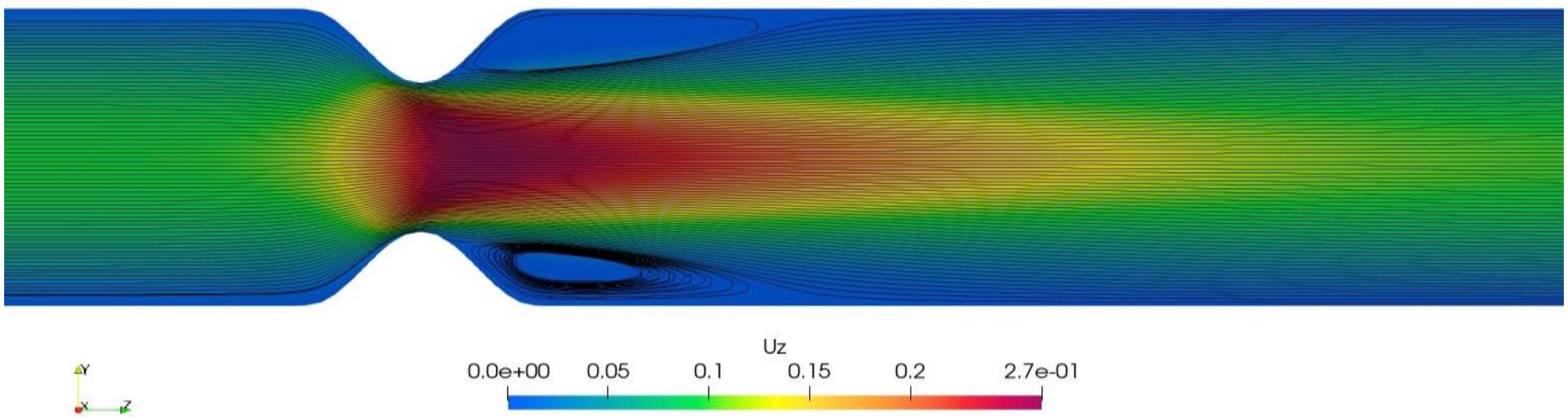

*(c)*

FIG. 6. Streamlines for blood flow through a 50% stenosis using (a) Newtonian modelling and micropolar modelling without an applied magnetic field with hematocrit of (b) $\varphi = 25\%$ and (c) $\varphi = 45\%$.

Figure 7 presents the vorticity contours for the Newtonian and the micropolar blood flow through a 50% stenosis using two hematocrit values, $\varphi = 25\%$ and $\varphi = 45\%$, with no external applied magnetic field. The vorticity shows an axisymmetric profile, with its maximum and minimum values appearing inside the stenotic region. The maximum vorticity value is located at the lower wall of the stenosis, while its minimum value occurs at the upper wall. Similar to the streamlines, the vorticity isolines are disturbed downstream of the stenosis. Here, the micropolar effect appears more pronounced, with the maximum and minimum vorticity values decreasing in absolute terms, both from the transition from the Newtonian to the micropolar profile and as the hematocrit increases. Also, the length of the disturbances downstream of the stenosis decreases. All of this was expected, as micropolar phenomena directly affect the vorticity through microrotation, ensuring that the total angular momentum remains constant.

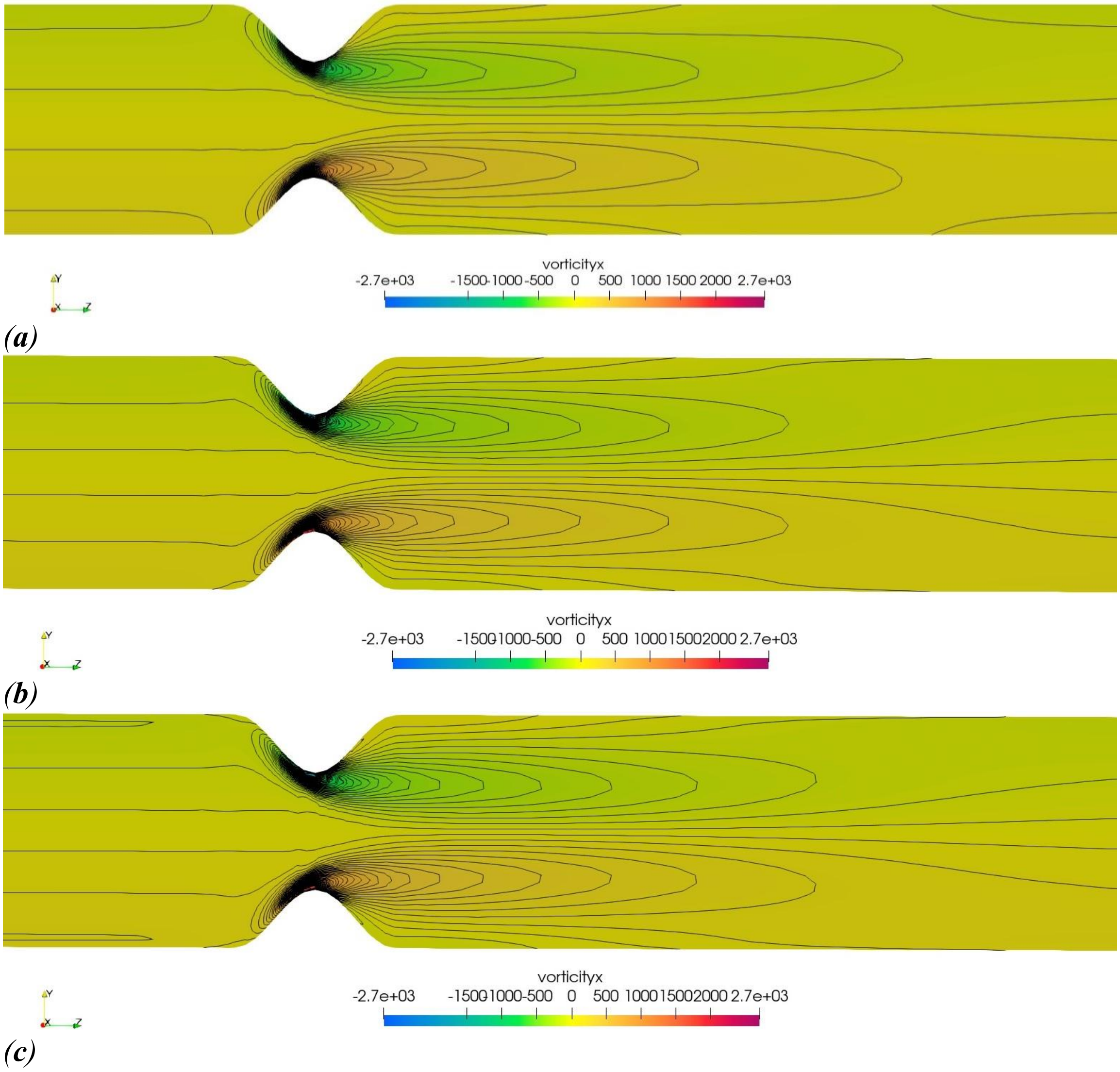

FIG. 7. Vorticity contour plots for blood flow through a 50% stenosis using (a) Newtonian modelling and micropolar modelling without an applied magnetic field with hematocrit of (b) $\varphi = 25\%$ and (c) $\varphi = 45\%$.

Figure 8 presents the microrotation contours for the micropolar blood flow through a 50% stenosis using two hematocrit values, $\varphi = 25\%$ and $\varphi = 45\%$, with no external applied magnetic field. As expected, the microrotation closely resembles the vorticity and exhibits an axisymmetric profile, with its maximum and minimum values situated within the stenotic region. The maximum microrotation value is located at the lower wall of the stenosis, while its minimum value occurs at the upper wall. The microrotation isolines are disrupted downstream of the stenosis. As the hematocrit increases, the maximum and minimum microrotation values are decreased in absolute terms, and the isolines become further disturbed.

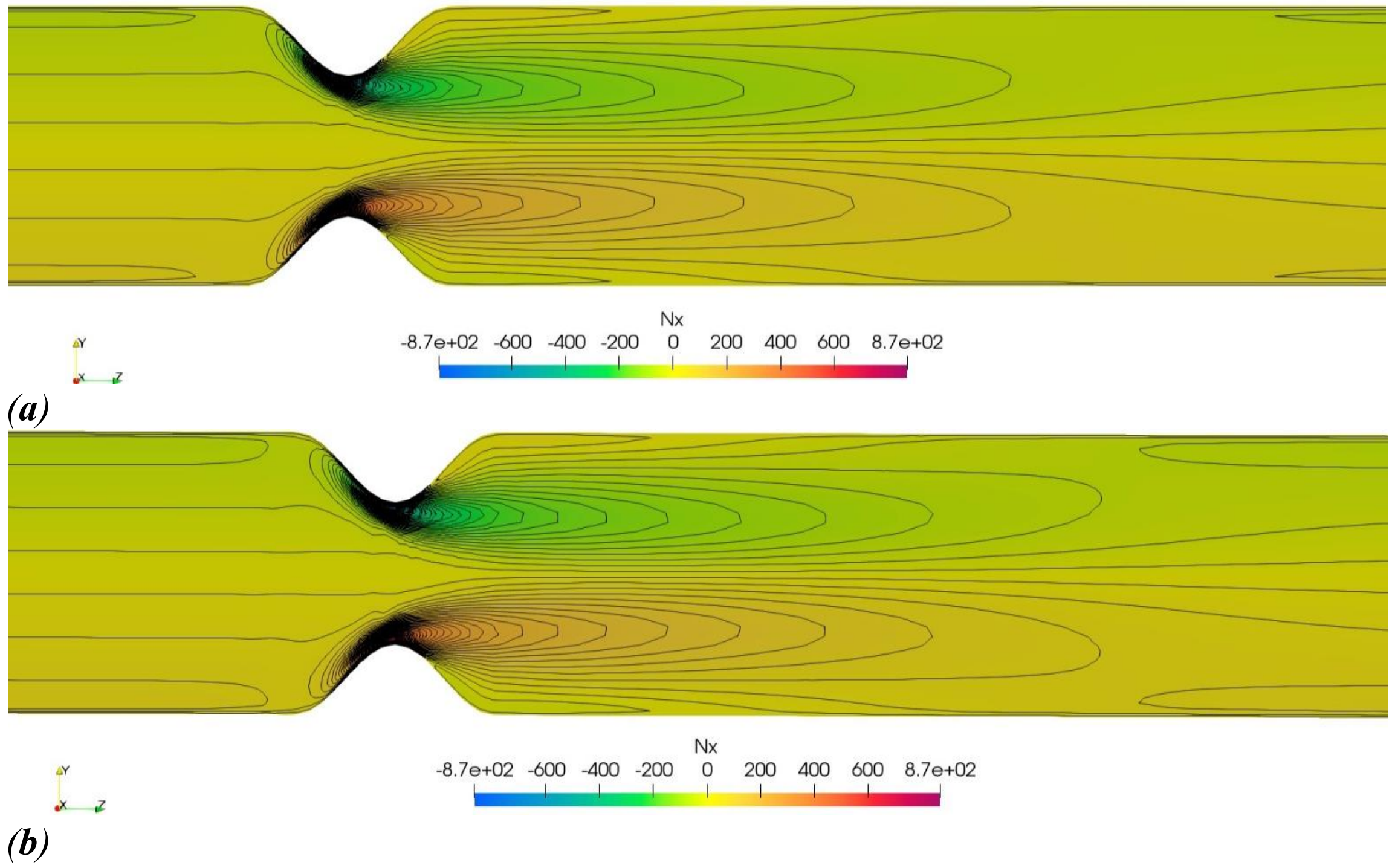

FIG. 8. Microrotation contour plots for the micropolar blood flow through a 50% stenosis without an applied magnetic field with hematocrit of (a) $\varphi = 25\%$ and (b) $\varphi = 45\%$.

Figure 9 illustrates the streamlines for the 50% stenosis using the MHD micropolar fluid theory both ignoring and acknowledging the MMR. The hematocrit is held constant at $\varphi = 25\%$, while the applied magnetic field is varied at $1\ T$, $3\ T$, and $8\ T$. It is evident that the streamlines for both micropolar blood flow and MHD micropolar blood flow through stenosis show no significant differences when the MMR effect is not considered, regardless of the strength of the applied magnetic field. This outcome was expected, as the Lorentz force minimally impacts the stenosis due to blood's relatively low electrical conductivity. However, when the MMR term is included, the maximum velocity within the stenotic region decreases - especially as the magnetic field strength increases (maximum decrease of 11.1% at $8\ T$) - and the vortices downstream are noticeably weakened. This behavior aligns with the previous studies by Aslani et al. [15, 34-36], which have demonstrated the dampening effect of micromagnetorotation.

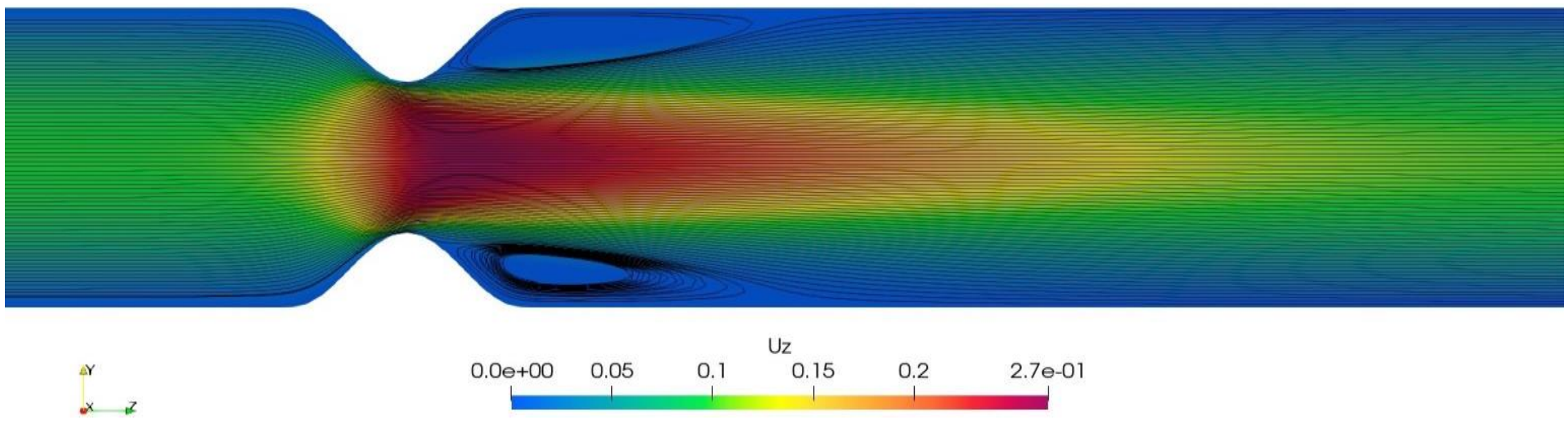

*(a.i)*

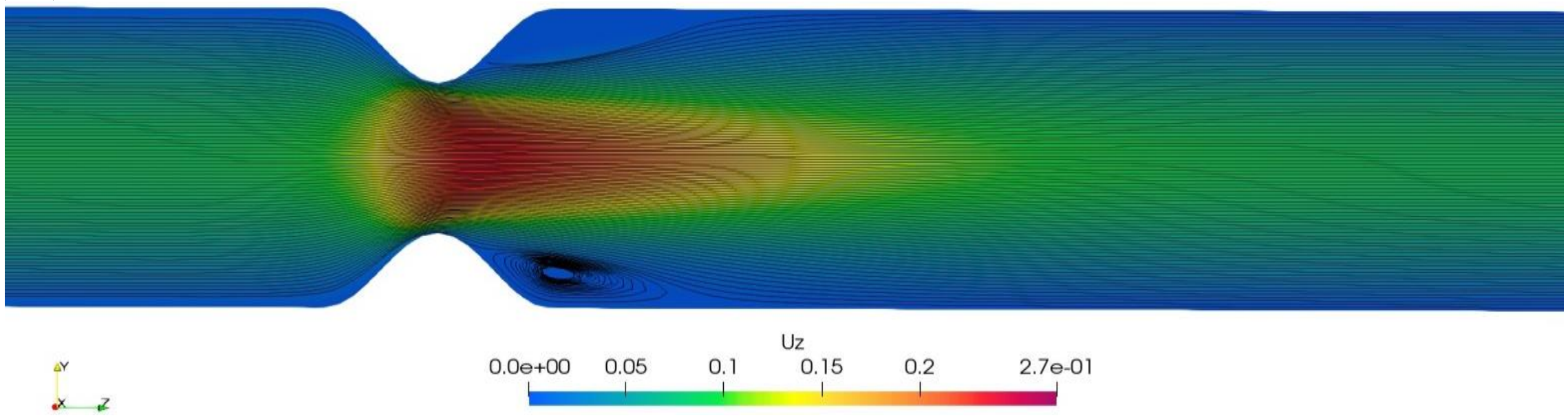

*(a.ii)*

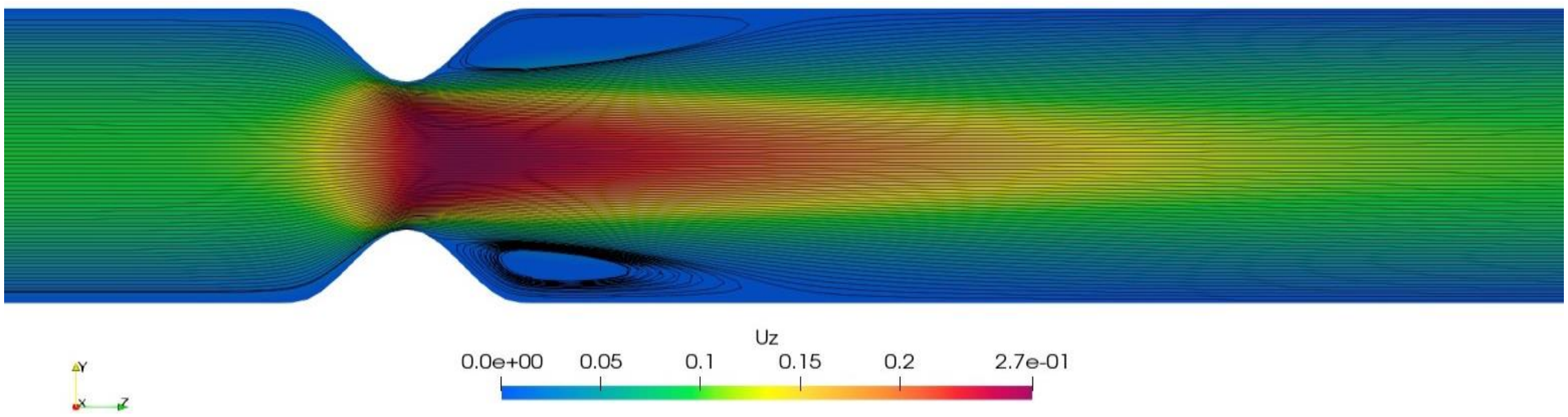

*(b.i)*

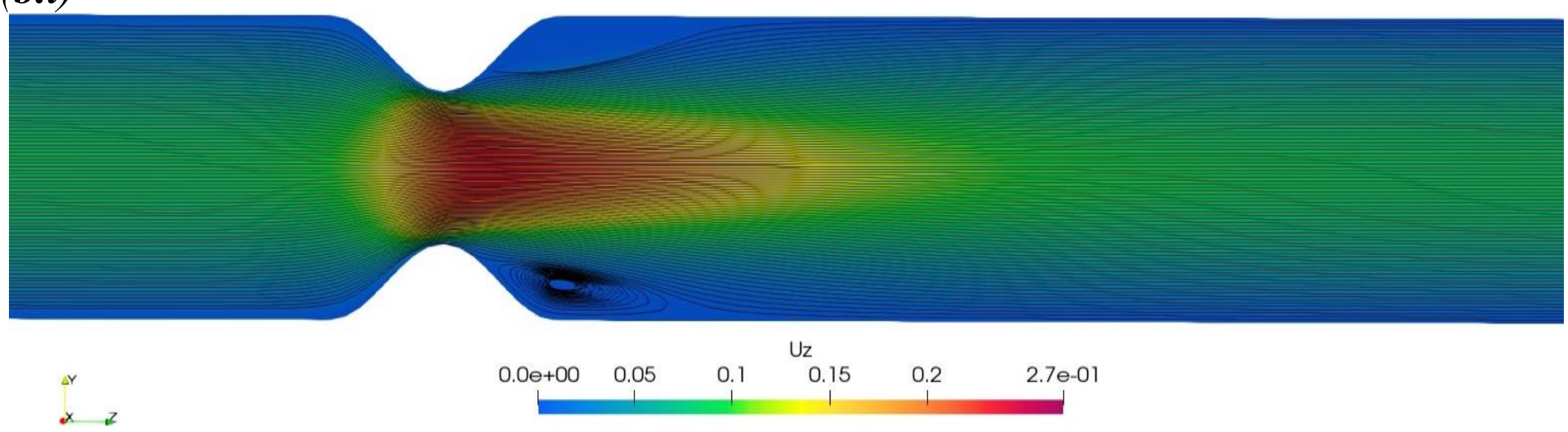

*(b.ii)*

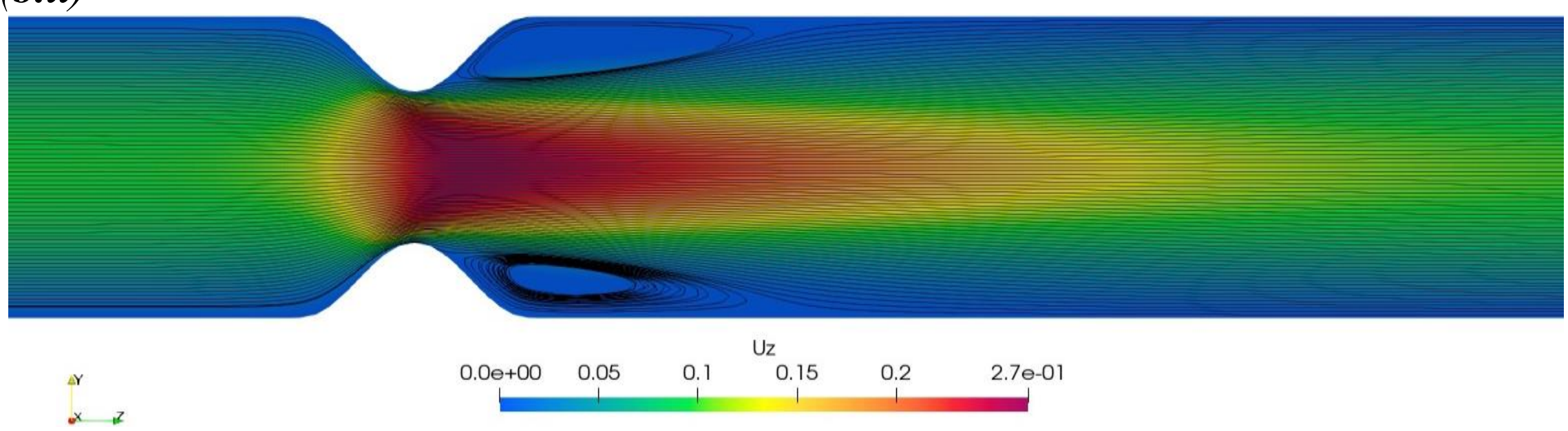

*(c.i)*

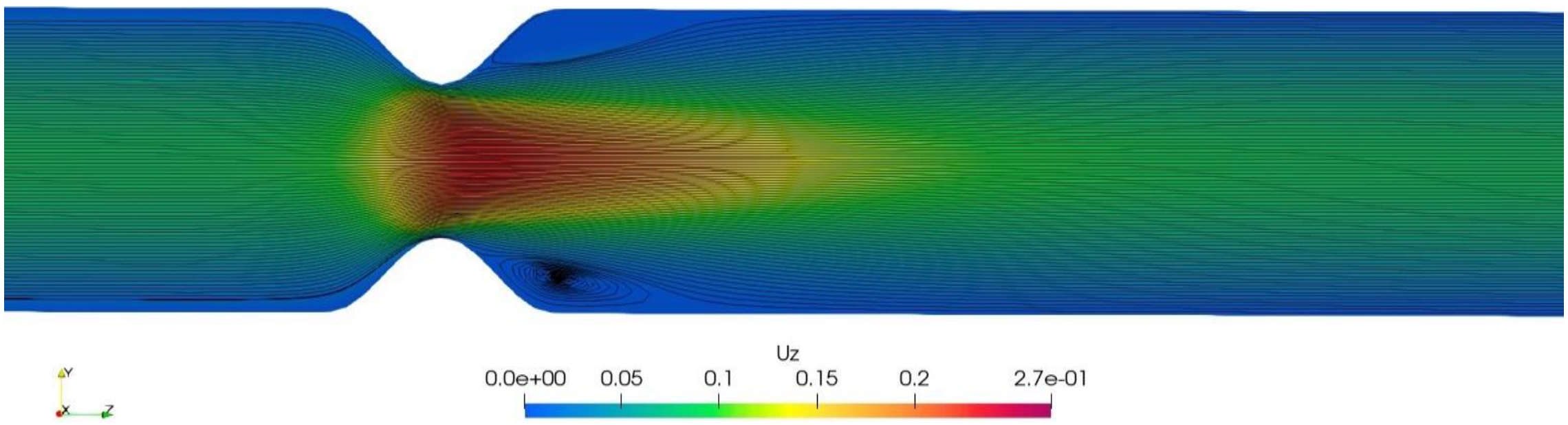

*(c.ii)*

FIG. 9. Streamlines for blood flow through a 50% stenosis using MHD micropolar modeling (i) without acknowledging MMR and (ii) considering MMR for an applied magnetic field of (a) $1\ T$, (b) $3\ T$, and (c) $8\ T$ and hematocrit of $\varphi = 25\%$.

Figure 10 presents the streamlines for the 50% stenosis using the MHD micropolar fluid theory, considering and disregarding MMR with a hematocrit of $\varphi = 45\%$. Once again, the applied magnetic field is varied at $1\ T$, $3\ T$, and $8\ T$. Similar to the scenario where $\varphi$ equals 25%, there are no noticeable differences between the streamlines of the micropolar blood flow and the MHD micropolar blood flow without MMR across all considered values of the applied magnetic field. In contrast, when the MMR term is included, the maximum velocity within the stenotic region decreases, and the vortices downstream of the stenotic region are dampened to a greater extent compared to the case where $\varphi$ equals 25%. Specifically, the maximum velocity is reduced by 22.2% compared to the case where $\varphi$ equals 25%, where the maximum reduction reached 11.1%. This outcome was expected, as the increase in hematocrit is directly related to the increase in the micropolar effect parameter, $\varepsilon$, which enhances the impact of micromagnetorotation.

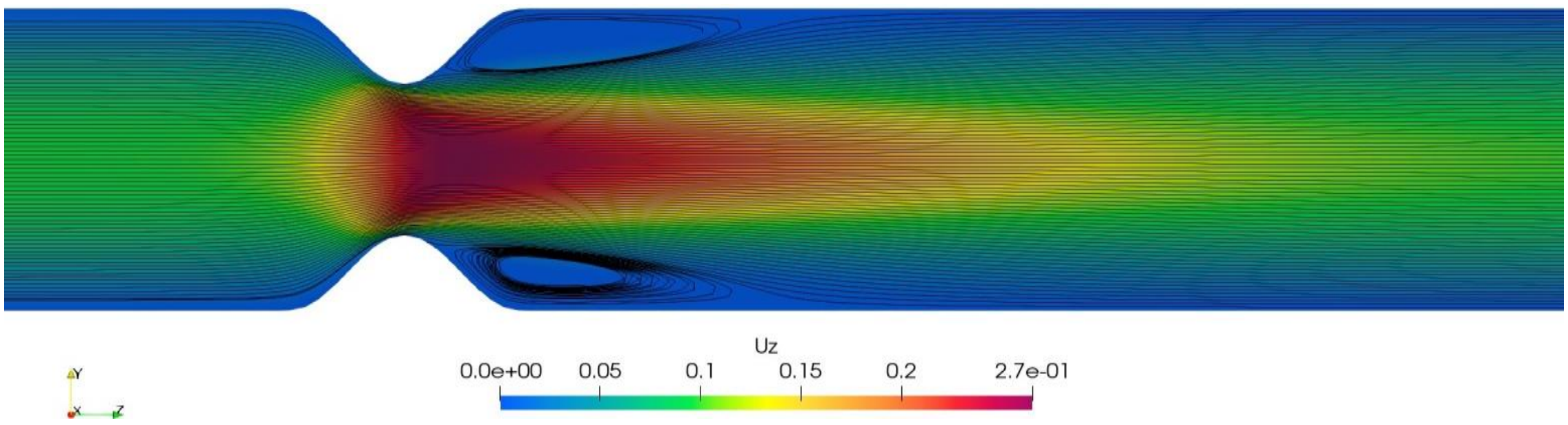

*(a.i)*

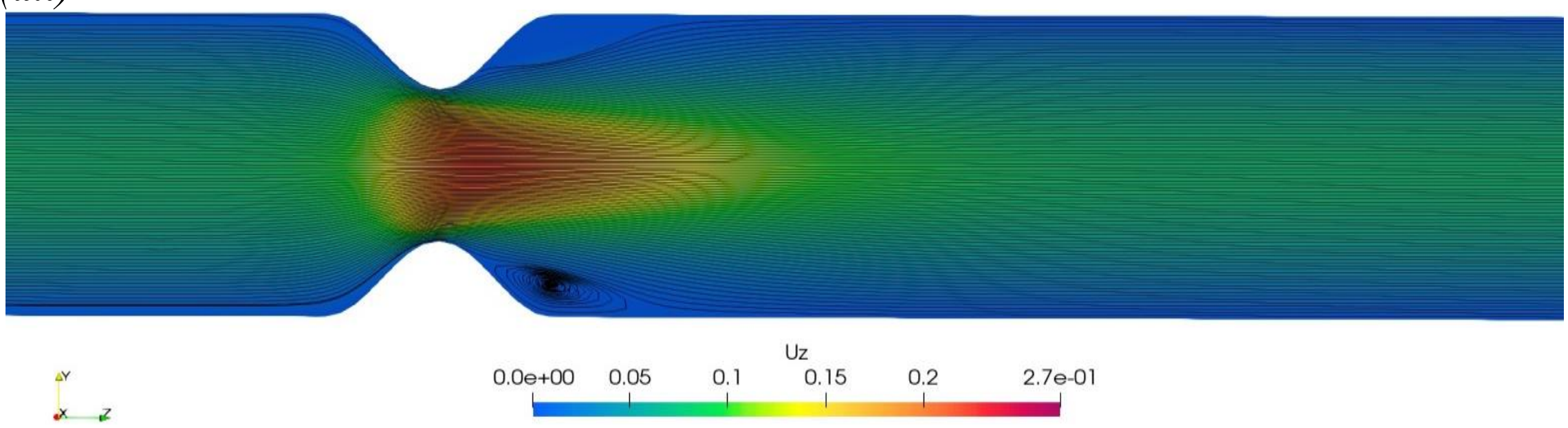

*(a.ii)*

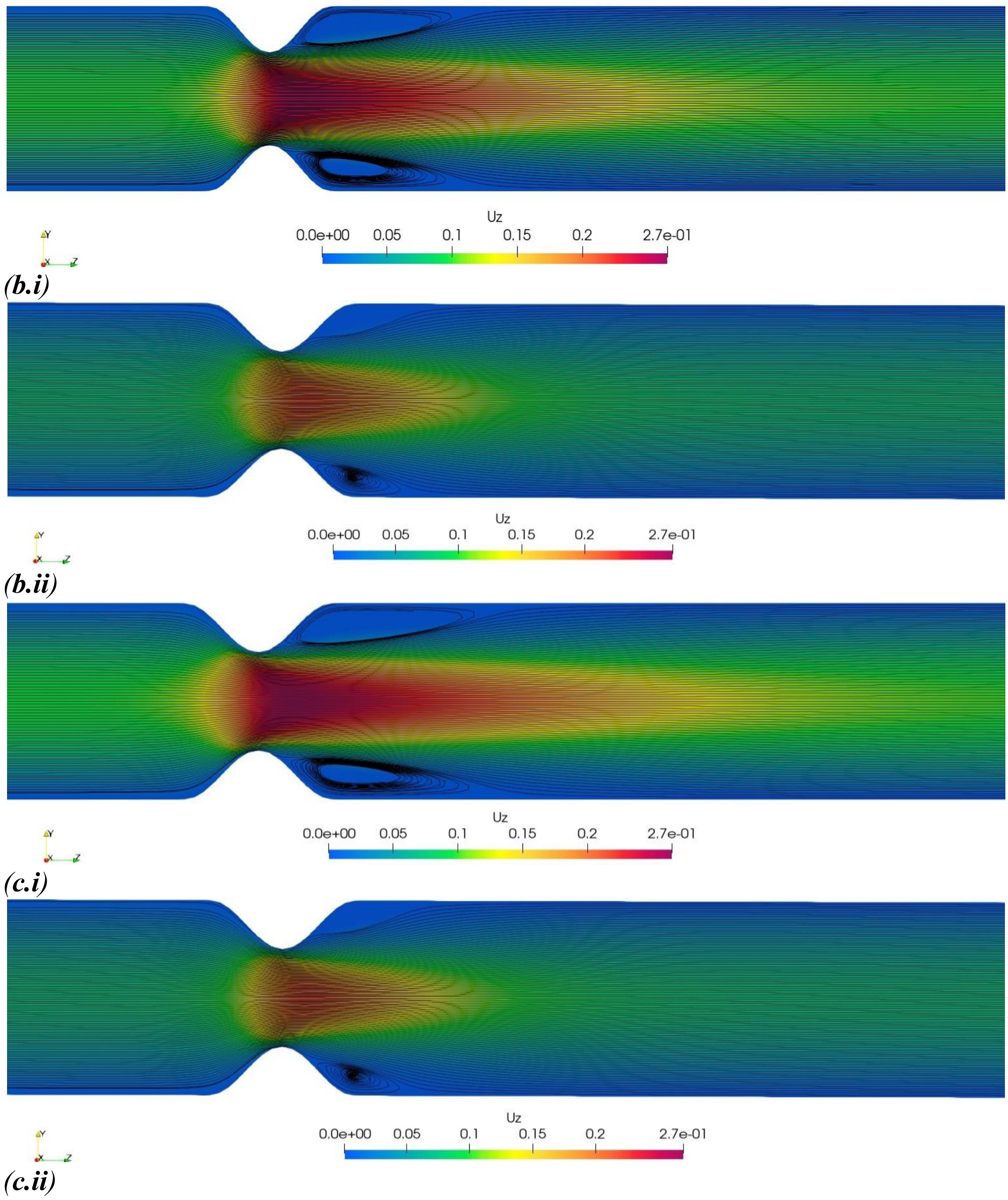

FIG. 10. Streamlines for blood flow through a 50% stenosis using MHD micropolar modeling (i) without acknowledging MMR and (ii) considering MMR for an applied magnetic field of (a) 1 $T$, (b) 3 $T$, and (c) 8 $T$ and hematocrit of $\varphi = 45\%$.

Figure 11 presents the vorticity contours for the 50% stenosis using the MHD micropolar fluid theory, both ignoring and acknowledging MMR. Here, hematocrit is held constant at $\varphi = 25\%$ and the applied magnetic field is varied at 1 $T$, 3 $T$, and 8 $T$. Similar to the streamlines, the vorticity contours for the micropolar blood flow and the MHD micropolar blood flow through stenosis without MMR show no significant differences, regardless of the strength of the applied magnetic field. However, when the MMR term is included, the maximum and minimum vorticity values decrease in absolute terms, with a reduction of 15% at 8 $T$. Furthermore, the length of the disturbances downstream of the stenosis decreases due to the dampening effect of the MMR.

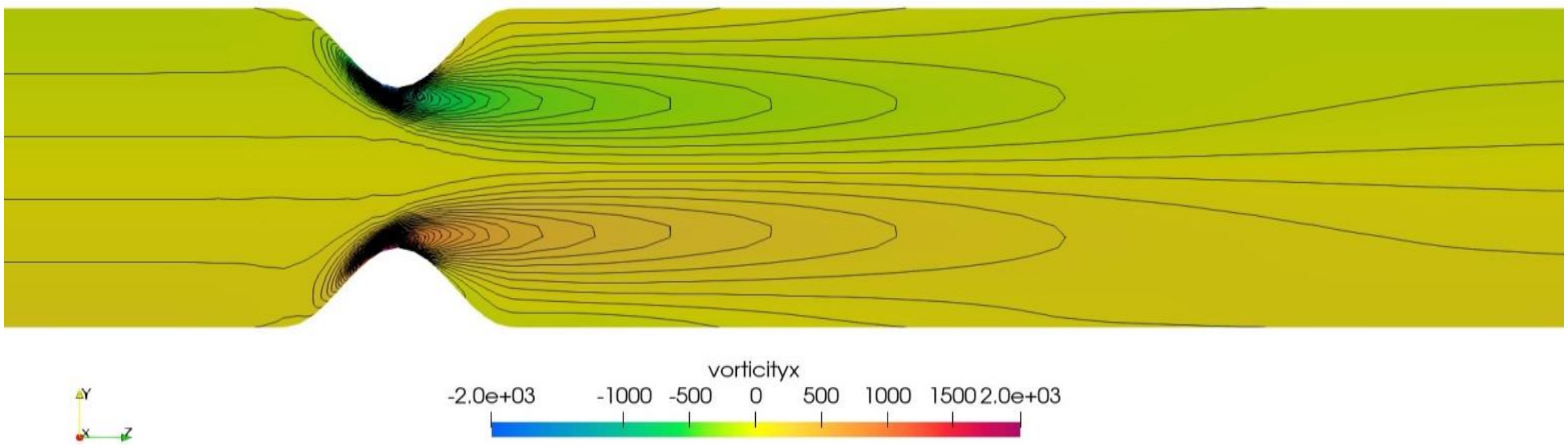

***(a.i)***

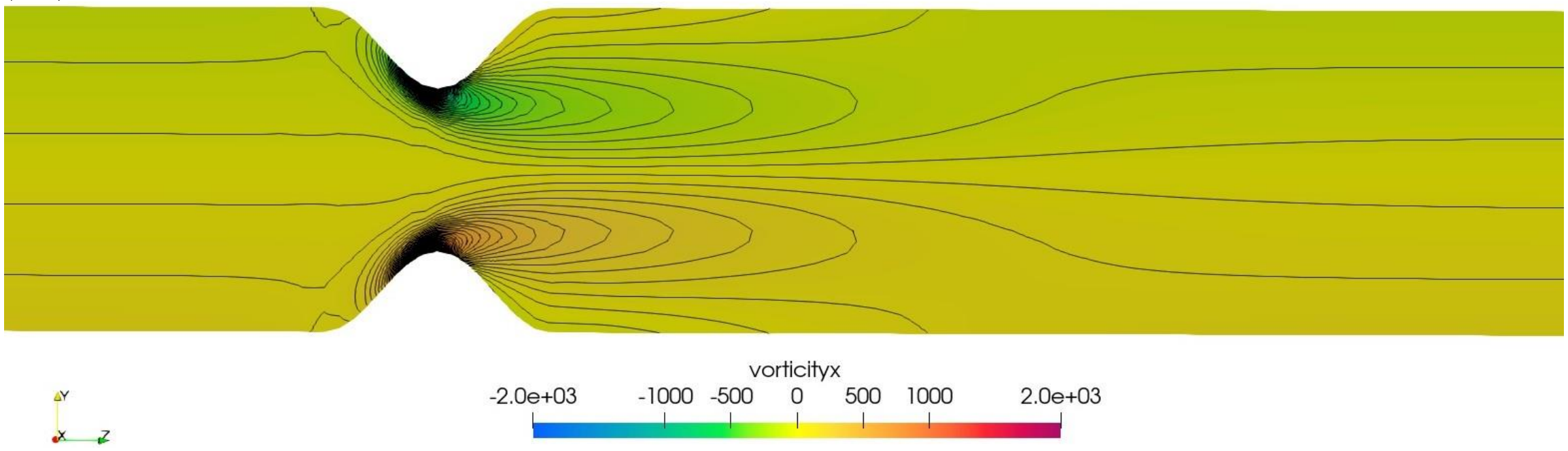

***(a.ii)***

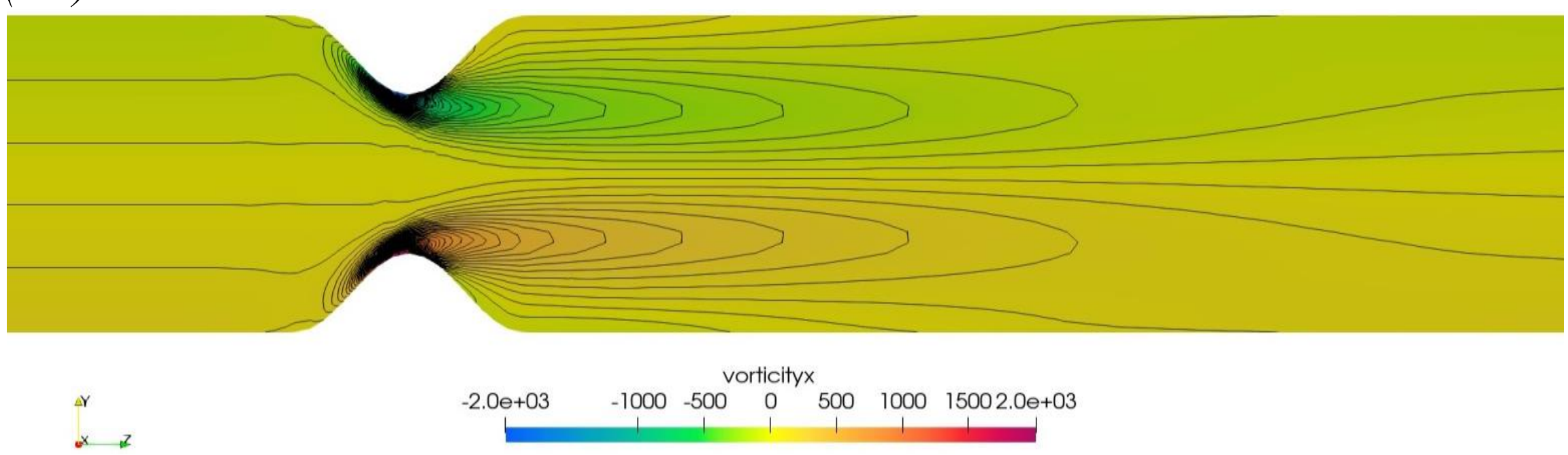

***(b.i)***

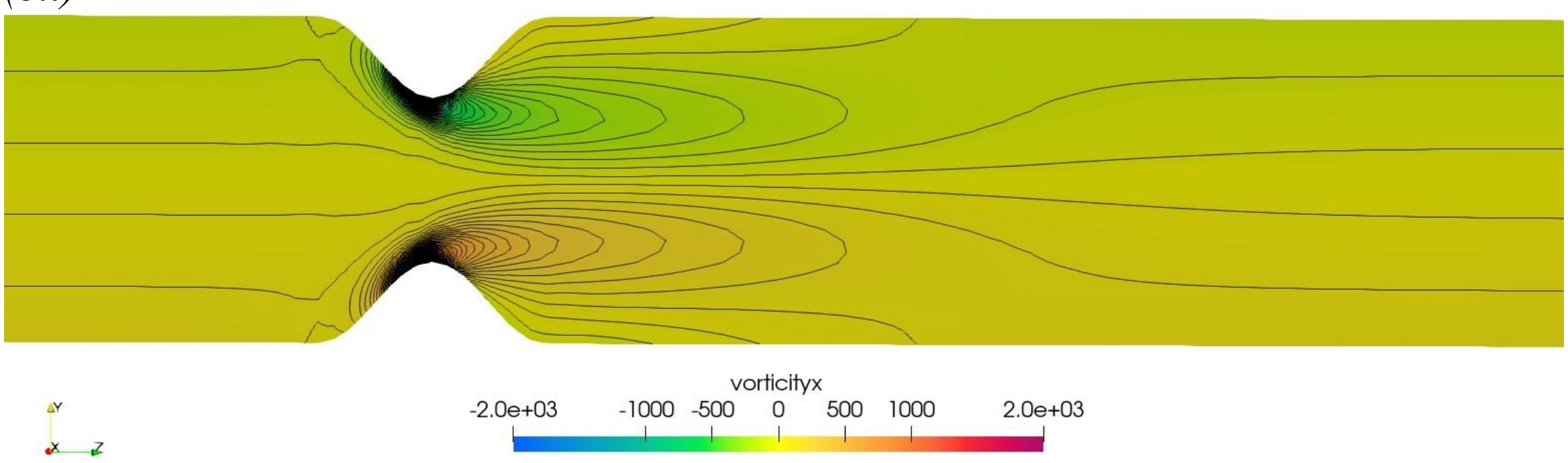

***(b.ii)***

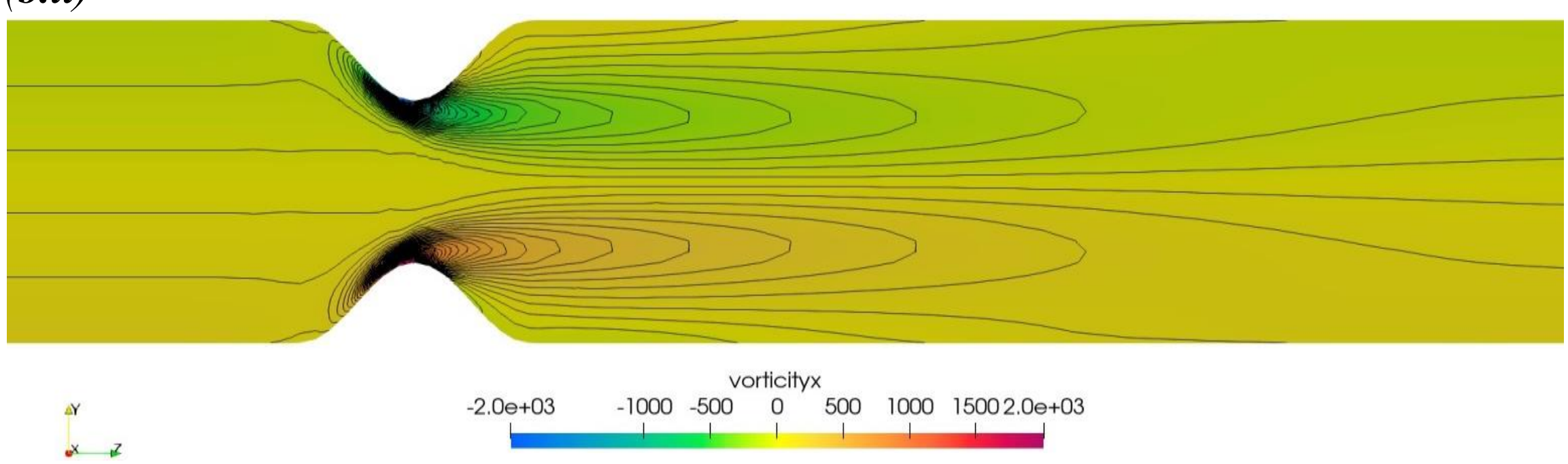

***(c.i)***

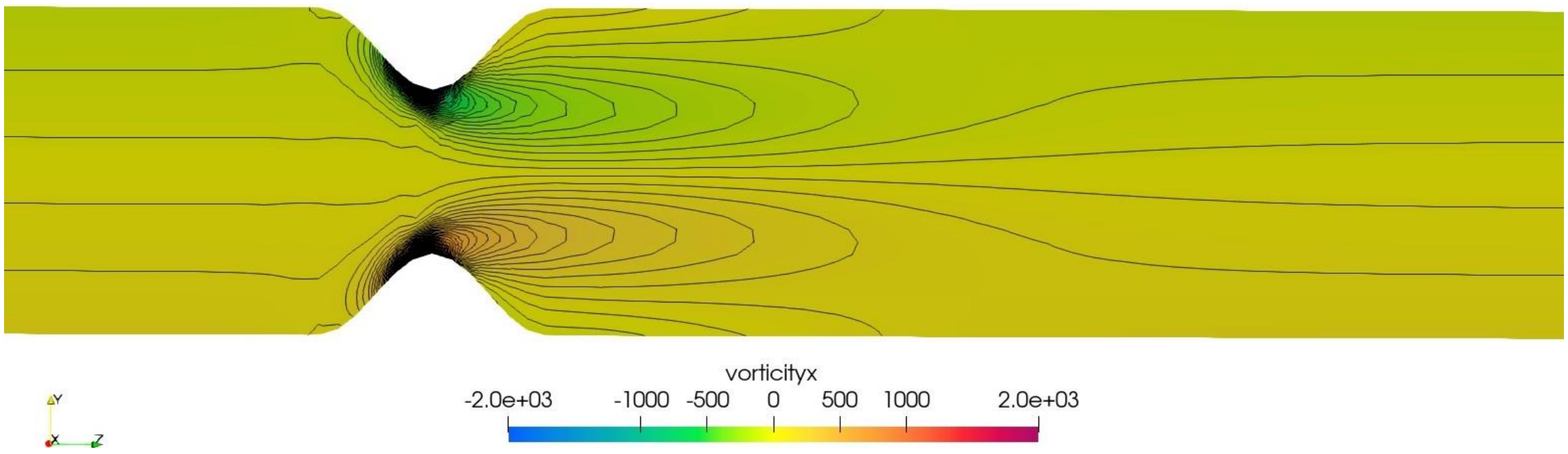

***(c.ii)***

FIG. 11. Vorticity contour plots for blood flow through a 50% stenosis using MHD micropolar modeling (i) without acknowledging MMR and (ii) considering MMR for an applied magnetic field of (a) 1 $T$, (b) 3 $T$, and (c) 8 $T$ and hematocrit of $\varphi = 25\%$.

Figure 12 illustrates the vorticity contours for the 50% stenosis using the MHD micropolar fluid theory, both considering and ignoring MMR with a hematocrit of $\varphi = 45\%$. As in the previous cases, the applied magnetic field is varied at 1 $T$, 3 $T$, and 8 $T$. Again, no significant differences exist between the vorticity contours of the micropolar blood flow and the MHD micropolar blood flow without MMR for all considered values of the applied magnetic field. However, due to the increase in hematocrit, when the MMR term is included, the maximum and minimum vorticity values decrease further, with a reduction of 22.22% at 8 $T$. Furthermore, the length of the disturbances downstream of the stenosis also decreases further.

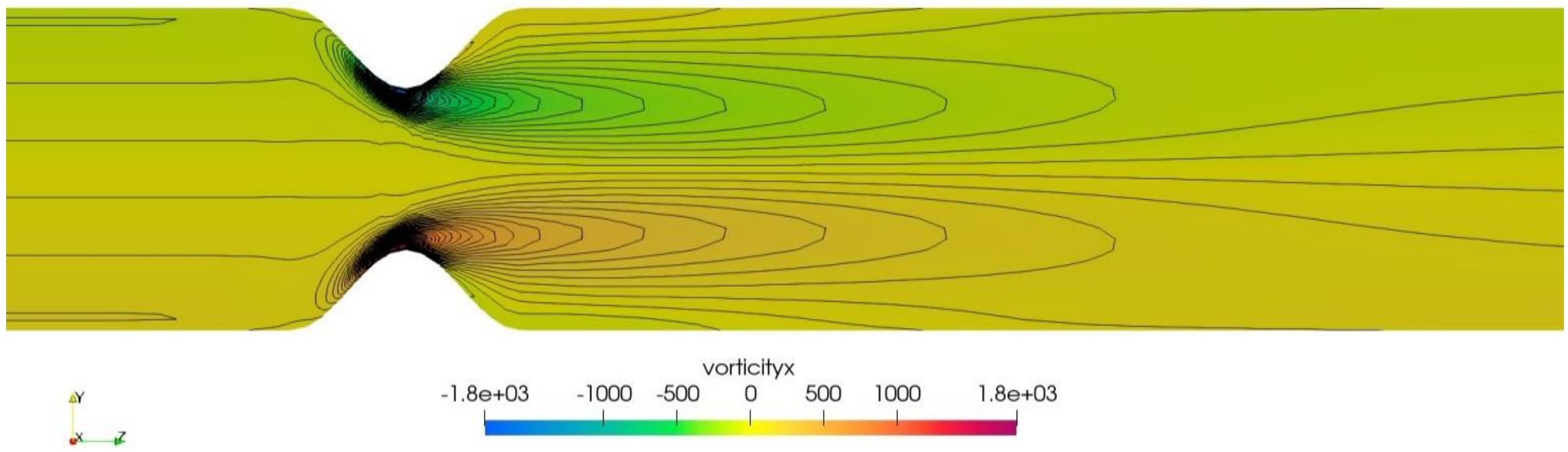

***(a.i)***

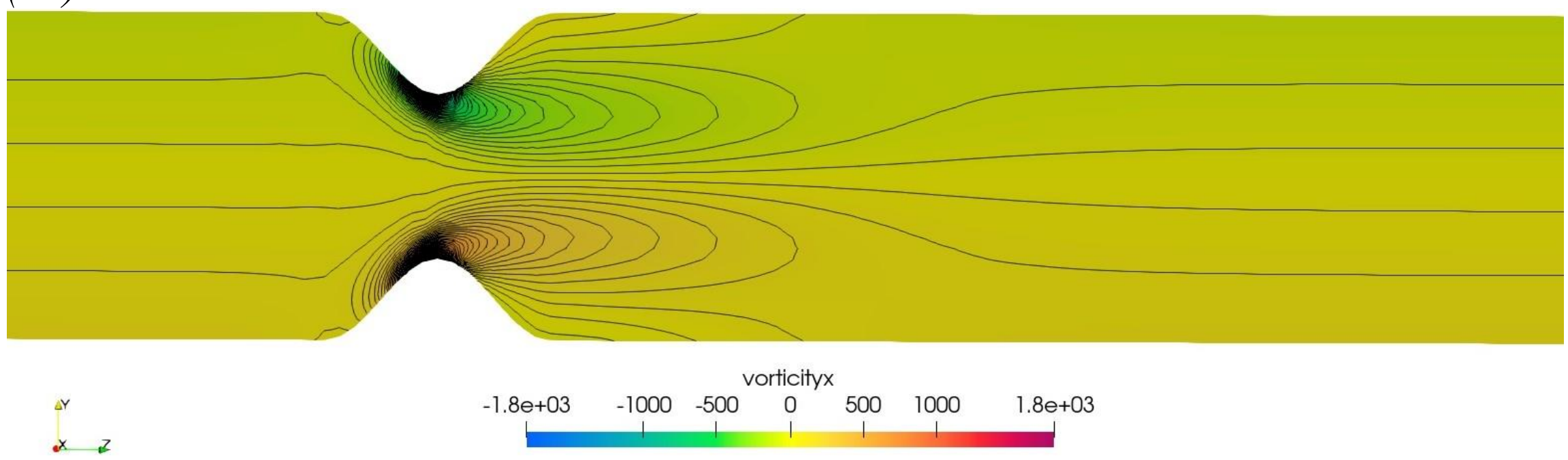

***(a.ii)***

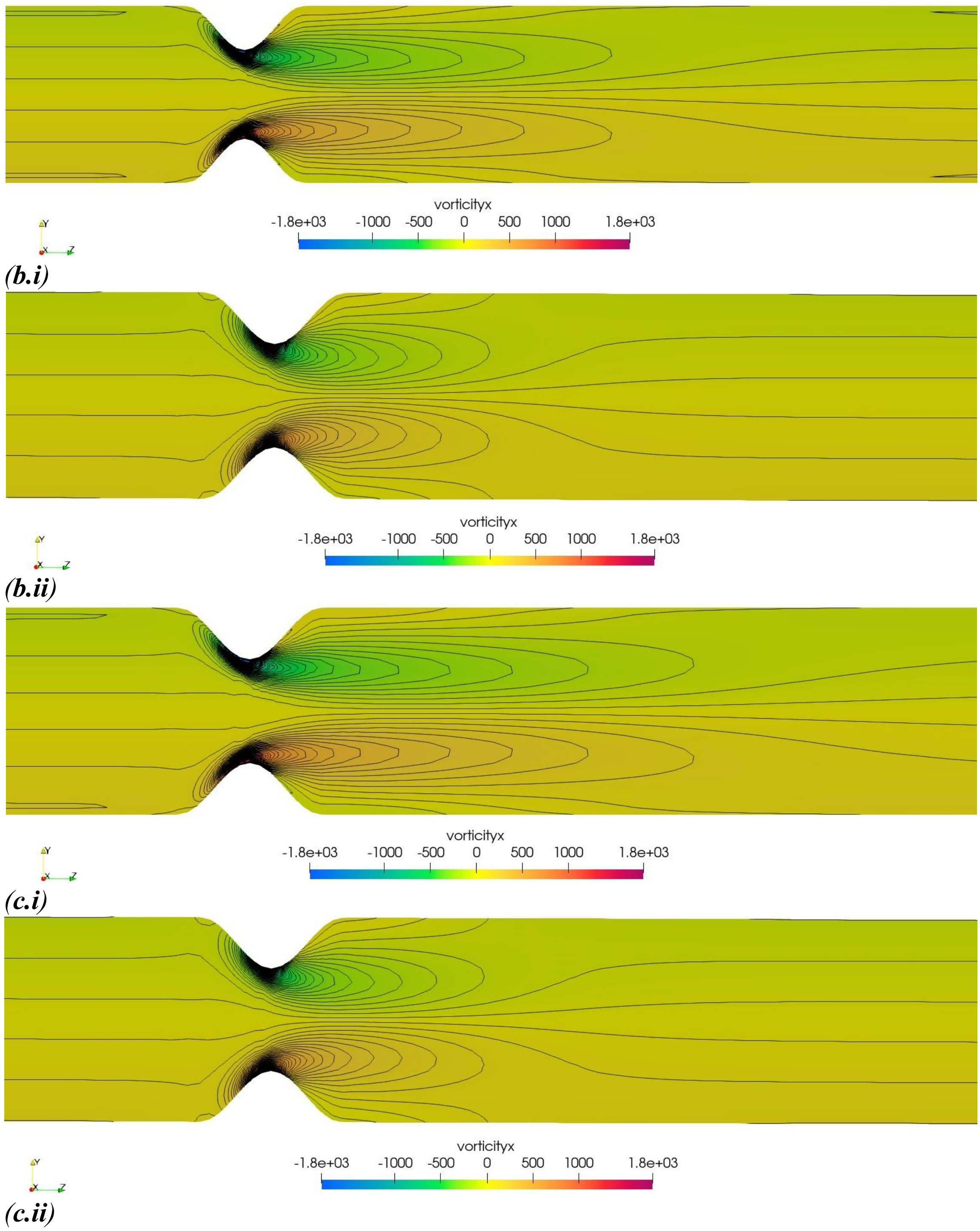

FIG. 12. Vorticity contour plots for blood flow through a 50% stenosis using MHD micropolar modeling (i) without acknowledging MMR and (ii) considering MMR for an applied magnetic field of (a) 1 $T$, (b) 3 $T$, and (c) 8 $T$ and hematocrit of $\varphi = 45\%$.

Figure 13 illustrates the microrotation contours for the 50% stenosis using the MHD micropolar fluid theory, both ignoring and acknowledging MMR. Once again, the hematocrit is held constant at $\varphi = 25\%$, and the applied magnetic field is varied at 1 $T$, 3 $T$, and 8 $T$. Similar to the vorticity contours, the microrotation contours for the micropolar blood flow and the MHD micropolar blood flow through stenosis without MMR show no significant differences, even as the strength of the applied magnetic field increases. However, when the MMR term is included, the maximum and minimum vorticity values decrease significantly, reducing by 94% at 1 $T$, 98% at 3 $T$, and 99% at 8 $T$. Additionally, the length of the disturbances downstream of the stenosis decreases substantially. Physically, these results are explained by the alignment of the erythrocytes with the externally applied magnetic field, which does not allow any internal rotation. Note that a common color scale was used for the streamlines and vorticity contours to enable direct comparison between cases, whereas separate color scales were employed for the microrotation contours due to the large difference in magnitude when MMR is included.

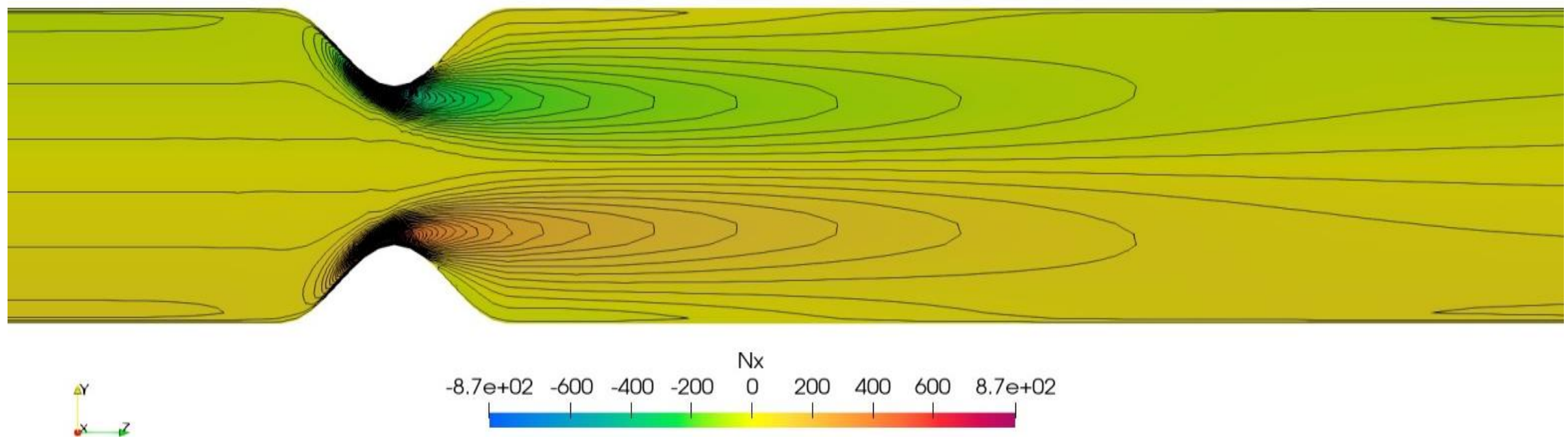

***(a.i)***

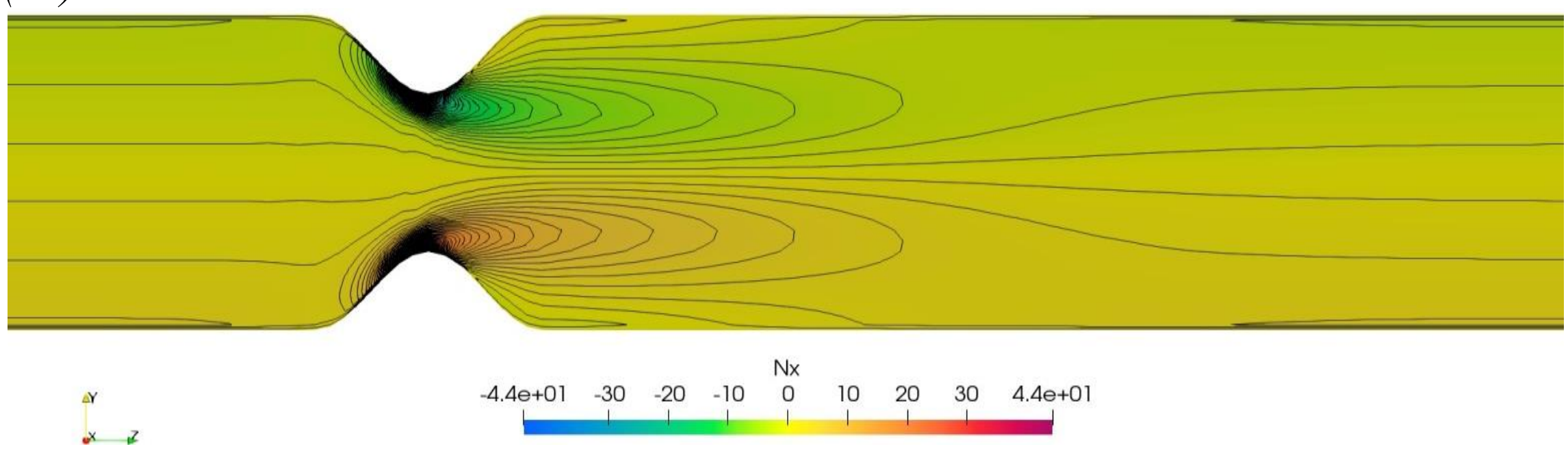

***(a.ii)***

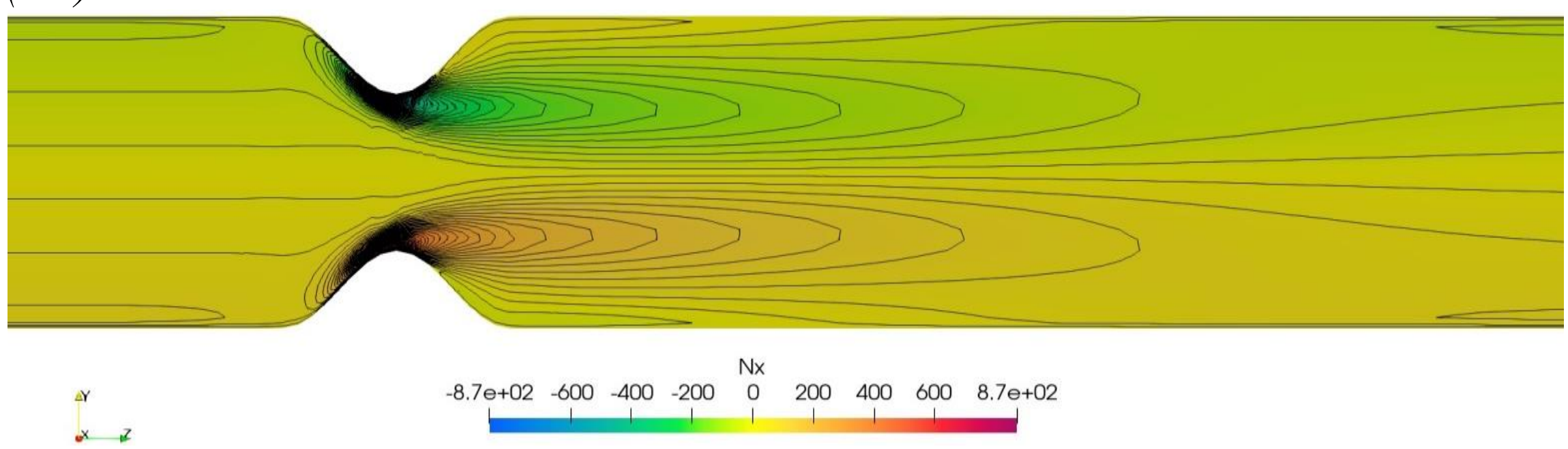

***(b.i)***

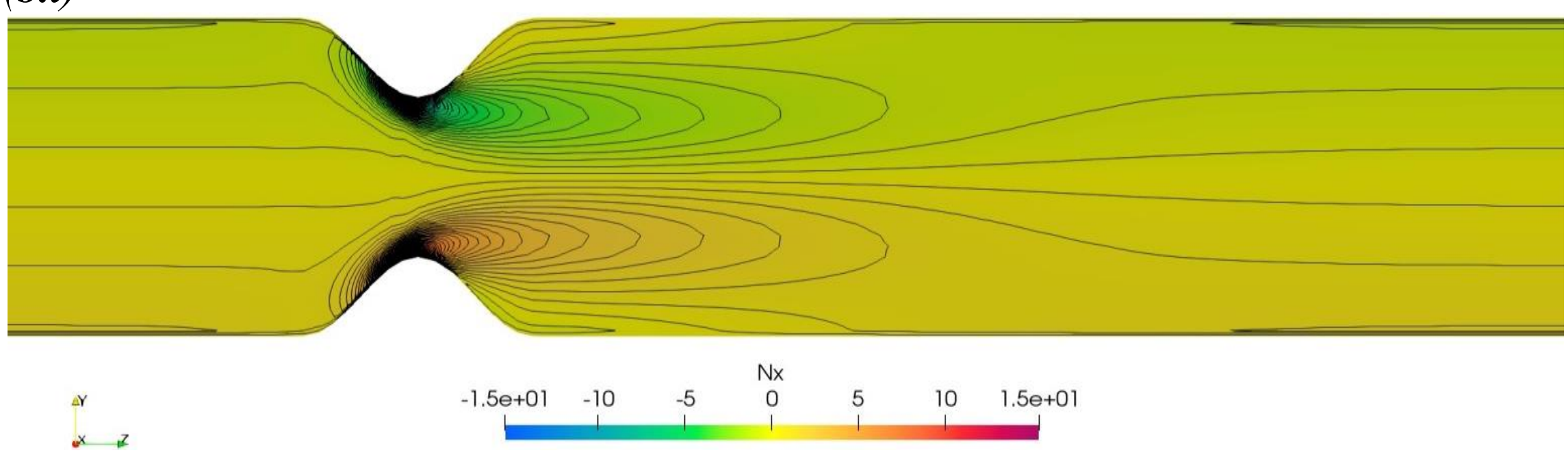

***(b.ii)***

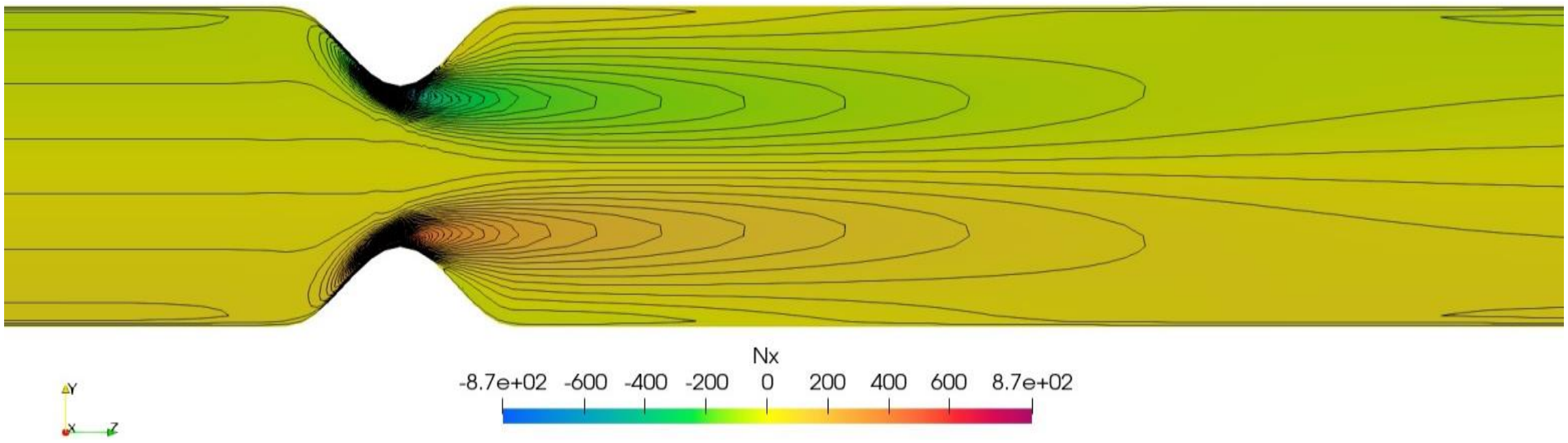

*(c.i)*

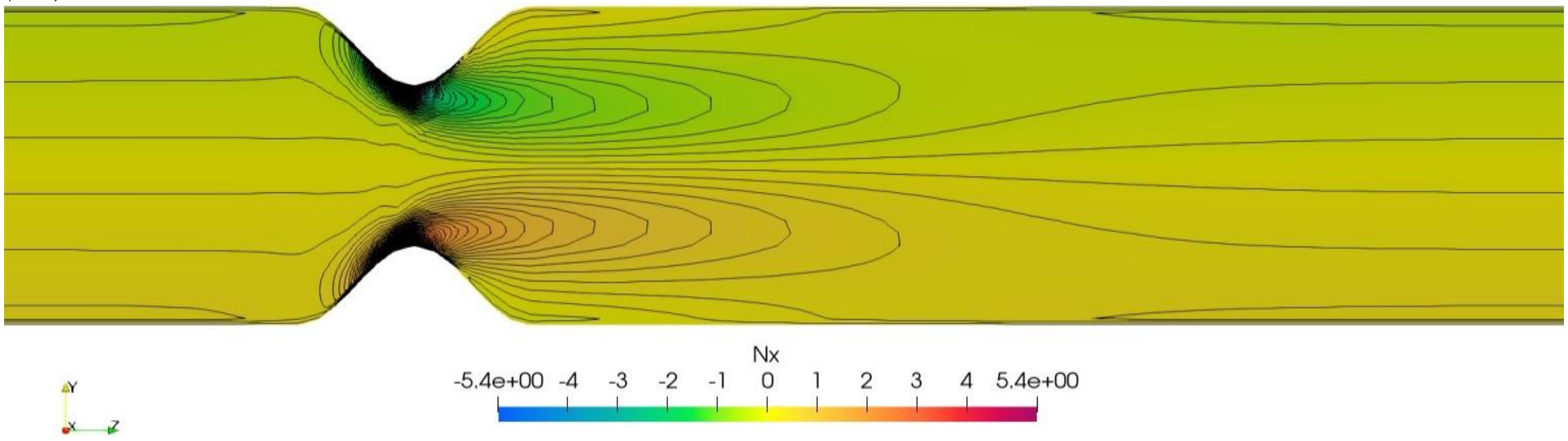

*(c.ii)*

FIG. 13. Microrotation contour plots for blood flow through a 50% stenosis using MHD micropolar modeling (i) without acknowledging MMR and (ii) considering MMR for an applied magnetic field of (a) 1 $T$, (b) 3 $T$, and (c) 8 $T$ and hematocrit of $\varphi = 25\%$.

Figure 14 presents the microrotation contours for the 50% stenosis using the MHD micropolar fluid theory, both considering and ignoring MMR with a hematocrit of $\varphi = 45\%$ and an applied magnetic field of 1 $T$, 3 $T$, and 8 $T$. Similar to the case of $\varphi = 25\%$, no significant differences are observed between the microrotation contours of the micropolar blood flow and the MHD micropolar blood flow without MMR for all considered values of the applied magnetic field. However, with the increase in hematocrit, including the MMR term, a significant decrease in the maximum and minimum microrotation values is found, which are reduced by 95% at 1 $T$, 98% at 3 $T$, and 99.4% at 8 $T$. Once again, the disturbances downstream of the stenosis are substantially disrupted.

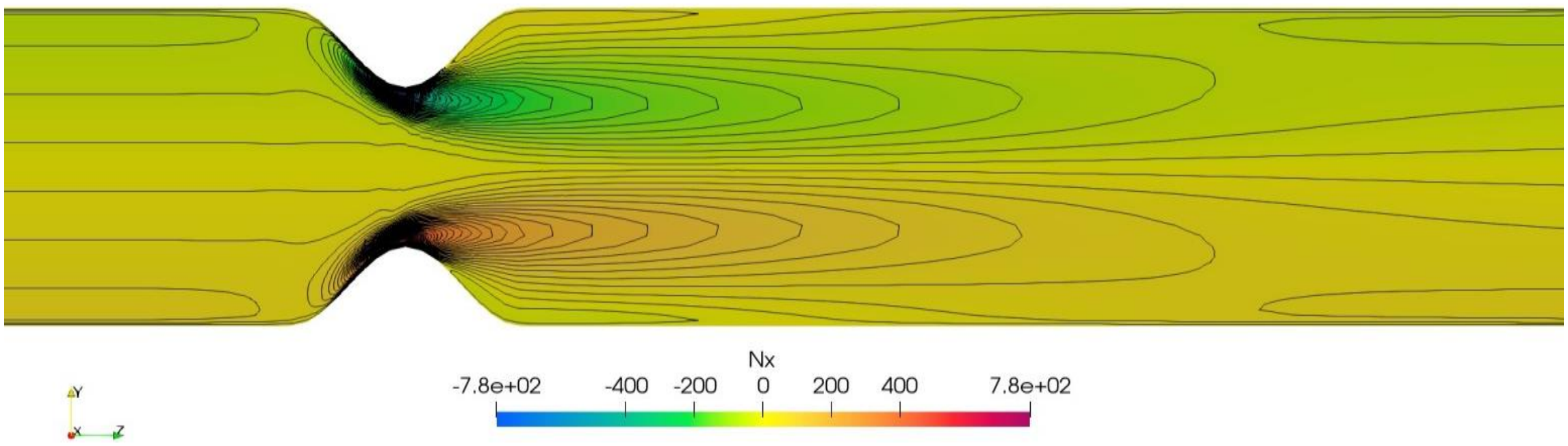

*(a.i)*

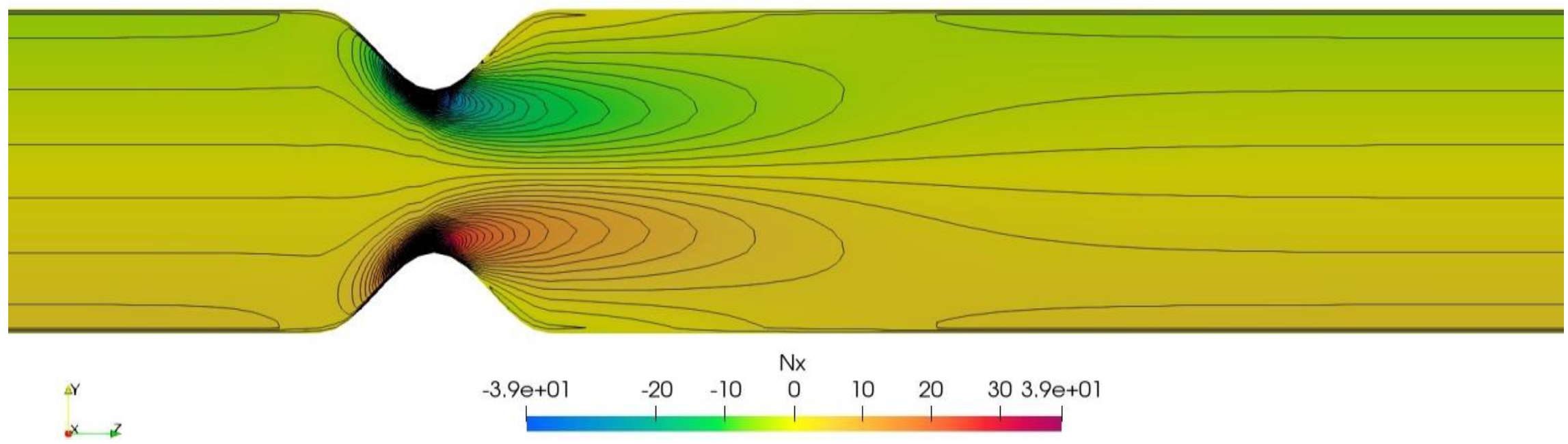

*(a.ii)*

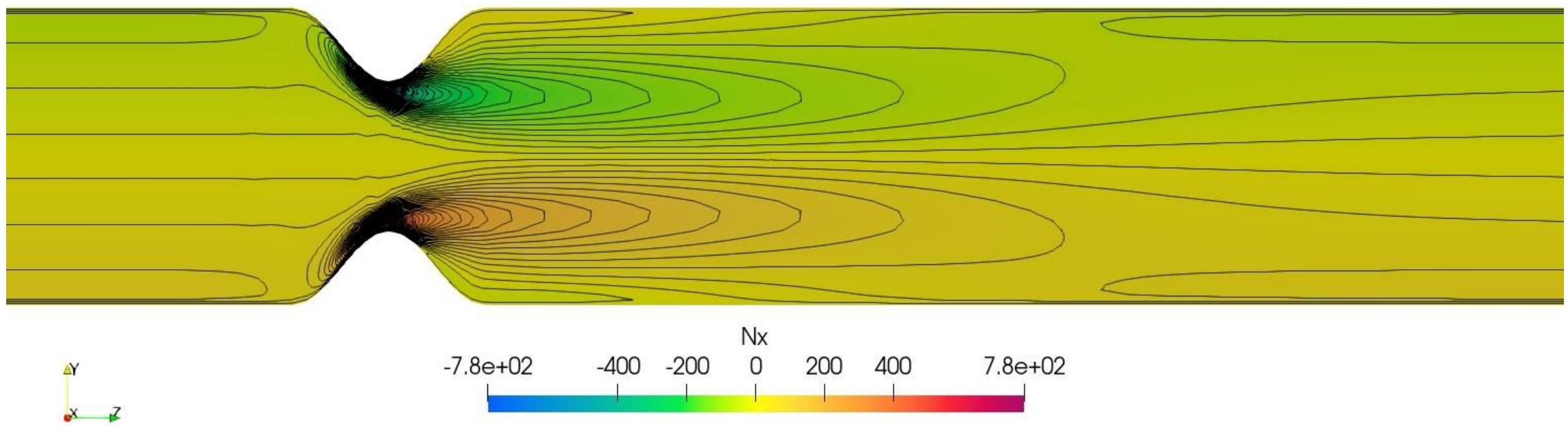

*(b.i)*

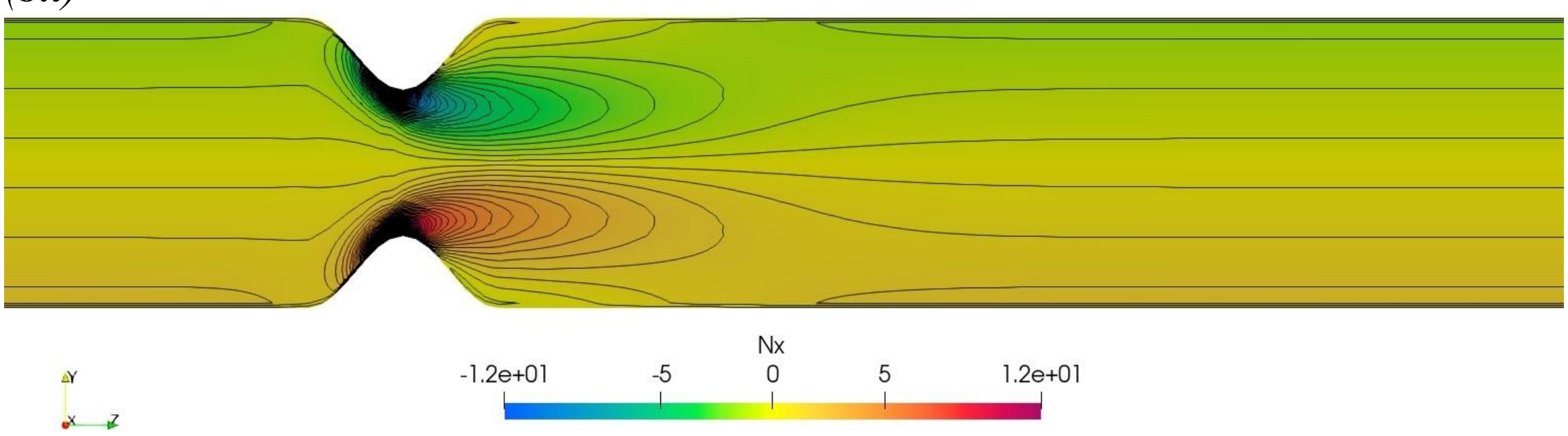

*(b.ii)*

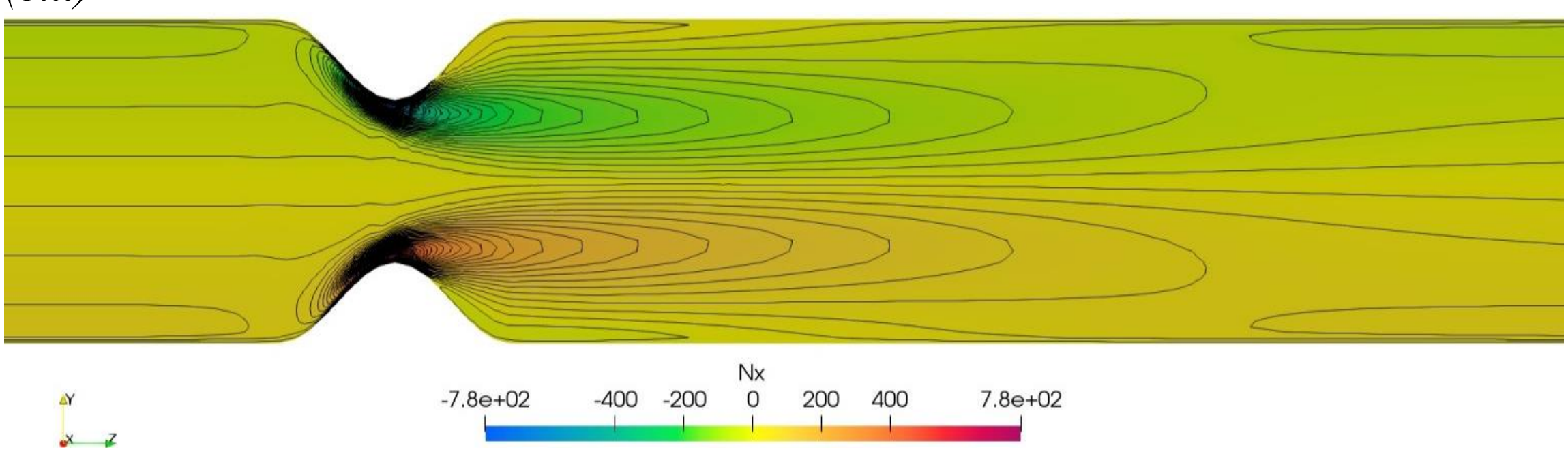

*(c.i)*

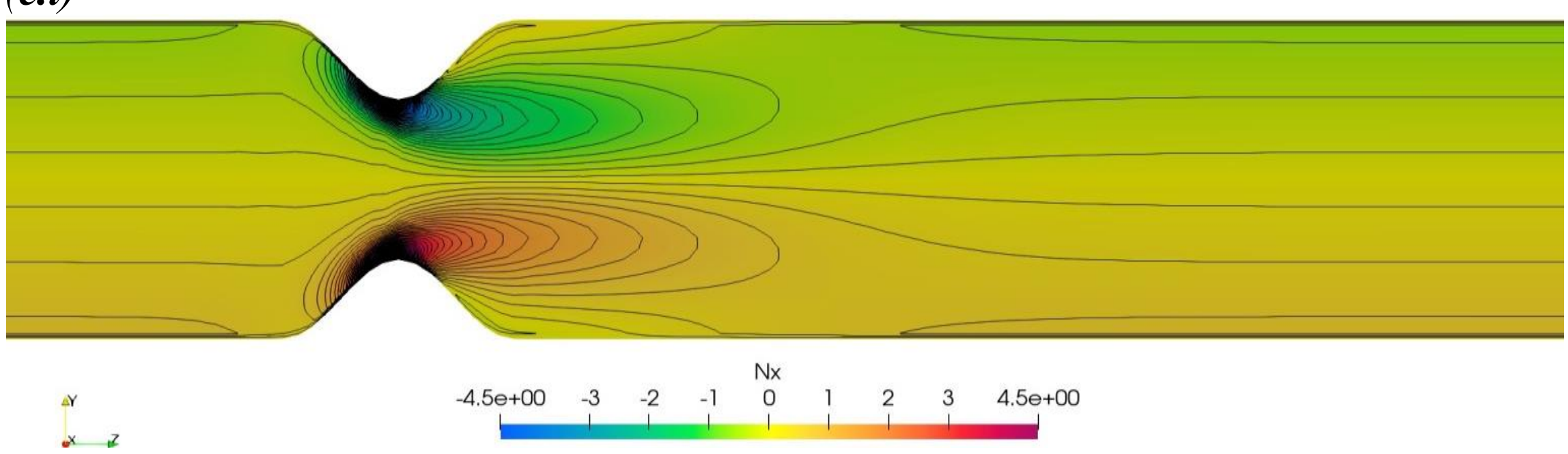

*(c.ii)*

FIG. 14. Microrotation contour plots for blood flow through a 50% stenosis using MHD micropolar modeling (i) without acknowledging MMR and (ii) considering MMR for an applied magnetic field of (a) 1 $T$, (b) 3 $T$, and (c) 8 $T$ and hematocrit of $\varphi = 45\%$.

In Figure 15, the velocity profile is presented for a 50% stenosis inside the stenotic region and downstream the latter (at $l = 0.02\ m$). The profiles are plotted for the Newtonian blood flow, the micropolar blood flow, the MHD blood flow without the MMR effect, and the MHD blood flow with the MMR effect included. Two hematocrit values are used, one at $\varphi = 25\%$ ($\varepsilon = 0.375$) and one at $\varphi = 45\%$ ($\varepsilon = 0.675$), while the applied magnetic field is varied at 1 $T$, 3 $T$, and 8 $T$. Immediately, one can notice a "plug-like" velocity profile within the stenosis, where the velocity becomes flat at the center, a common phenomenon in arterial blood flows, especially in moderate to severe stenoses, such as this one. Physically, in a stenosis, the fluid accelerates, and the viscous forces (which create the parabolic shape) have less influence due to the high velocity. This results in a flatter velocity profile, especially at the center of the stenotic region.

Similar to the streamlines, the differences observed in velocity between Newtonian and micropolar profiles are negligible both inside and downstream of the stenotic region, even when hematocrit increases. This was expected due to the relatively high $\lambda$ value corresponding to the size of the artery, which minimizes any micropolar effects on velocity (see the relative discussion in Section II.C). Once again, no significant differences are observed between the velocity profiles of micropolar blood flow and MHD micropolar blood flow without MMR, for all considered values of the applied magnetic field and hematocrit, both within and downstream of the stenosis. However, when the MMR term is included, the velocity decreases, with a maximum reduction of 28% at 8 $T$, and $\varphi = 45\%$ inside the stenosis and 30% at 8 $T$, and $\varphi = 45\%$ downstream of the stenosis. As expected, these differences become larger as the hematocrit increases.

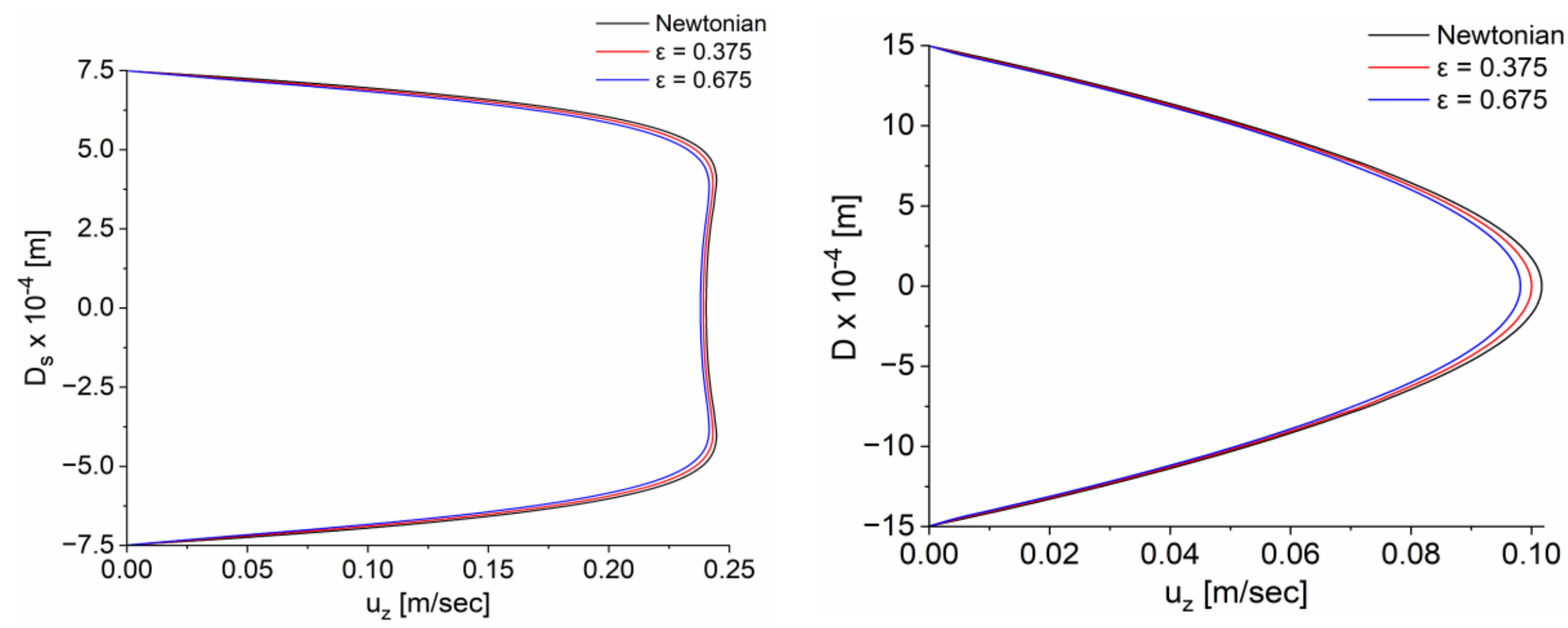

*(a)*

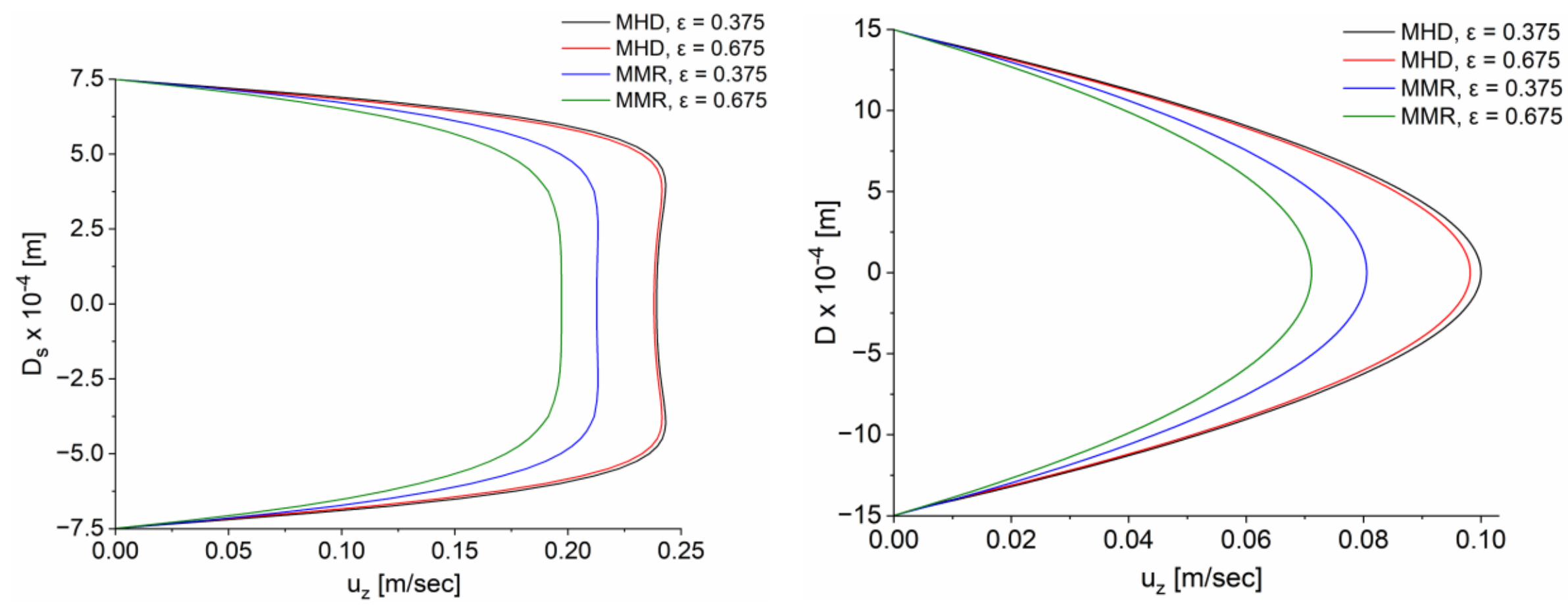

*(b)*

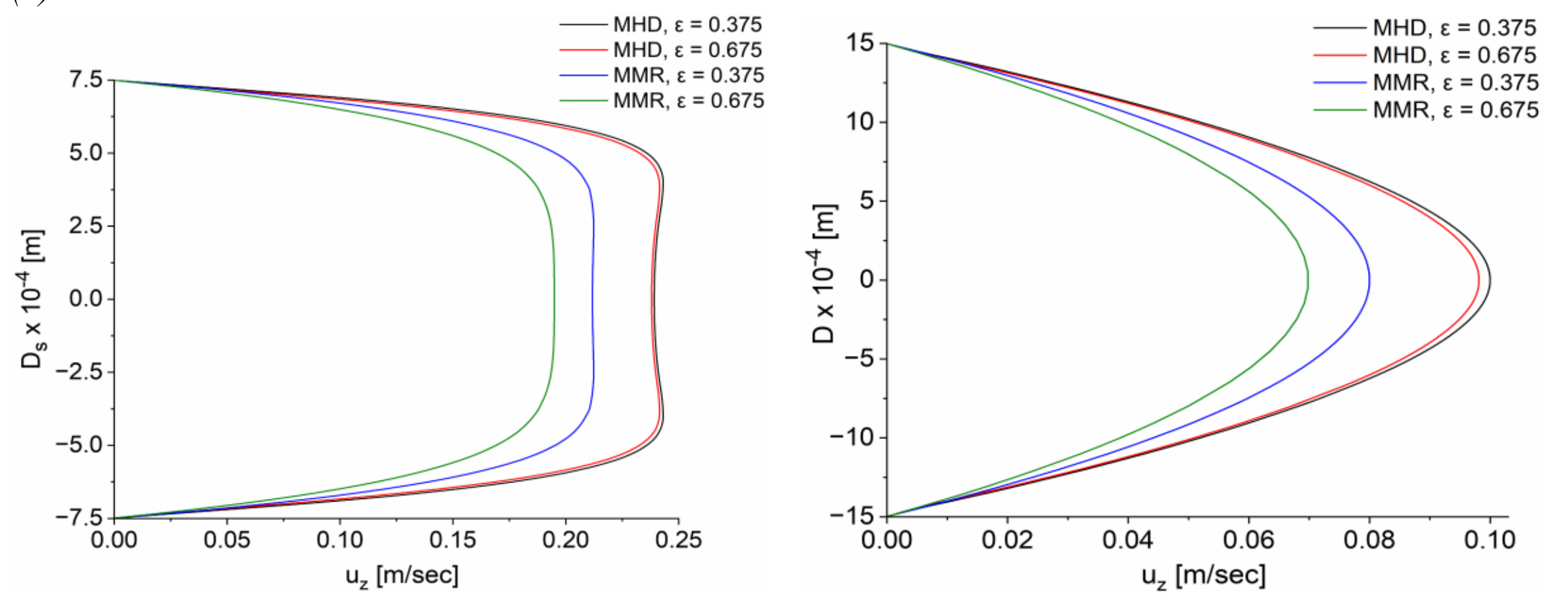

*(c)*

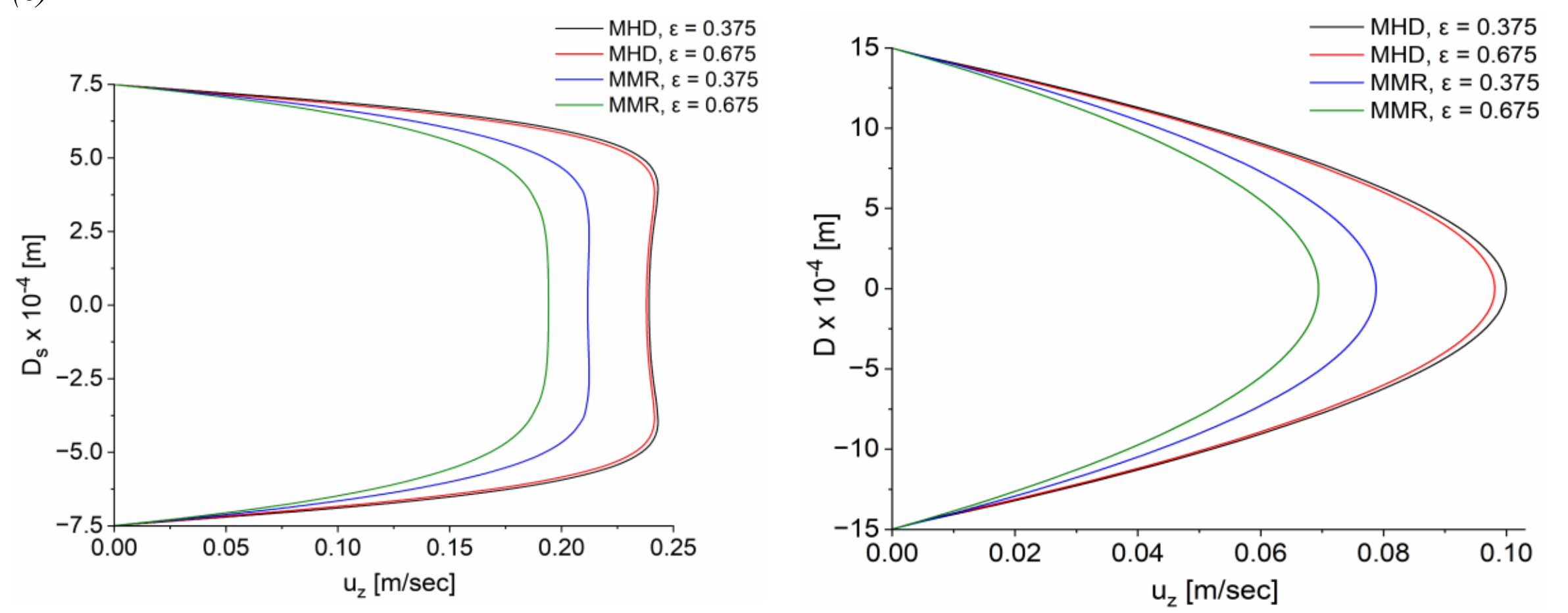

*(d)*

FIG. 15. Velocity profiles at the center of the stenotic region (left) and downstream the latter (right) with a 50 % stenosis for (a) the Newtonian and micropolar blood flow without an applied magnetic field and (b) the micropolar blood flow by ignoring and considering the MMR term with a magnetic field of 1 $T$, (c) 3 $T$ and (d) 8 $T$. The hematocrit is varied at $\varphi =$ 25% and $\varphi = 45\%$ ($\varepsilon = 0.375$ and $\varepsilon = 0.675$, respectively).

In Figure 16, the microrotation profile is illustrated for a 50% stenosis inside the stenotic region and downstream the latter (at $l = 0.02\,m$). Similar to the velocity, microrotation is plotted for the Newtonian blood flow, the micropolar blood flow, the MHD blood flow without the MMR effect, and the MHD blood flow with the MMR effect included. Two hematocrit values are used, one at $\varphi = 25\%$ ($\varepsilon = 0.375$) and one at $\varphi = 45\%$ ($\varepsilon = 0.675$), while the applied magnetic field is varied at $1\ T$, $3\ T$, and $8\ T$. It is immediately evident that the microrotation profile is somewhat disrupted within the stenosis due to the velocity flattening at the center. The increase in hematocrit slightly decreases microrotation both inside and outside the stenosis, although the difference is small due to the relatively high $\lambda$ value. Furthermore, applying the external magnetic field without considering the MMR term does not result in any noticeable differences in microrotation. This occurs because the Lorentz force has a minimal effect on blood flow due to blood's low electrical conductivity. However, when the impact of MMR is considered, microrotation significantly decreases in both the stenotic region and outside of it, with a reduction of nearly 99.9% for the magnetic field of 8 $T$, for both hematocrit values. It is evident that the internal rotation of the erythrocytes almost "freezes" because they are oriented parallel to the applied magnetic field.

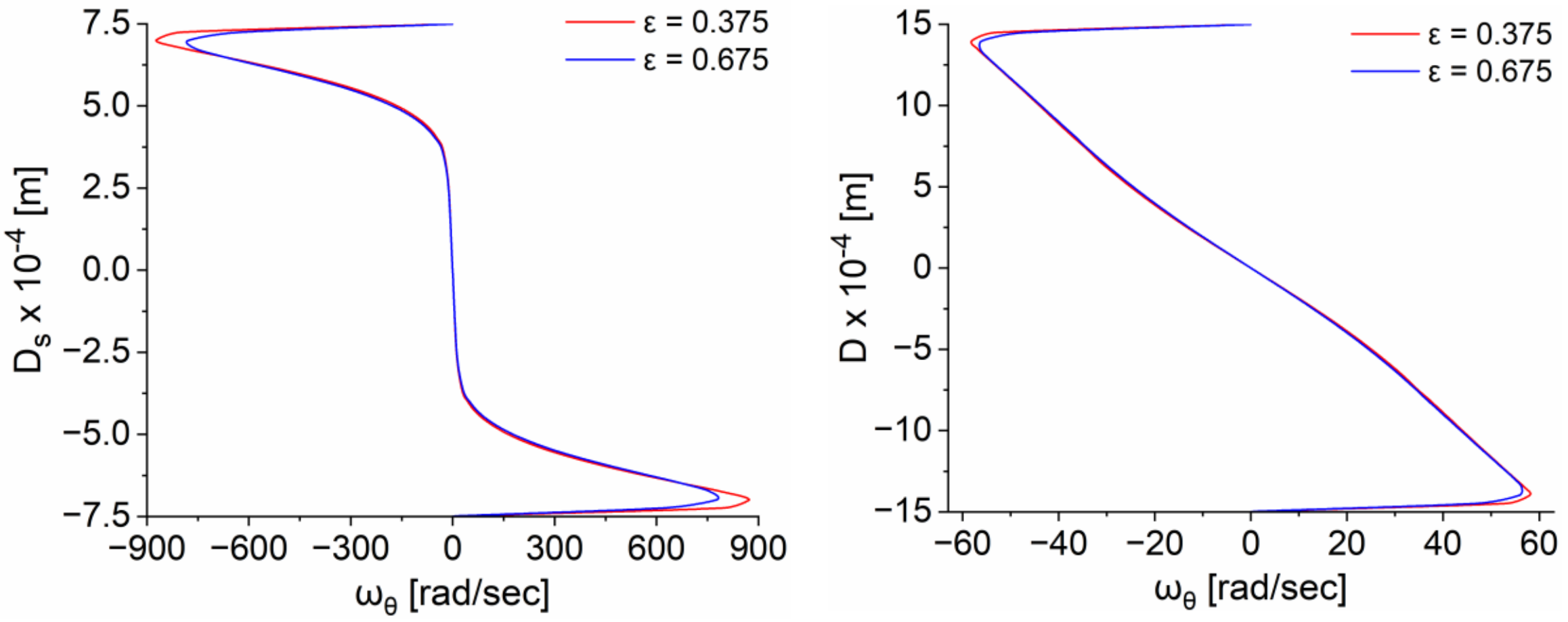

*(a)*

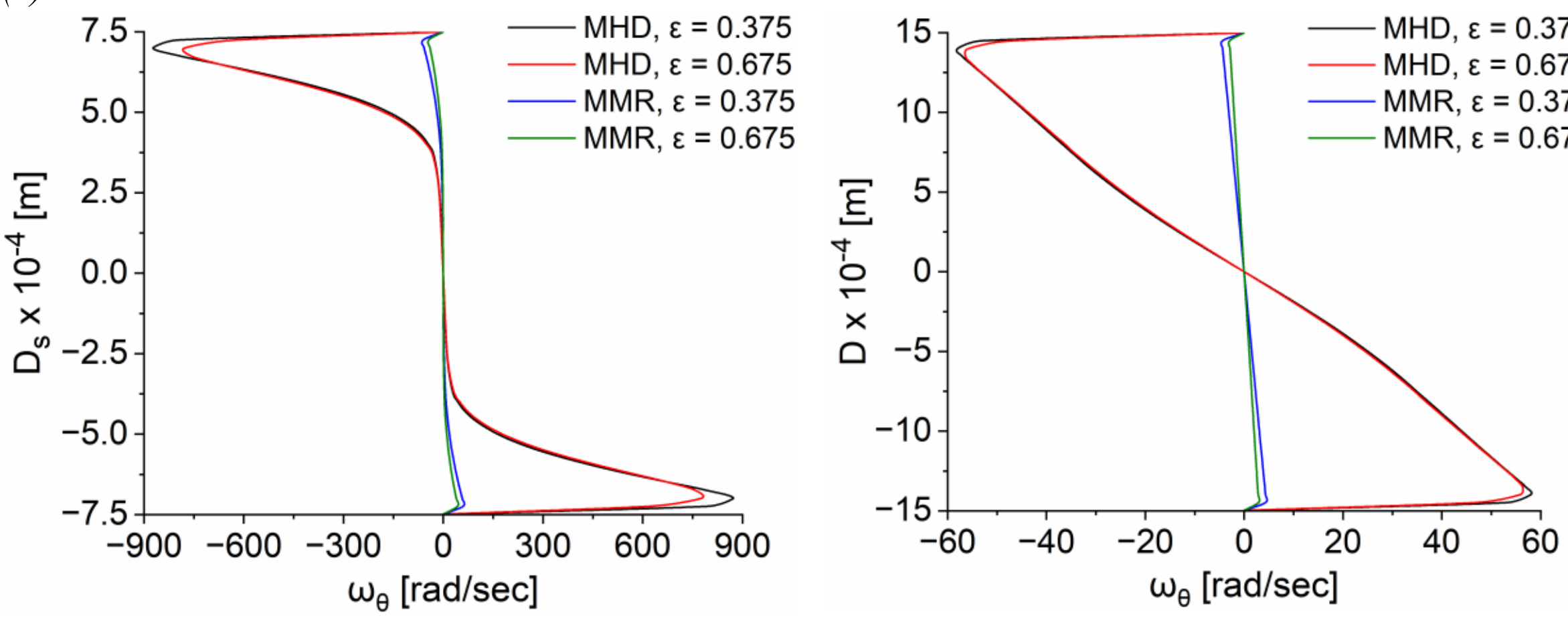

*(b)*

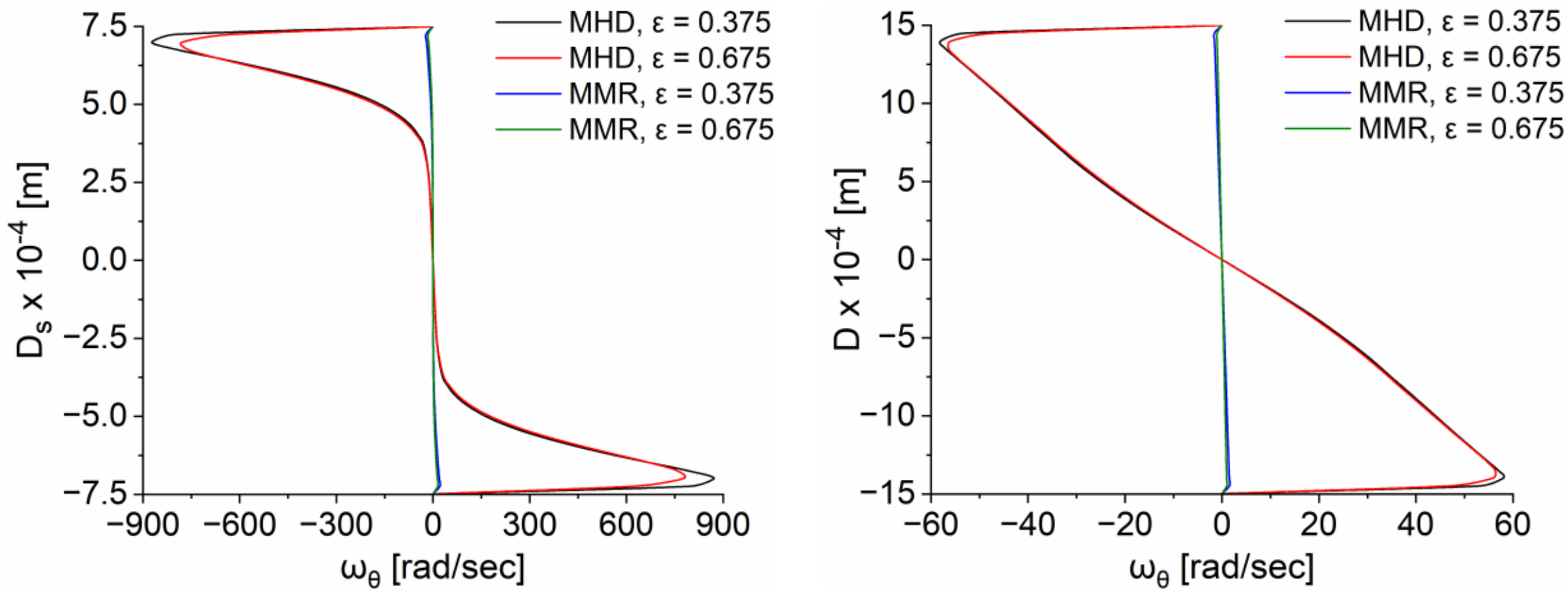

*(c)*

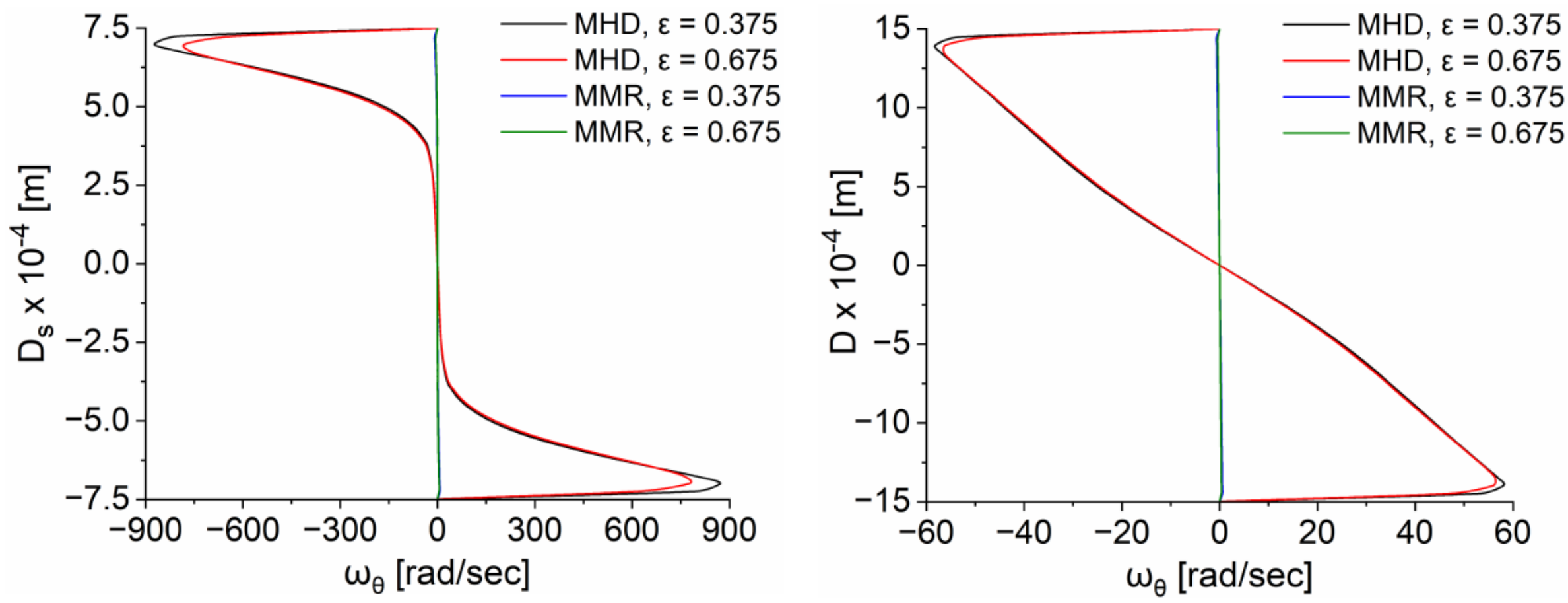

*(d)*

FIG. 16. Microrotation profiles at the center of the stenotic region (left) and downstream the latter (right) with a 50 % stenosis for (a) the Newtonian and micropolar blood flow without an applied magnetic field and (b) the micropolar blood flow by ignoring and considering the MMR term with a magnetic field of 1 $T$ , (c) 3 $T$ and (d) 8 $T$ . The hematocrit is varied at $\varphi = 25\%$ and $\varphi = 45\%$ ($\varepsilon = 0.375$ and $\varepsilon = 0.675$, respectively).

Figure 17 shows the axial variation of the wall shear stress (WSS) $\tau_w$ for the 50% stenosis. In all cases, $\tau_w$ attains its maximum inside the stenosis and decays rapidly in the post-stenotic region, where the flow separates and the small recirculation zone forms. For a given flow rate, the micropolar model produces higher peak $\tau_w$ values than the Newtonian model, in agreement with classical micropolar stenosis studies (e.g., the paper of Hogan and Henriksen[16]) where the no-spin boundary condition for the microrotation ($\delta = 0$) leads to an enhanced near-wall shear layer. This behavior is consistent with our choice of the microrotation being zero at the walls, which increases the effective resistance to shear. The magnitude and localization of the $\tau_w$ peak inside the stenosis are comparable with previous numerical studies of idealized stenoses, which also report sharply elevated $\tau_w$ confined to the constricted section and much lower values upstream and downstream (see the papers: [53-56]).

When a static magnetic field is applied, the MMR contribution reduces both the peak and the downstream oscillations of $\tau_w$ relative to the purely hydrodynamic micropolar case. This damping becomes more pronounced as the field strength increases from 1 to 8 $T$ and as the hematocrit (and thus $\varepsilon$) increases. Physically, the MMR torque tends to align the erythrocytes with the external magnetic field, attenuating the near-wall velocity gradients and leading to a smoother, less oscillatory $\tau_w$ distribution along the stenosis. Despite this reduction, the $\tau_w$ levels remain higher than in the corresponding Newtonian flow, which is consistent with the stronger microstructural interactions represented by the micropolar fluid theory.

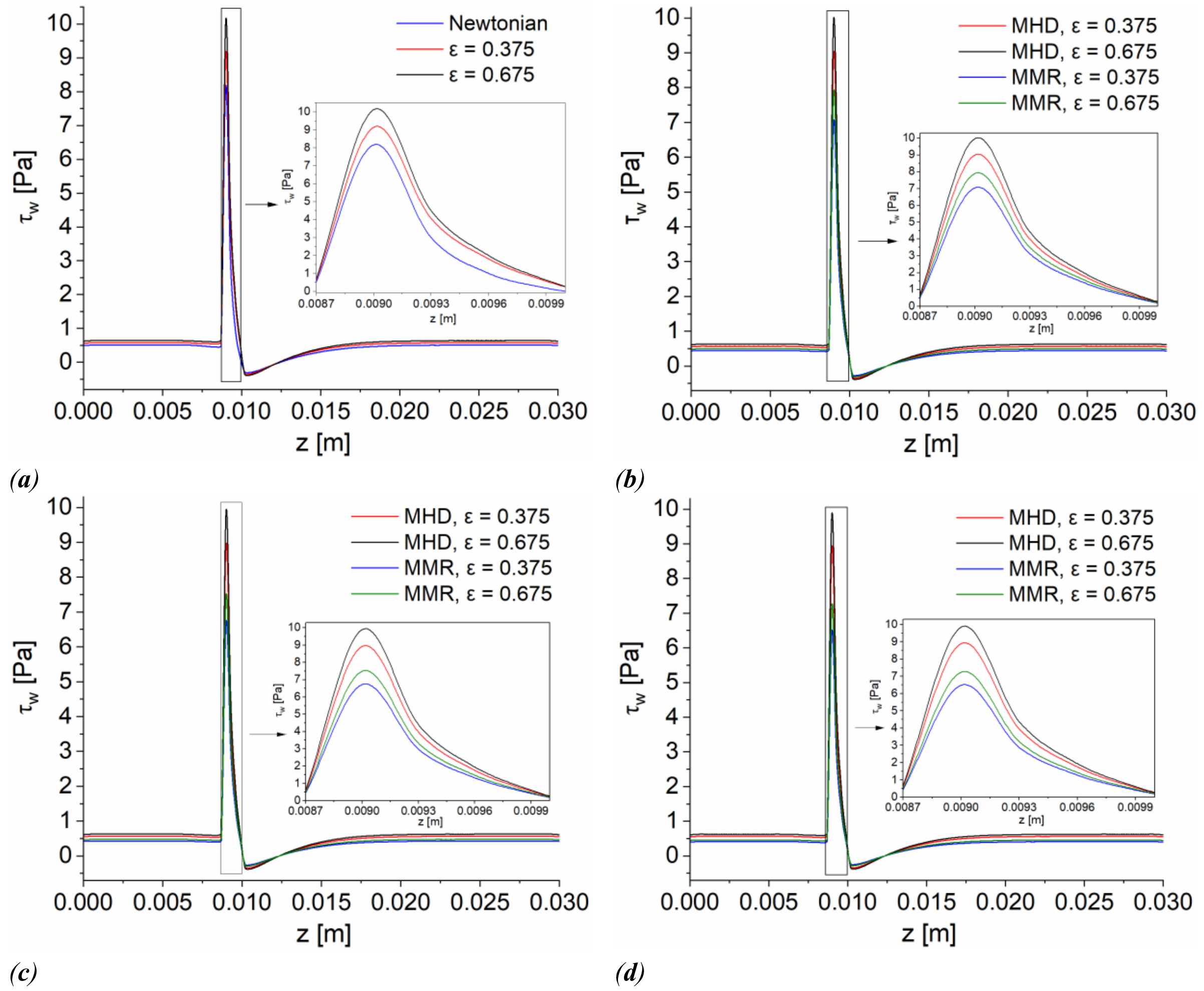

FIG. 17. Wall shear stress $\tau_w$ for the 50% stenosis for (a) the Newtonian and micropolar blood flow without an applied magnetic field and (b) the micropolar blood flow by ignoring and considering the MMR term with a magnetic field of 1 $T$ , (c) 3 $T$ and (d) 8 $T$ . The hematocrit is varied at $\varphi = 25\%$ and $\varphi = 45\%$ ($\varepsilon = 0.375$ and $\varepsilon = 0.675$, respectively). The rectangle region indicates the area magnified in the inset, which highlights the differences in the peak wall shear stress near the stenosis throat.

Figure 18 presents the corresponding pressure drop along the centerline for the 50% stenosis. In all cases, the pressure decreases monotonically through the geometry, with the steepest gradient being inside the stenosis and an almost linear variation in the upstream and downstream sections. As expected, the micropolar model predicts a slightly larger overall pressure drop than the Newtonian model for the same inlet conditions, reflecting the larger effective viscosity and the additional couple-stress dissipation. Similar increases in pressure loss with non-Newtonian blood modelling have been reported in idealized coronary stenoses [53-55, 57]. Including the MMR effect further increases the total pressure drop across the stenosis, especially at higher hematocrit and stronger magnetic fields. In this case, the Lorentz force and the MMR torque oppose the axial motion, reducing the core velocity and requiring a larger axial pressure gradient to sustain the same flow rate.

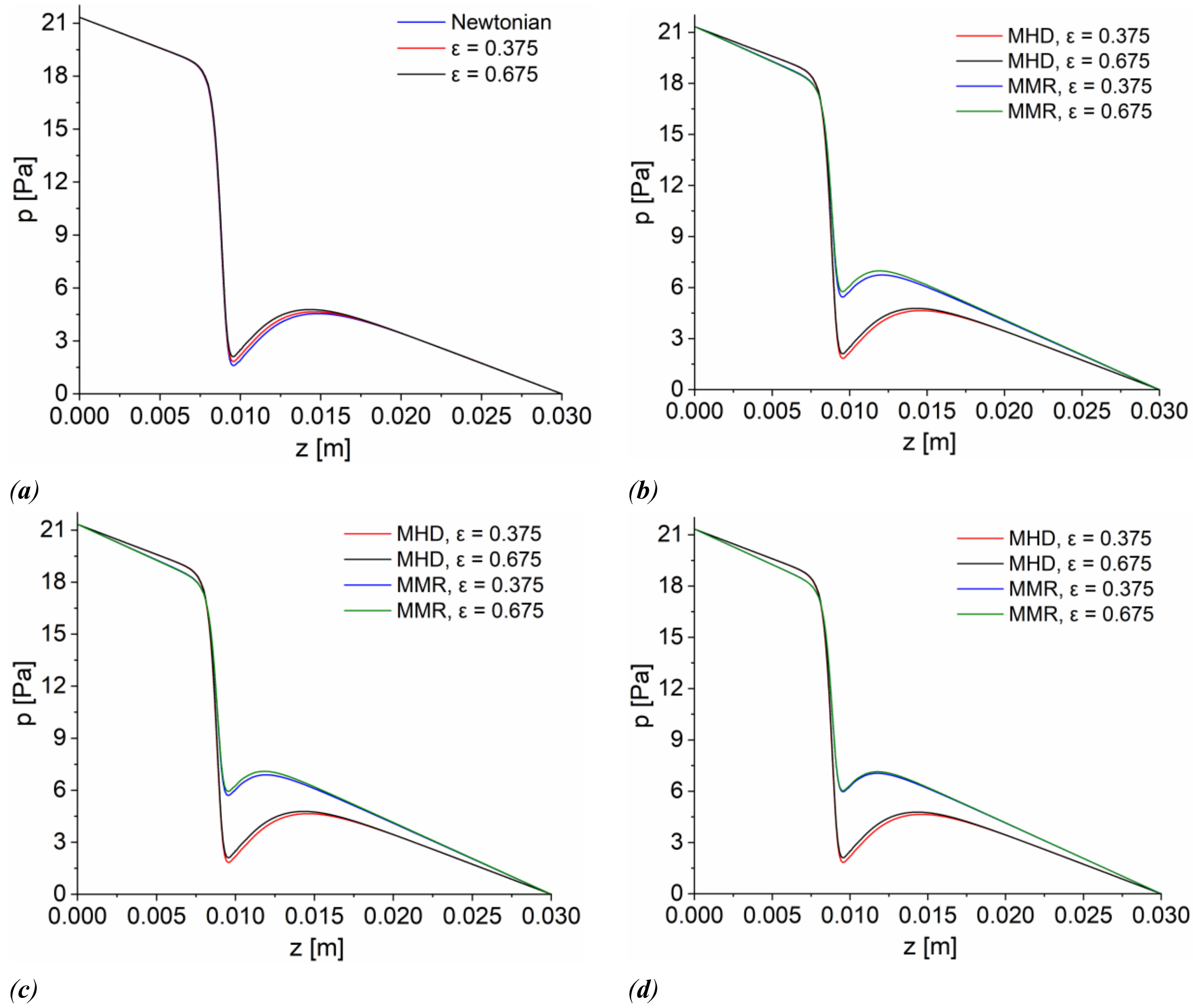

FIG. 18. Pressure drop for the 50% stenosis for (a) the Newtonian and micropolar blood flow without an applied magnetic field and (b) the micropolar blood flow by ignoring and considering the MMR term with a magnetic field of 1 $T$ , (c) 3 $T$ and (d) 8 $T$ . The hematocrit is varied at $\varphi = 25\%$ and $\varphi = 45\%$ ($\varepsilon = 0.375$ and $\varepsilon = 0.675$, respectively).

### B. Results for 80% stenosis

Figure 19 illustrates the streamlines for the Newtonian and micropolar blood flow through an 80% stenosis using two hematocrit values, $\varphi = 25\%$ and $\varphi = 45\%$. No external magnetic field is applied to the flow. As expected, due to the high stenosis degree, the flow velocity reaches its maximum value at the position where the stenosis is most tight, while it is almost zero downstream of the stenosis. It should be noted, however, that the maximum velocity within the stenosis is lower than that of the 50% stenosis, as this particular stenosis is slightly longer (see Section II.A. Downstream of the stenosis, right at the end, two vortices are formed—this time almost symmetrical, but smaller than those in the 50% stenosis, again due to the shorter length of the latter. In contrast to the results of the 50% stenosis, the micropolar effects here are significant, becoming more apparent as the hematocrit increases. The maximum velocity within the stenosis decreases by 11.11% when $\varphi = 25\%$, and by 16.67% when $\varphi = 45\%$. The vortices are nearly braked as hematocrit increases. This occurs because the high degree of stenosis significantly reduces the diameter of the artery, making any micropolar phenomena more apparent.

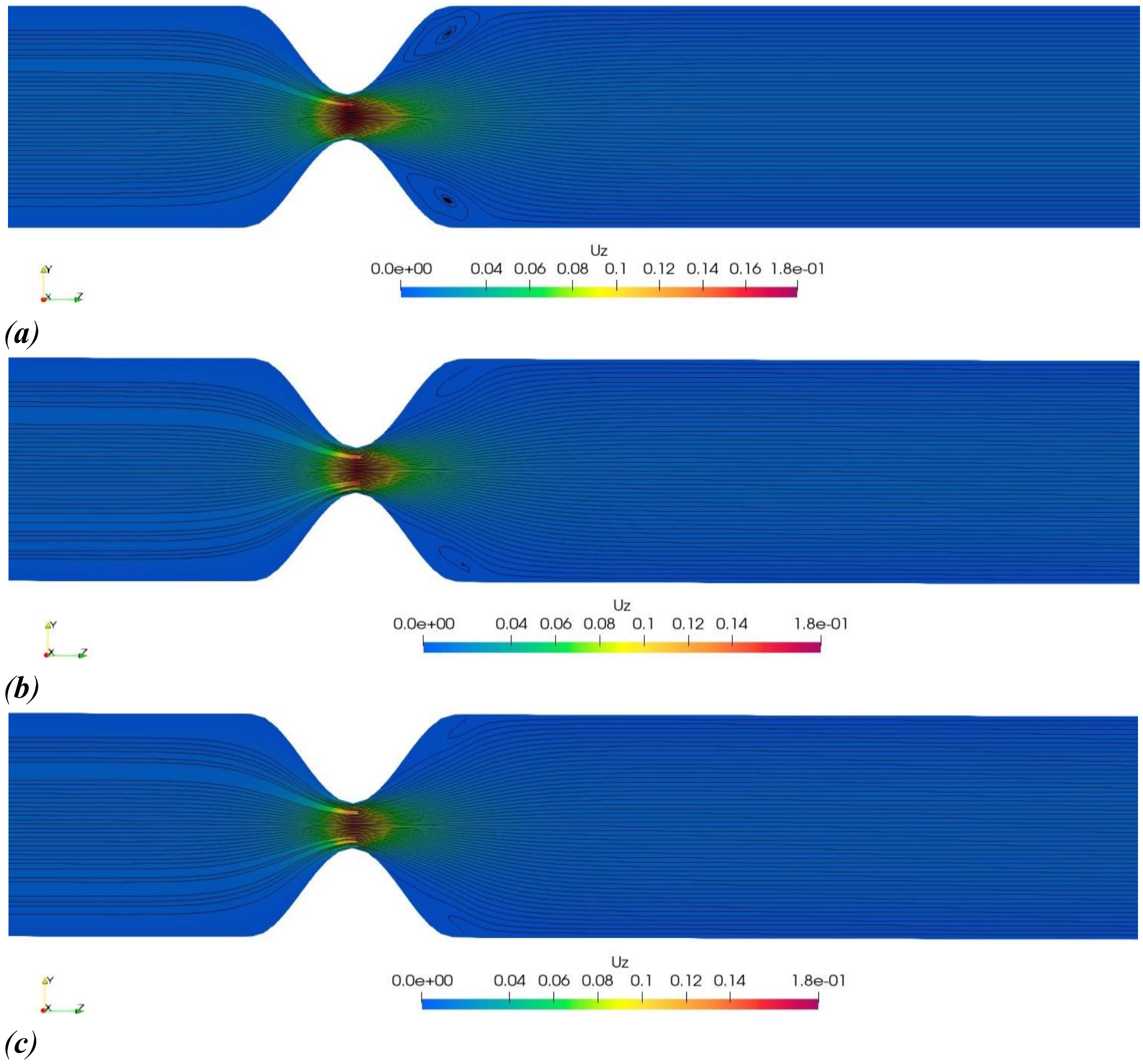

FIG. 19. Streamlines for blood flow through an 80% stenosis using (a) Newtonian modelling and micropolar modelling without an applied magnetic field with hematocrit of (b) $\varphi = 25\%$ and (c) $\varphi = 45\%$.

Figure 20 shows the vorticity contours for both Newtonian and micropolar blood flow through an 80% stenosis, using two hematocrit values: $\varphi = 25\%$ and $\varphi = 45\%$, with no external magnetic field applied. Similar to the 50% stenosis case, the vorticity exhibits an axisymmetric profile, with maximum and minimum values found within the stenotic region. The maximum vorticity value is at the stenosis's lower wall, while the minimum value is at the upper wall. Due to the high stenosis degree, the vorticity isolines are concentrated within the stenosis, with some small vortices appearing downstream. The maximum and minimum vorticity values decrease in absolute terms during the transition from the Newtonian to the micropolar profile and as the hematocrit increases, reaching the values of 23% when $\varphi = 25\%$, and by 30% when $\varphi = 45\%$.

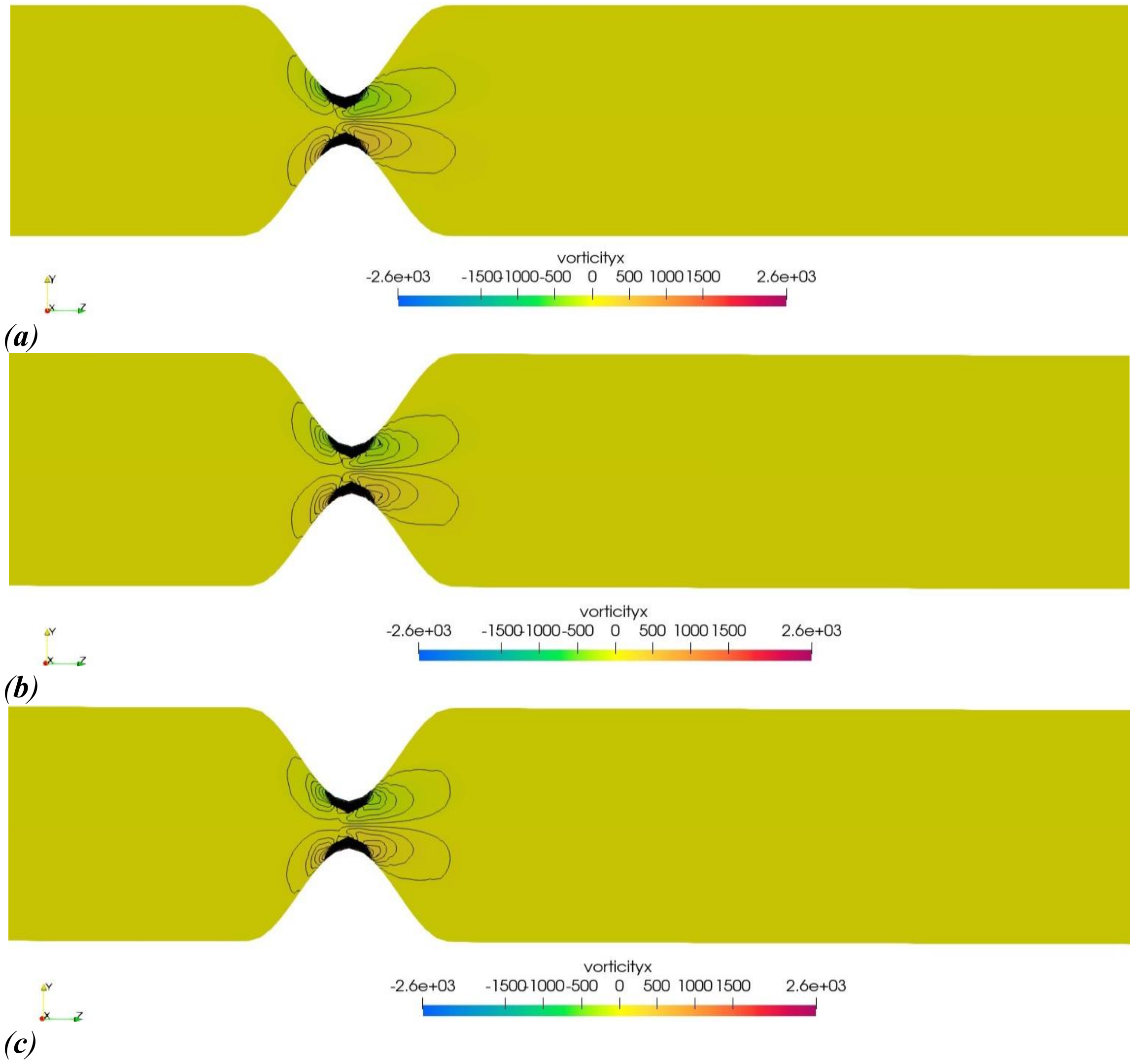

FIG. 20. Vorticity contour plots for blood flow through an 80% stenosis using (a) Newtonian modelling and micropolar modelling without an applied magnetic field with hematocrit of (b) $\varphi = 25\%$ and (c) $\varphi = 45\%$.

Figure 21 illustrates the microrotation contours for both Newtonian and micropolar blood flow through an 80% stenosis, utilizing two hematocrit levels: $\varphi = 25\%$ and $\varphi = 45\%$, in the absence of an external magnetic field. As expected, the microrotation displays an axisymmetric pattern, with its maximum and minimum values located within the stenotic region—similar to the vorticity distribution. The maximum microrotation value occurs along the lower wall of the stenosis, while the minimum appears along the upper wall. Once again, due to the high degree of stenosis, the microrotation isolines are concentrated within the stenosis, while small vortices form just downstream. As expected, as the hematocrit increases, the maximum and minimum microrotation values decrease in magnitude.

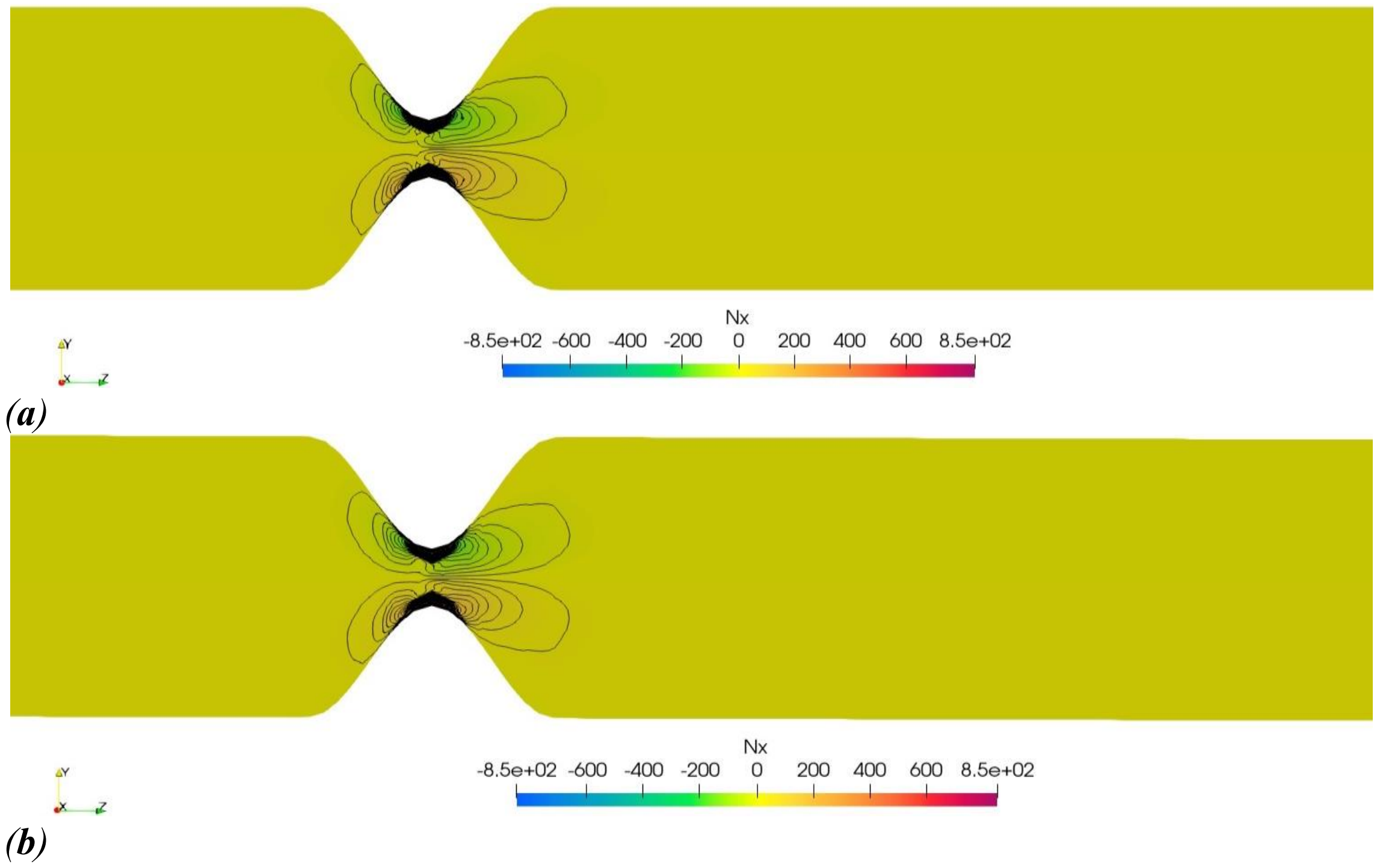

FIG. 21. Microrotation contour plots for blood flow through an 80% stenosis using micropolar modelling without an applied magnetic field with hematocrit of (a) $\varphi = 25\%$ and (b) $\varphi = 45\%$.

Figure 22 presents the streamlines for the 80% stenosis based on the MHD micropolar fluid model, both with and without considering the MMR effect. The hematocrit level is fixed at $\varphi = 25\%$, while the strength of the applied magnetic field varies at 1 $T$, 3 $T$, and 8 $T$. As observed in the 50% stenosis case, the streamlines for micropolar and MHD micropolar blood flow show minimal differences when the MMR effect is neglected, regardless of the magnetic field intensity. However, when the MMR effect is included, the maximum velocity within the stenotic region is reduced—up to 12.5% at 8 $T$ —and the disturbances downstream of the stenosis are significantly diminished. These findings further validate the damping effect of micromagnetorotation.

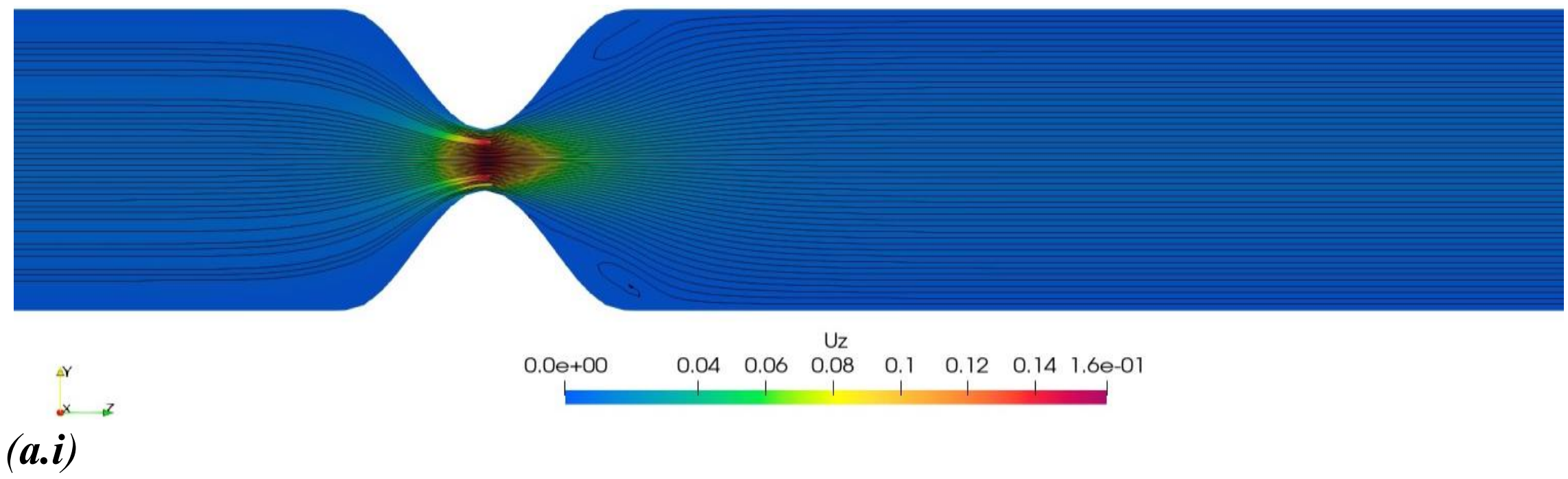

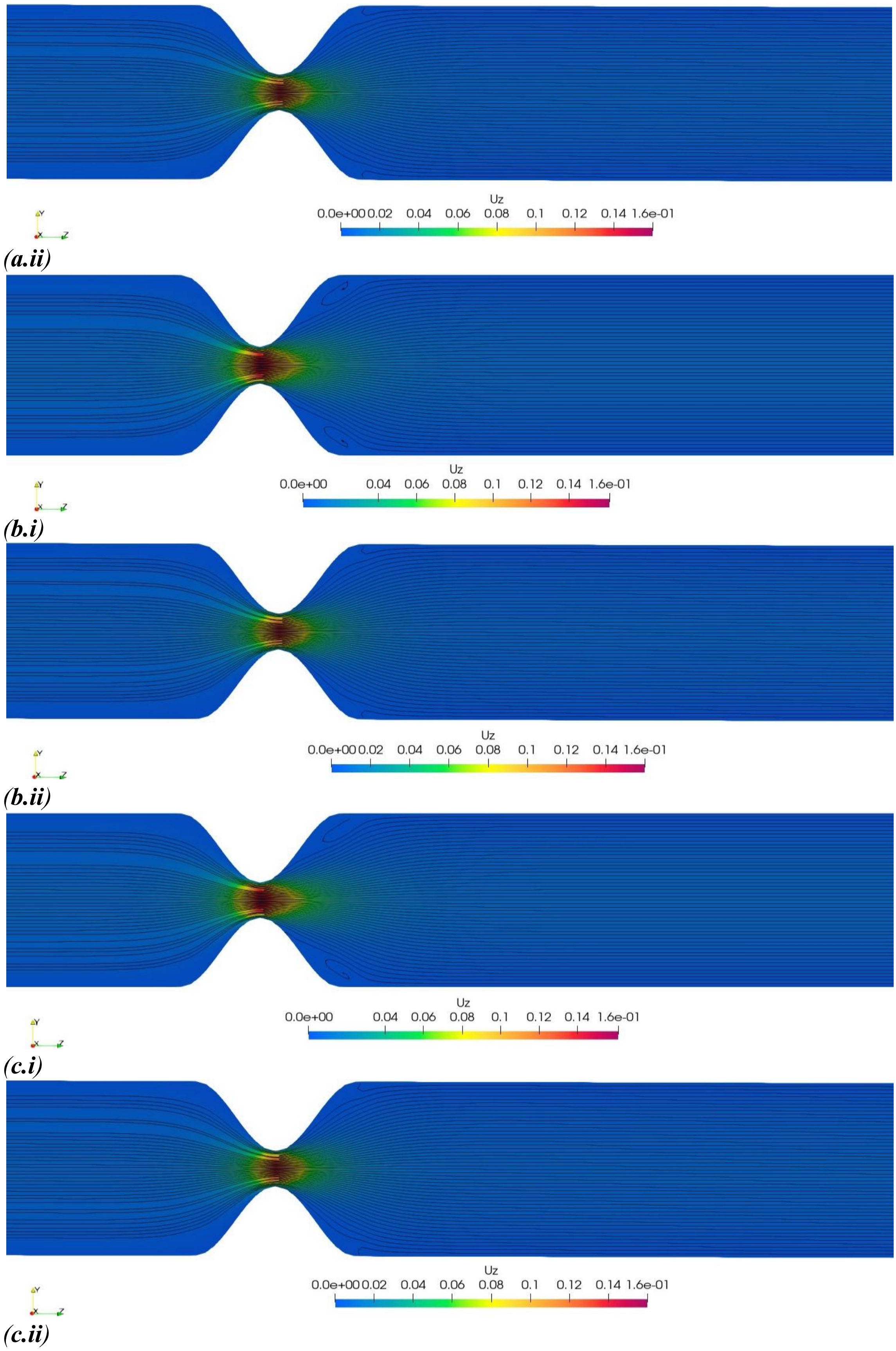

FIG. 22. Streamlines for blood flow through an 80% stenosis using MHD micropolar modeling (i) without acknowledging MMR and (ii) considering MMR for an applied magnetic field of (a) 1 $T$, (b) 3 $T$, and (c) 8 $T$ and hematocrit of $\varphi = 25\%$.

Figure 23 illustrates the streamlines for the 80% stenosis using the MHD micropolar fluid theory, considering and disregarding MMR with a hematocrit of $\varphi = 45\%$ and applied magnetic field of $1\ T$, $3\ T$, and $8\ T$. Similar to the scenario where $\varphi$ equals $25\%$ and the stenotic region equals 50%, no noticeable differences exist between the streamlines of the micropolar blood flow and the MHD micropolar blood flow without MMR across all considered values of the applied magnetic field. When the MMR term is included, the maximum velocity within the stenotic region decreases by 20%, and the disturbances downstream of the stenosis are nearly braked.

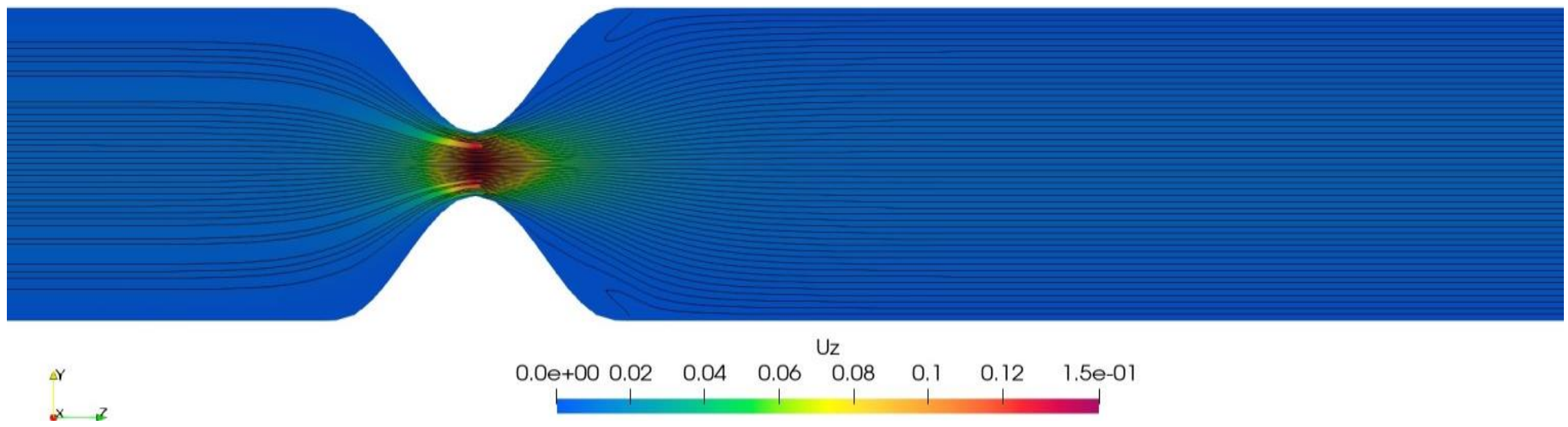

***(a.i)***

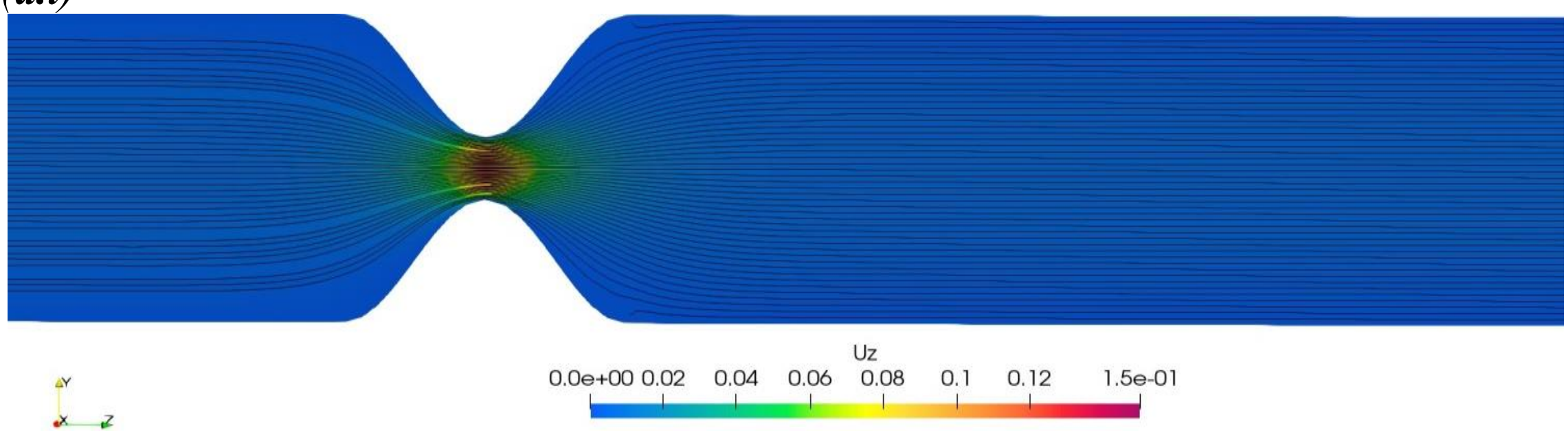

***(a.ii)***

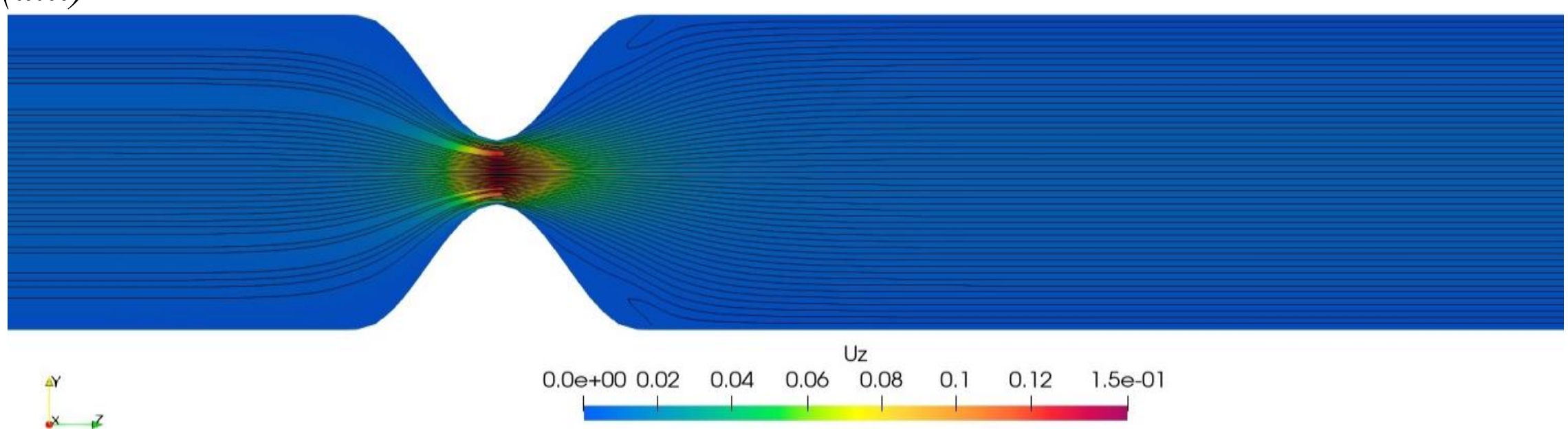

***(b.i)***

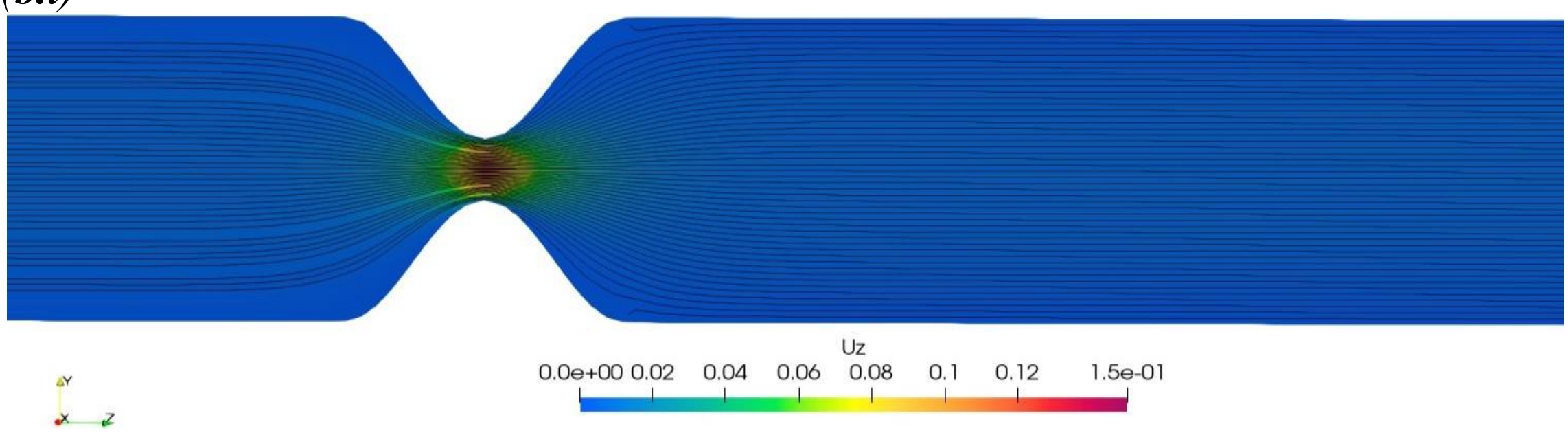

***(b.ii)***

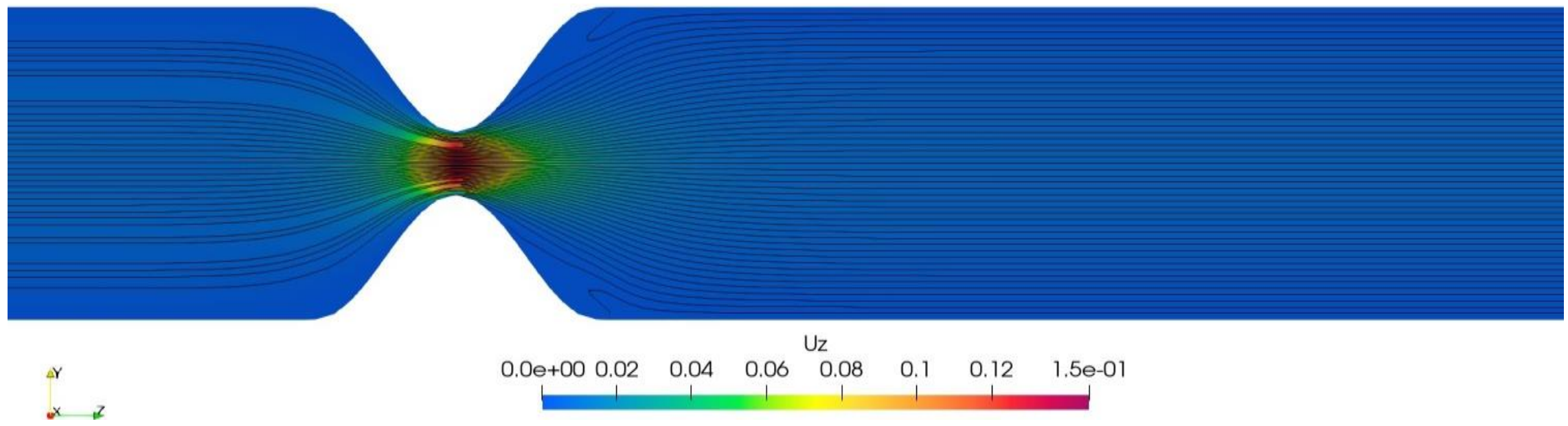

***(c.i)***

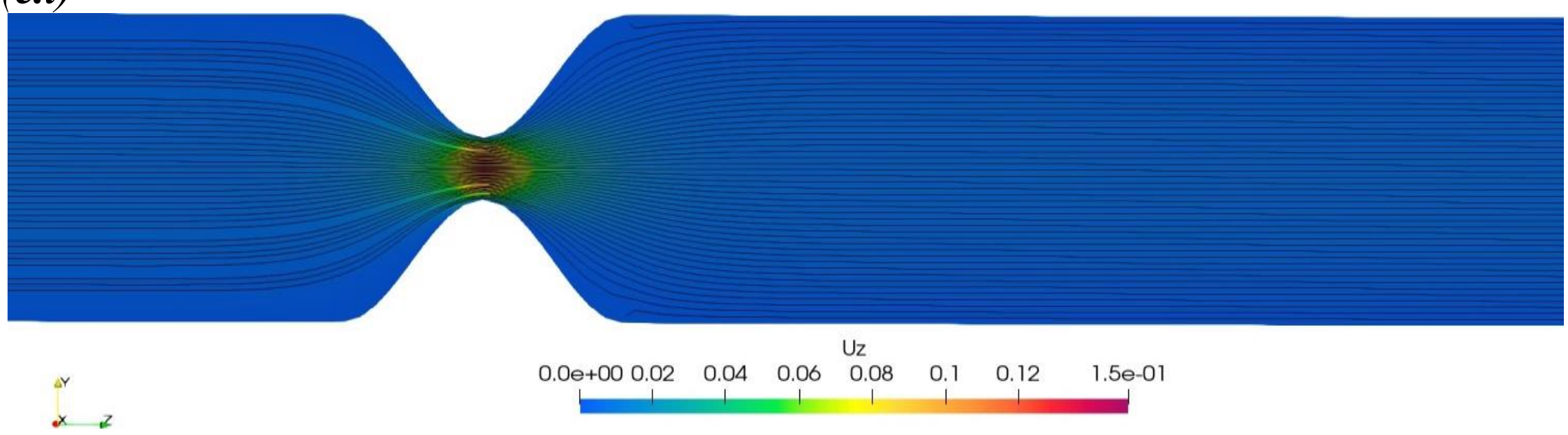

***(c.ii)***

FIG. 23. Streamlines for blood flow through an 80% stenosis using MHD micropolar modeling (i) without acknowledging MMR and (ii) considering MMR for an applied magnetic field of (a) 1 $T$, (b) 3 $T$, and (c) 8 $T$ and hematocrit of $\varphi = 45\%$.

Figure 24 presents the vorticity contours for the 80% stenosis using the MHD micropolar fluid theory, both ignoring and acknowledging MMR. Here, hematocrit is held constant at $\varphi = 25\%$ and the applied magnetic field is varied at 1 $T$, 3 $T$, and 8 $T$. Similar to all previous cases, the vorticity contours for the micropolar blood flow and the MHD micropolar blood flow through stenosis without MMR show no significant differences, regardless of the magnetic field intensity. However, when the MMR term is included, the maximum and minimum vorticity values decrease in magnitude, with a 5% reduction, while the small disturbances downstream of the stenosis are slightly diminished.

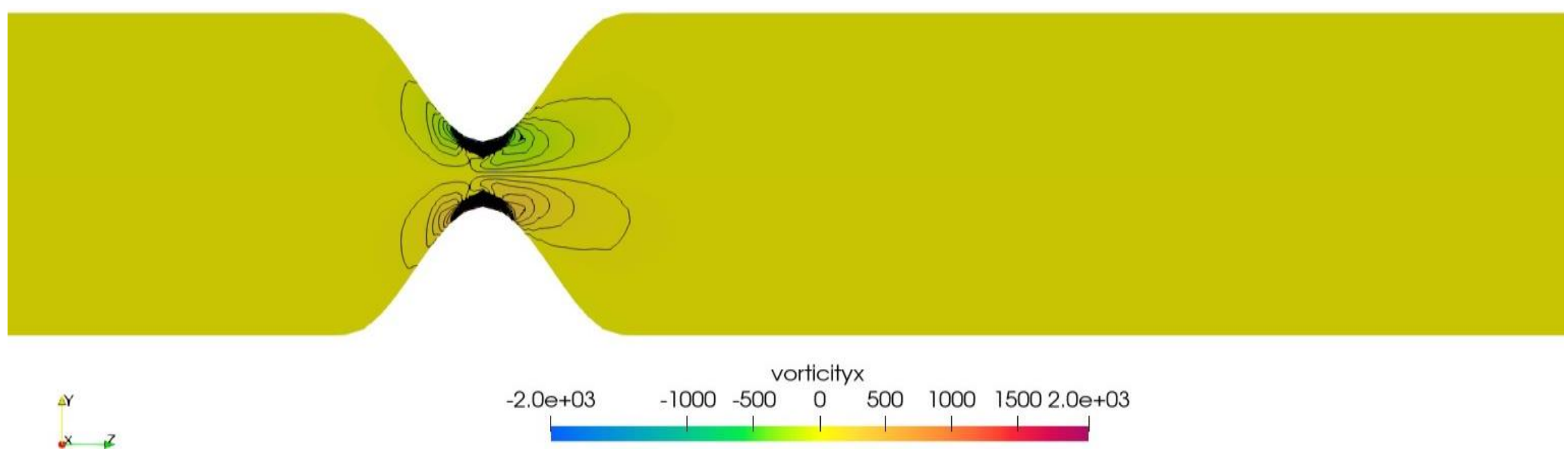

***(a.i)***

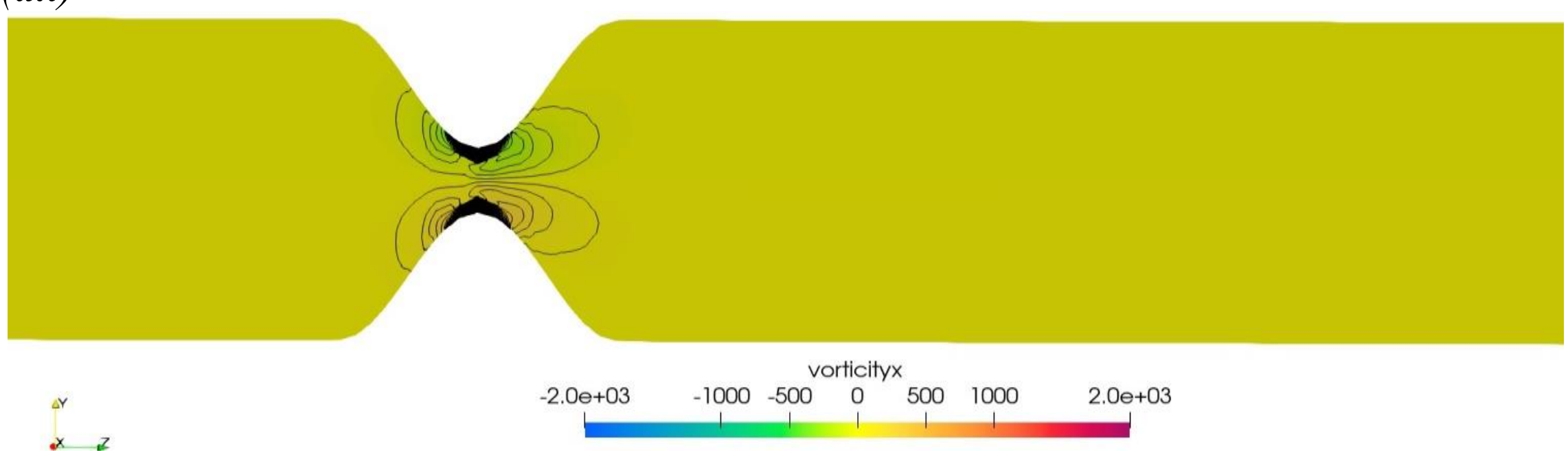

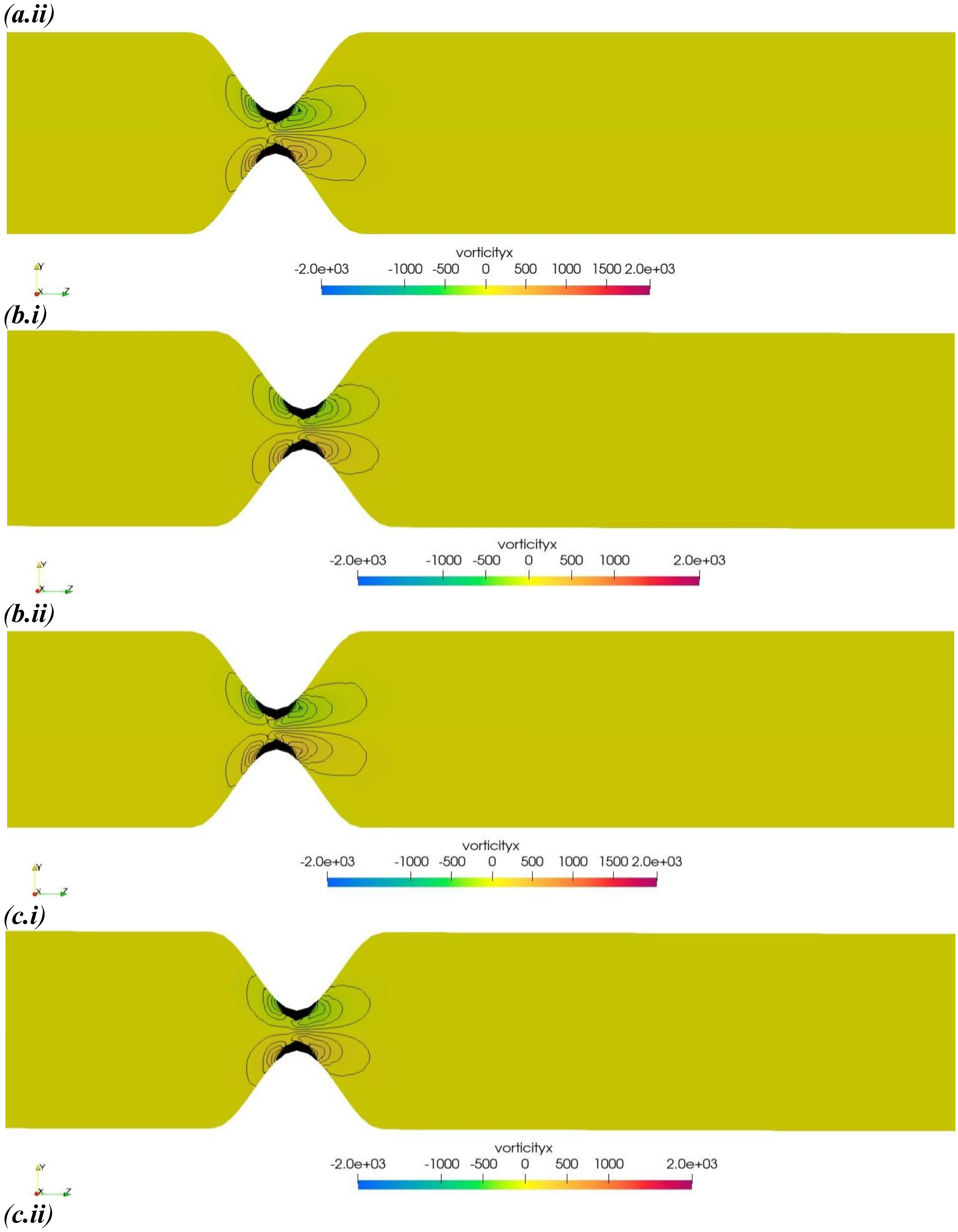

FIG. 24. Vorticity contour plots for blood flow through an 80% stenosis using MHD micropolar modeling (i) without acknowledging MMR and (ii) considering MMR for an applied magnetic field of (a) 1 $T$, (b) 3 $T$, and (c) 8 $T$ and hematocrit of $\varphi = 25\%$.

Figure 25 illustrates the vorticity contours for 50% stenosis using the MHD micropolar fluid theory, considering and ignoring MMR, with a hematocrit of $\varphi = 45\%$ and applied magnetic fields of 1 $T$, 3 $T$, and 8 $T$. Once again, no significant differences exist between the vorticity contours of micropolar blood flow and MHD micropolar blood flow without MMR for all applied magnetic field values considered. However, due to the increase in hematocrit, when the MMR term is included, the maximum and minimum vorticity values decrease further, showing an 11.11% reduction. Additionally, in the case of $\varphi = 45\%$, the length of the disturbances downstream of the stenosis slightly decreases.

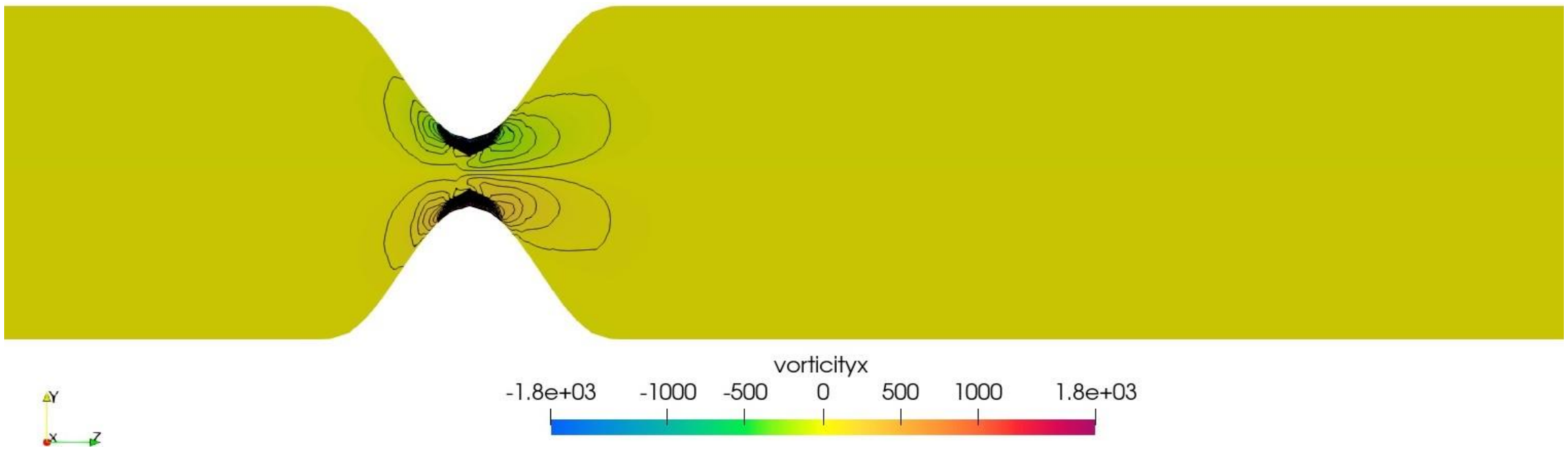

***(a.i)***

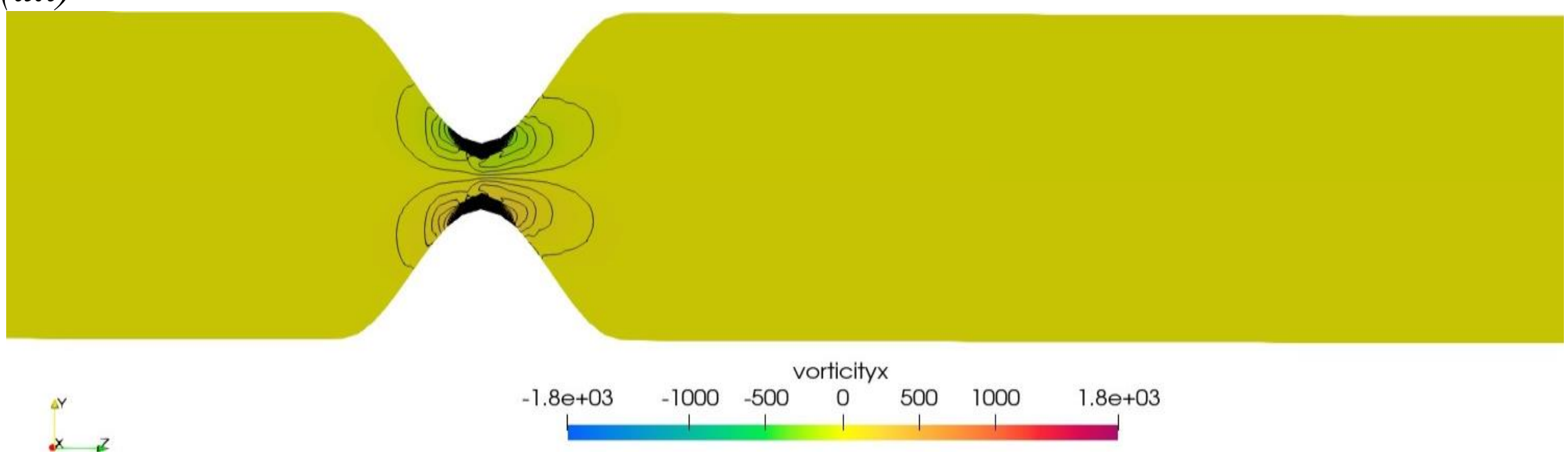

***(a.ii)***

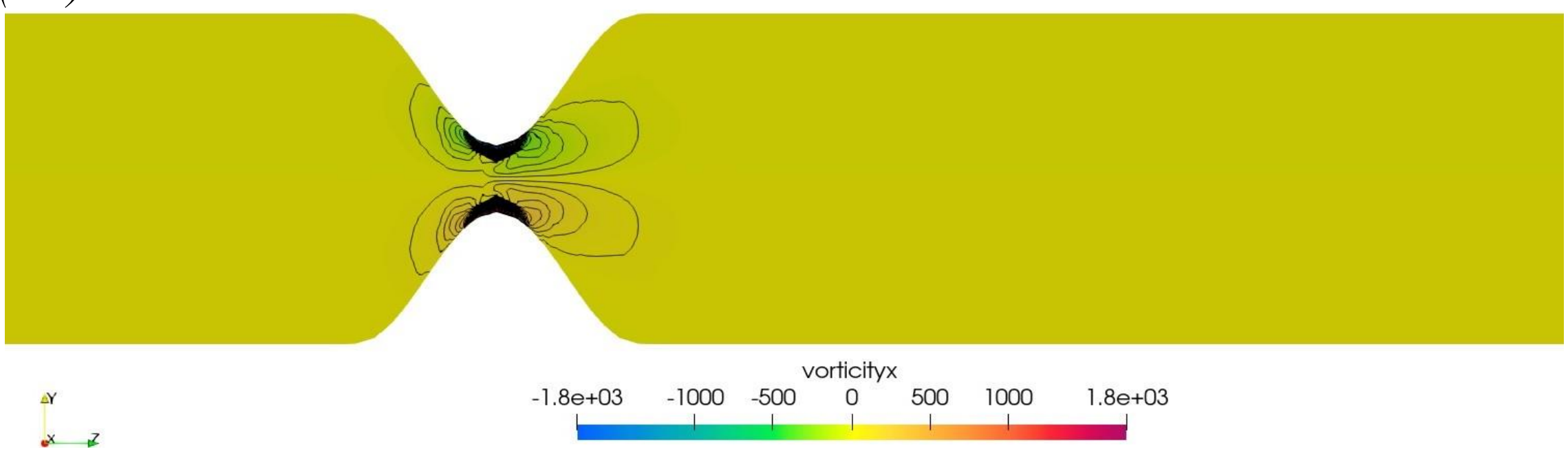

***(b.i)***

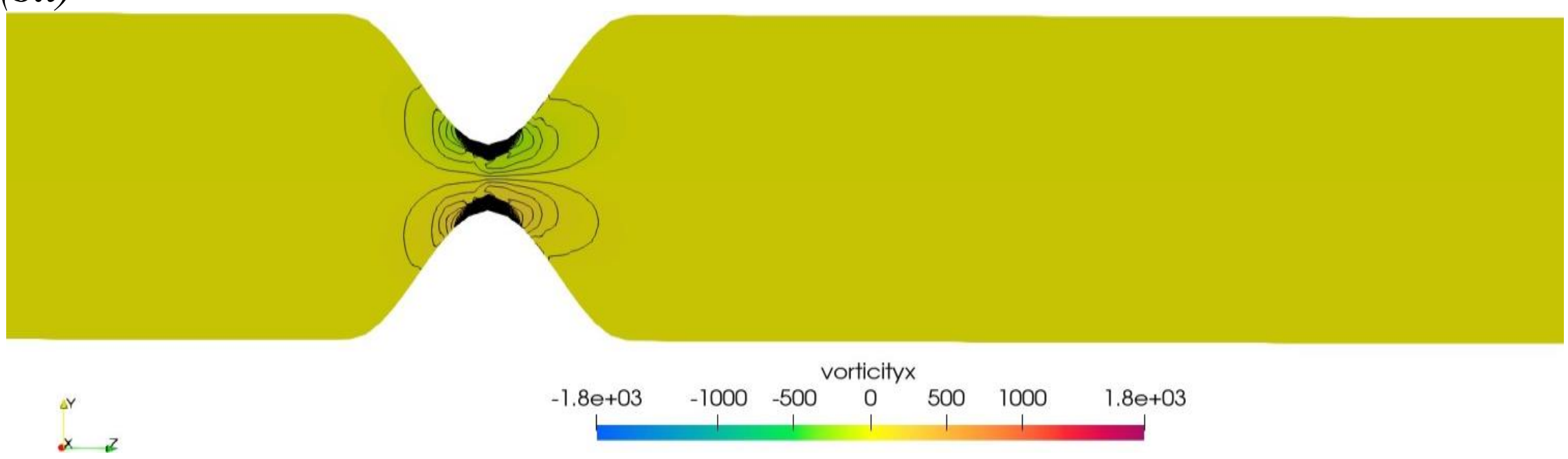

***(b.ii)***

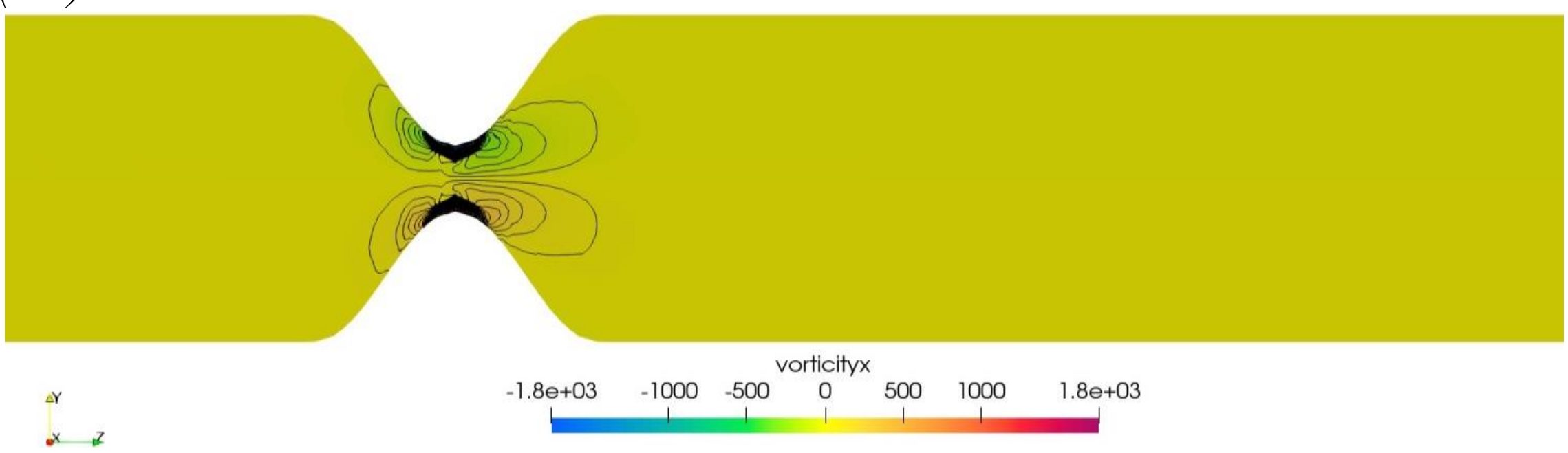

***(c.i)***

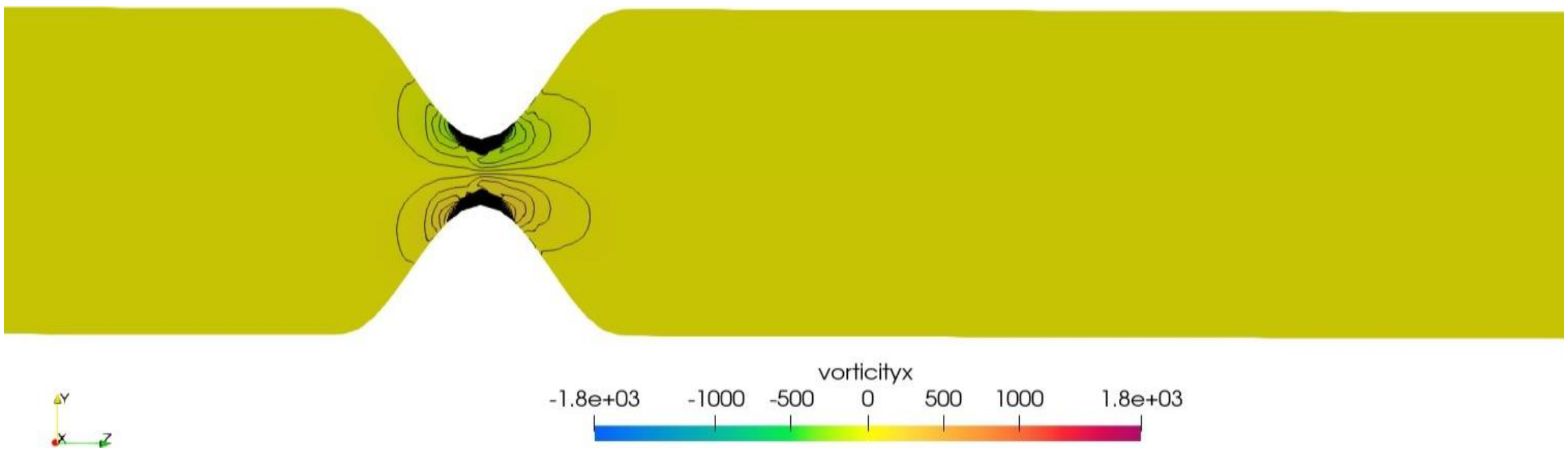

*(c.ii)*

FIG. 25. Vorticity contour plots for blood flow through an 80% stenosis using MHD micropolar modeling (i) without acknowledging MMR and (ii) considering MMR for an applied magnetic field of (a) 1 $T$, (b) 3 $T$, and (c) 8 $T$ and hematocrit of $\varphi = 45\%$.

Figure 26 presents the microrotation contours for an 80% stenosis modeled using the MHD micropolar fluid theory, both with and without considering MMR. As always, the hematocrit level is maintained at $\varphi = 25\%$, while the strength of the applied magnetic field is varied at 1 $T$, 3 $T$, and 8 $T$. As with the vorticity contours, the microrotation contours for micropolar and MHD micropolar blood flow show minimal differences when MMR is not taken into account, regardless of the magnetic field intensity. However, once the MMR term is incorporated, the maximum and minimum microrotation values substantially reduce by magnitude - by approximately 94% at 1 T, 98% at 3 T, and 99% at 8 T. Moreover, the length of the disturbances downstream of the stenosis significantly decreases. These findings further validate that the erythrocytes align with the externally applied magnetic field, preventing any internal rotation.

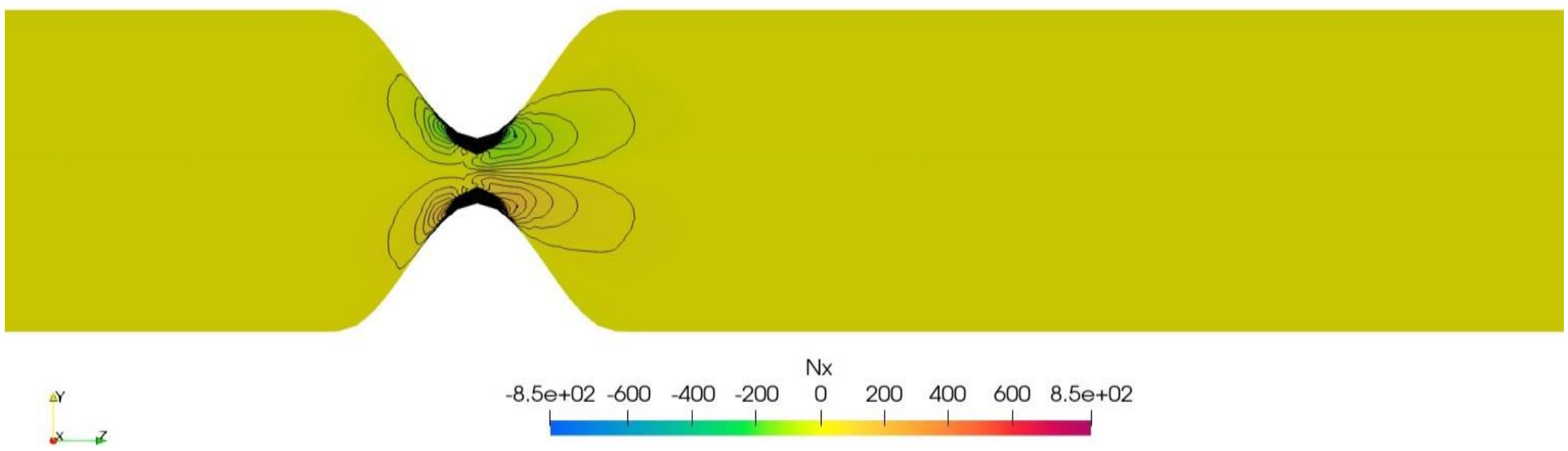

*(a.i)*

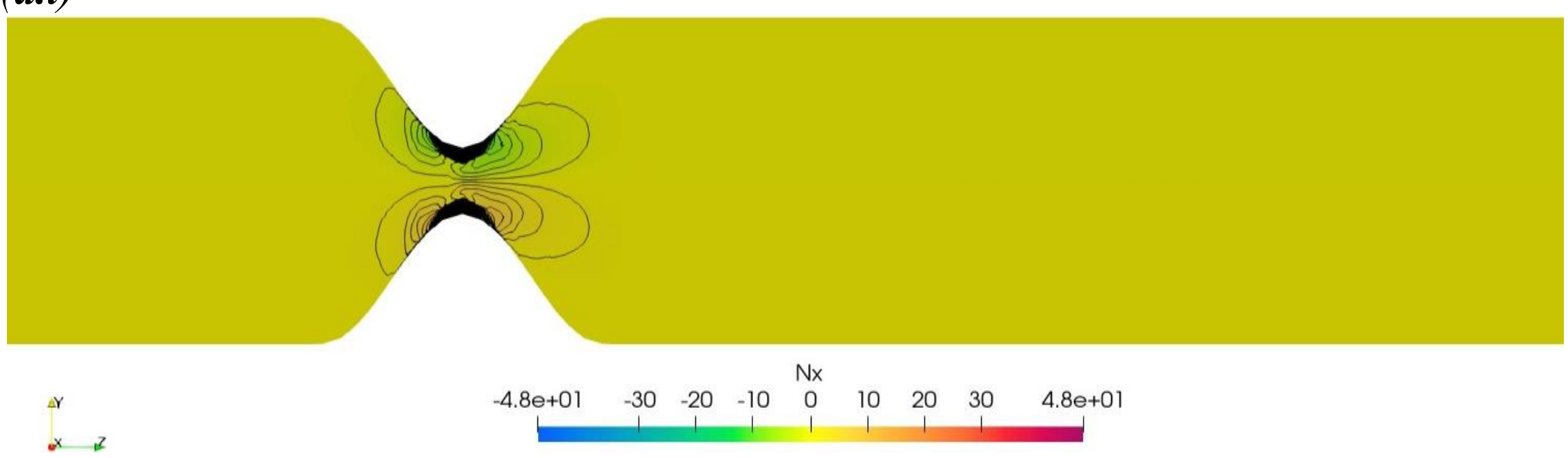

*(a.ii)*

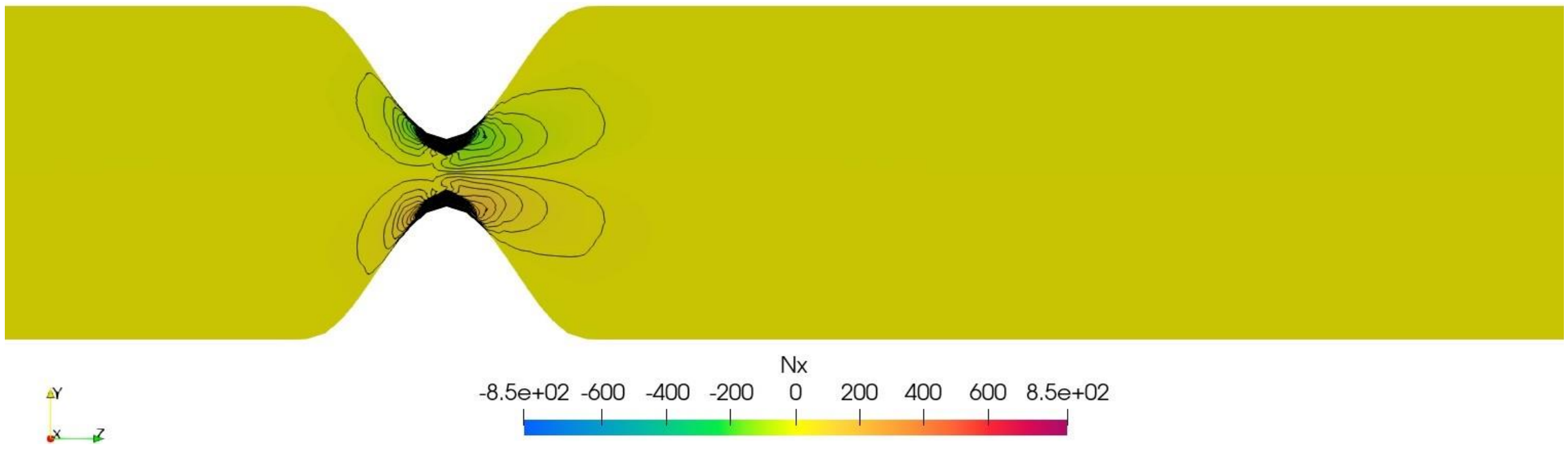

***(b.i)***

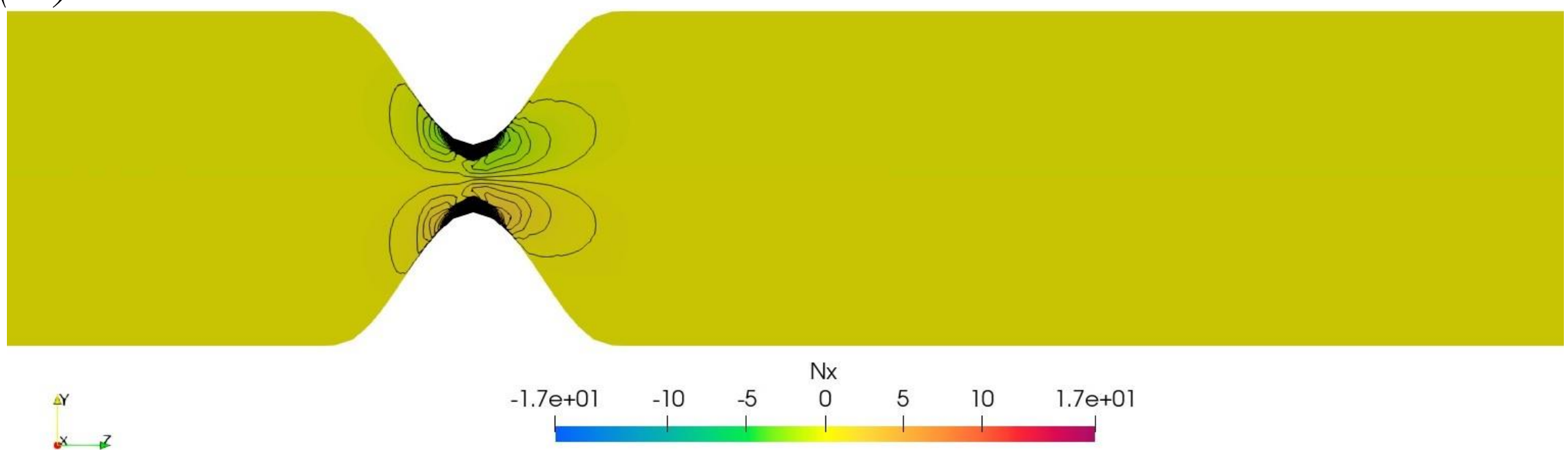

***(b.ii)***

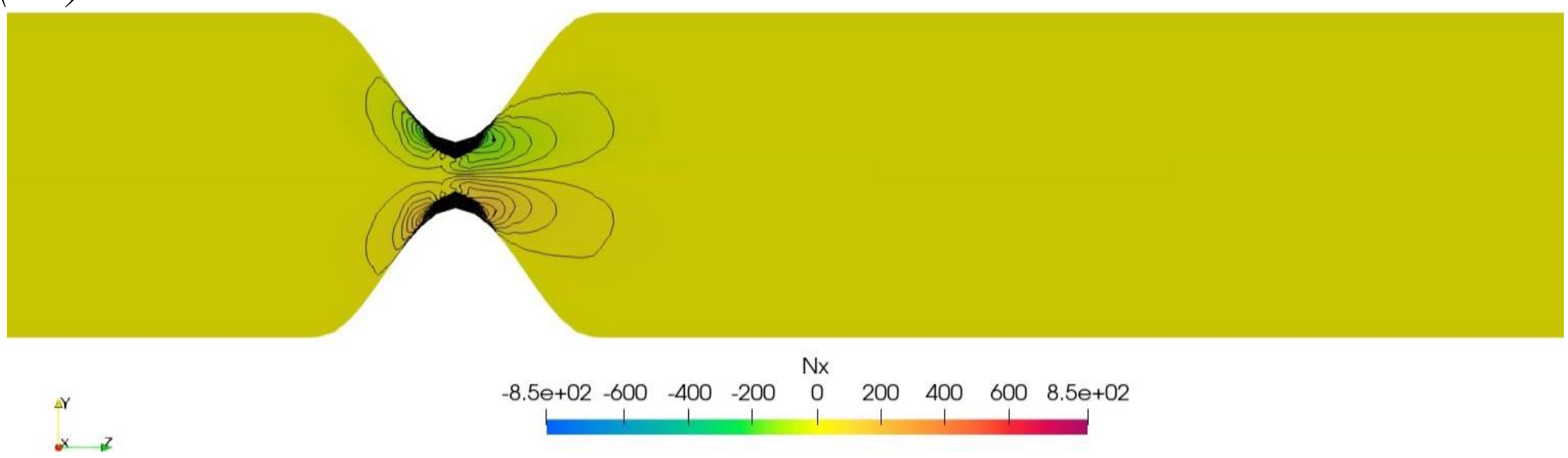

***(c.i)***

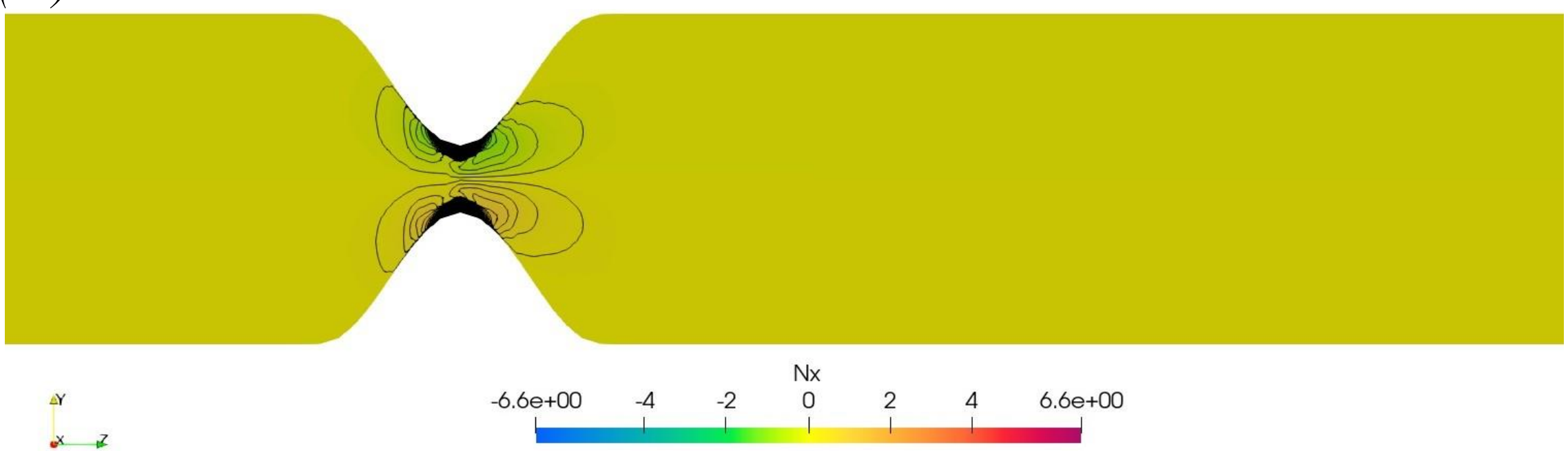

***(c.ii)***

FIG. 26. Microrotation contour plots for blood flow through an 80% stenosis using MHD micropolar modeling (i) without acknowledging MMR and (ii) considering MMR for an applied magnetic field of (a) 1 $T$, (b) 3 $T$, and (c) 8 $T$ and hematocrit of $\varphi = 25\%$.

Figure 27 illustrates the microrotation contours for 80% stenosis using the MHD micropolar fluid theory, comparing scenarios with and without MMR, for a hematocrit of $\varphi = 45\%$ and applied magnetic fields of 1 $T$, 3 $T$, and 8 $T$. Similar to the case where $\varphi$ is 25% and the stenotic region is 50%, no significant differences are observed between the microrotation of the micropolar blood flow and the MHD micropolar blood flow without MMR for all values of the applied magnetic field. However, as the hematocrit increases, including the MMR term leads to a substantial reduction in both the maximum and minimum microrotation values, which decrease by 94% at 1 $T$, 98% at 3 $T$, and 99.5% at 8 $T$. Additionally, the disturbances downstream of the stenosis are slightly diminished.

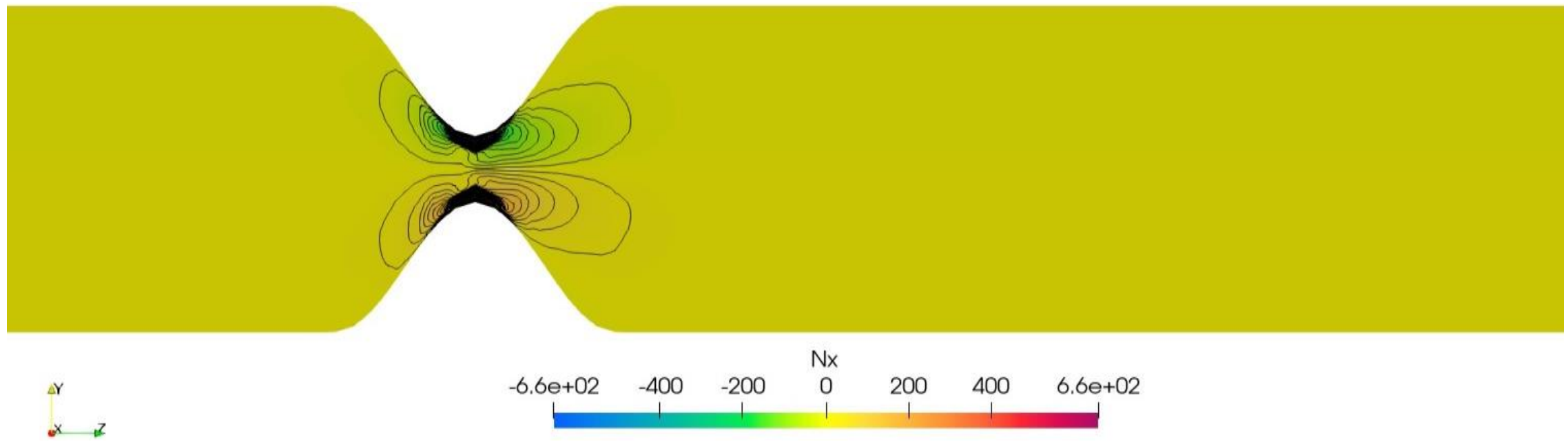

*(a.i)*

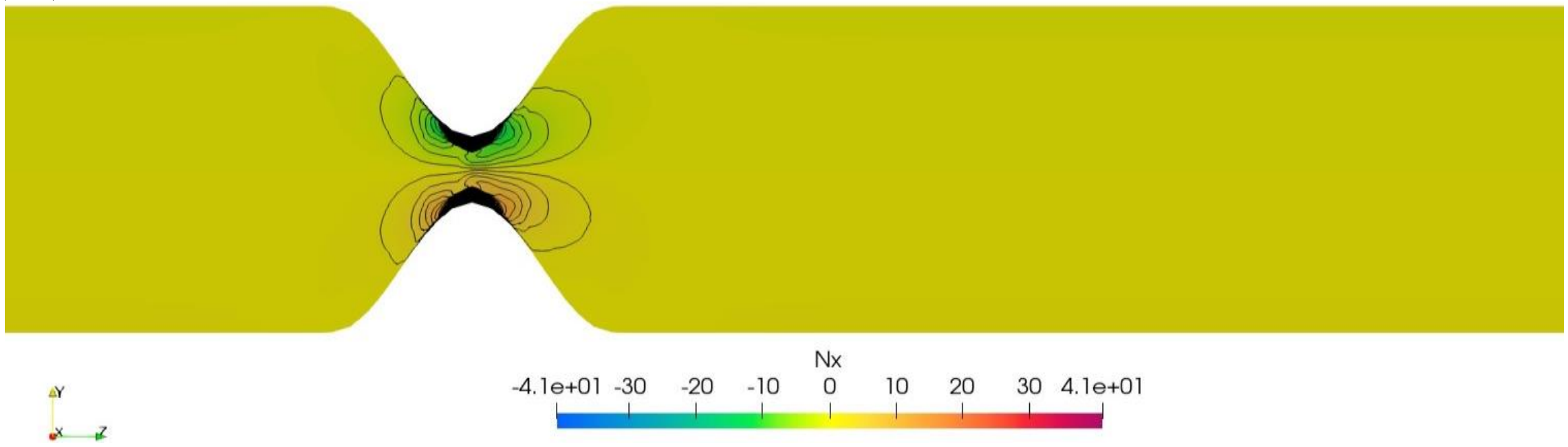

*(a.ii)*

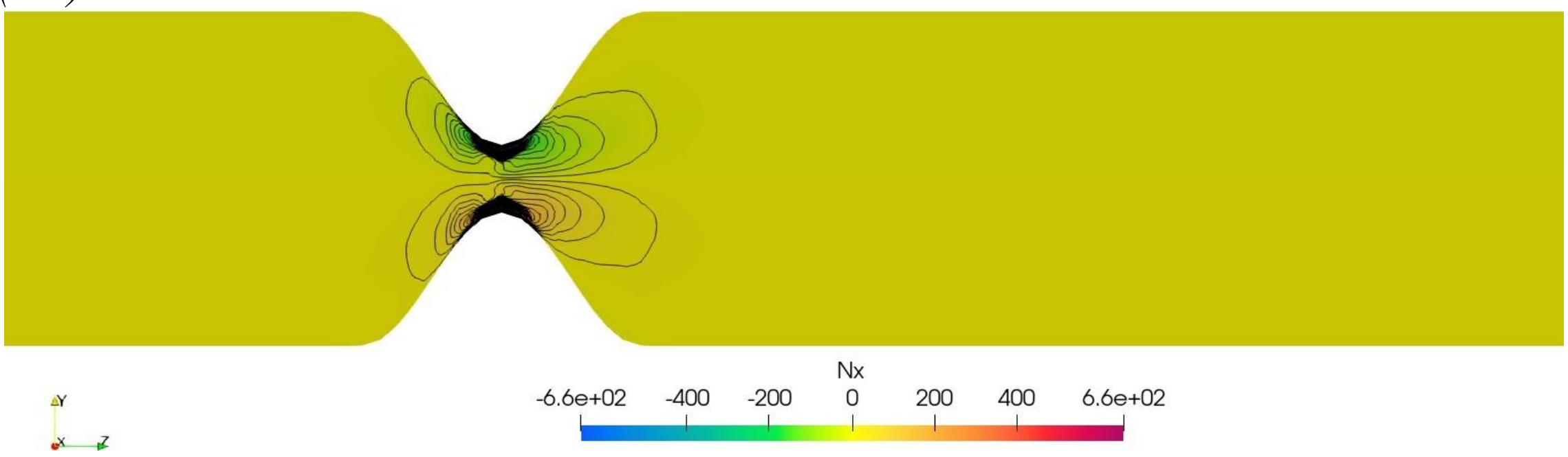

*(b.i)*

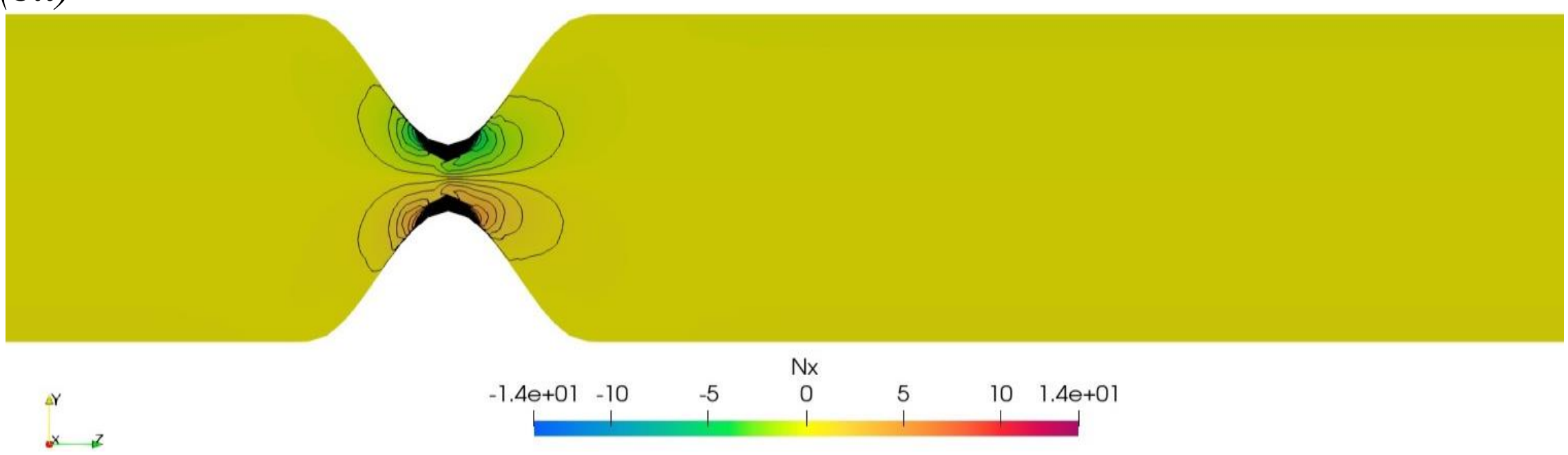

*(b.ii)*

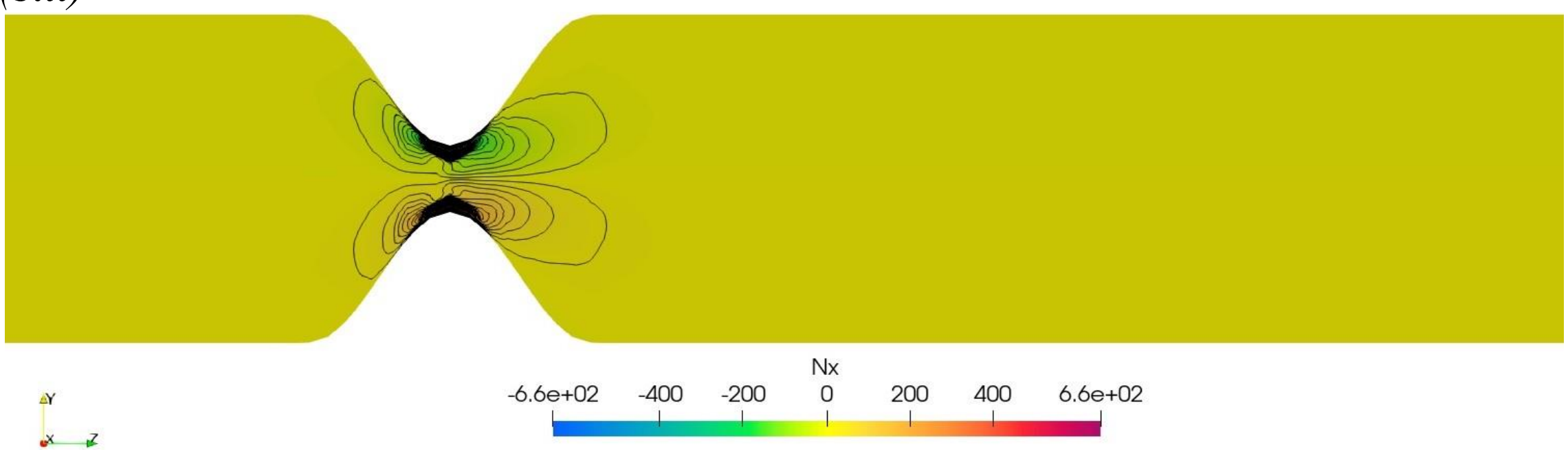

*(c.i)*

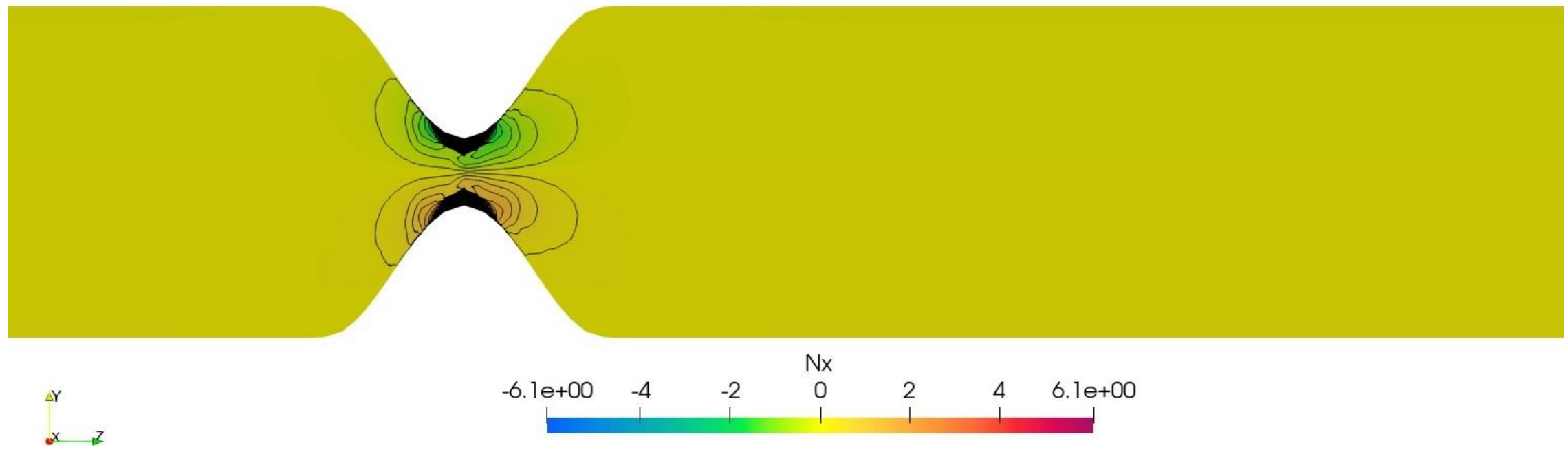

*(c.ii)*

FIG. 27. Microrotation contour plots for blood flow through an 80% stenosis using MHD micropolar modeling (i) without acknowledging MMR and (ii) considering MMR for an applied magnetic field of (a) 1 $T$, (b) 3 $T$, and (c) 8 $T$ and hematocrit of $\varphi = 45\%$.

In Figure 28, the velocity profile is illustrated for the 80% stenosis within the stenotic region and downstream the latter (at $l = 0.02\ m$). The profiles are plotted for the Newtonian blood flow, the micropolar blood flow, the MHD blood flow without the MMR effect, and the MHD blood flow with the MMR effect included. Two hematocrit values are used, one at $\varphi = 25\%$ ($\varepsilon = 0.375$) and one at $\varphi = 45\%$ ($\varepsilon = 0.675$), while the applied magnetic field is varied at 1 $T$, 3 $T$, and 8 $T$. Similar to the 50% stenosis case, the velocity exhibits a blunted parabolic shape, which is not completely flat, due to the longer length of the 80% stenosis. Due to the high degree of stenosis, the velocity downstream significantly decreases by 10 times compared to the 50% stenosis.

Compared to the 50% stenosis, the micropolar effect becomes evident, with the velocity decreasing by 27% both within and downstream of the stenotic region during the transition from the Newtonian to the micropolar profile, coinciding with an increase in hematocrit. This is due to the smaller $\lambda$ values that are associated with severe stenoses, where the radius of the artery decreases significantly (see the relative discussion in Section II.C). Once again, no significant differences are observed between the velocity profiles of micropolar blood flow and MHD micropolar blood flow without MMR, for all values of the applied magnetic field and hematocrit considered, both within and downstream of the stenosis. However, when the MMR term is included, the velocity decreases, with a maximum reduction of 25% at 8 $T$, and $\varphi = 45\%$ inside the stenosis and 30% at 8 $T$, and $\varphi = 45\%$ downstream of the stenosis. As expected, these differences become larger as the hematocrit increases.

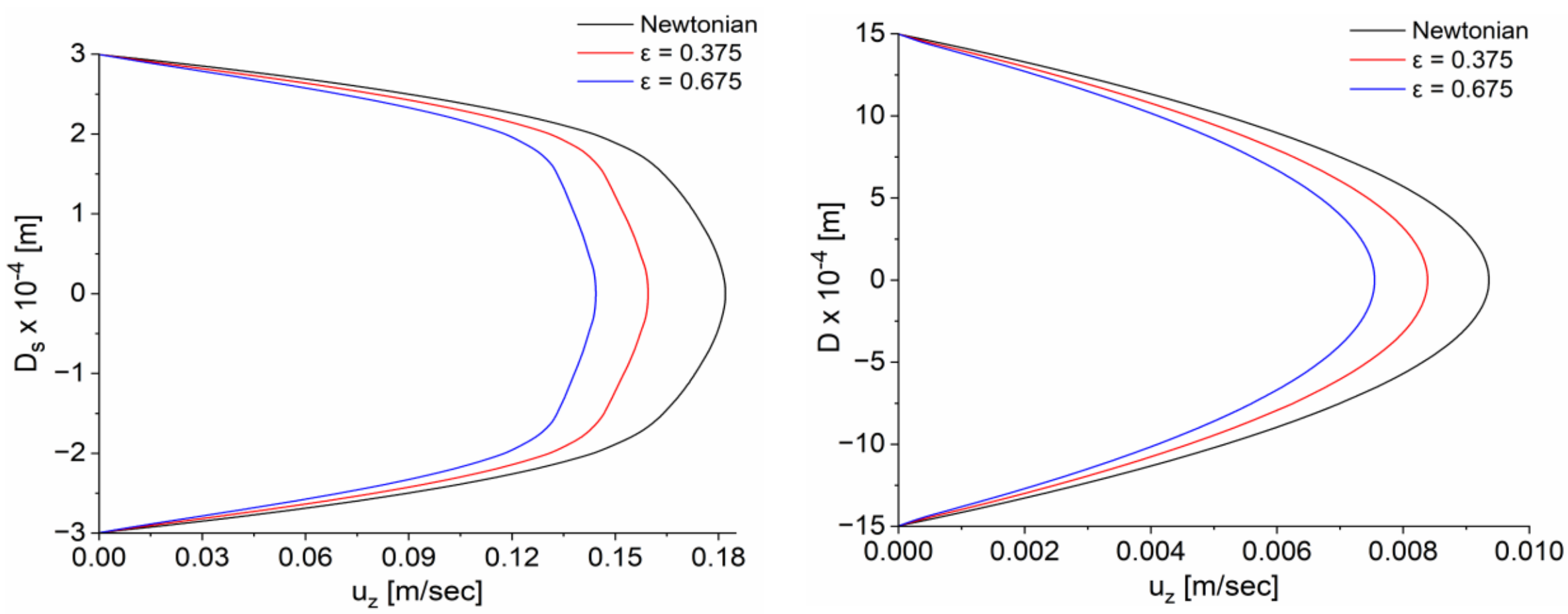

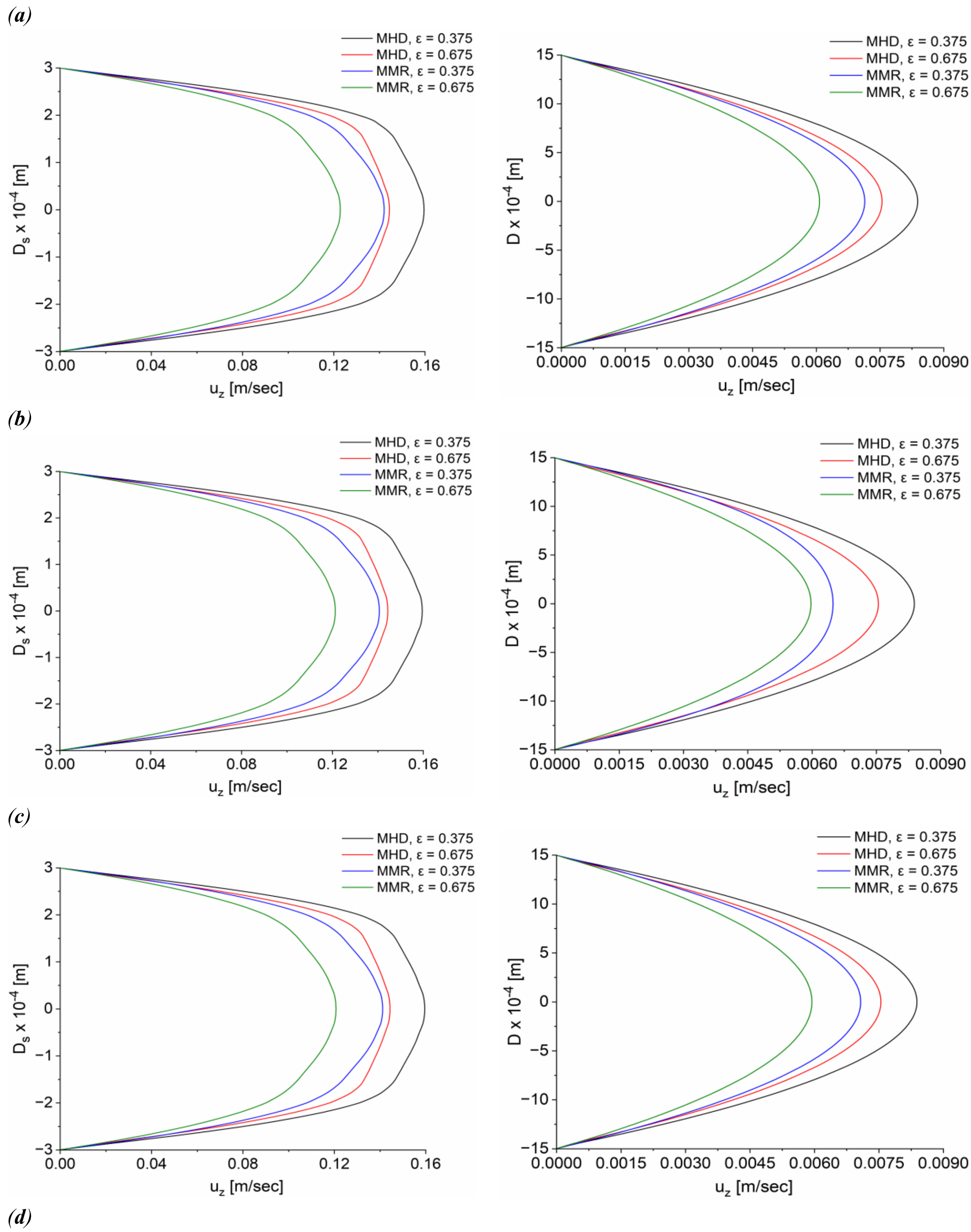

FIG. 28. Velocity profiles at the center of the stenotic region (left) and downstream the latter (right) with an 80% stenosis for (a) the Newtonian and micropolar blood flow without an applied magnetic field and (b) the micropolar blood flow by ignoring and considering the MMR term with a magnetic field of 1 $T$ , (c) 3 $T$ and (d) 8 $T$ . The hematocrit is varied at $\varphi =$ 25% and $\varphi = 45\%$ ($\varepsilon = 0.375$ and $\varepsilon = 0.675$, respectively).

In Figure 29, the microrotation profile is shown for the 80% stenosis within the stenotic region and downstream (at $l = 0.02\ m$). Similar to the velocity profile, microrotation is plotted for different types of blood flow: Newtonian, micropolar, MHD blood flow without the MMR effect, and MHD blood flow with the MMR effect. Two hematocrit values are used $\varphi = 25\%$ ($\varepsilon = 0.375$) and $\varphi = 45\%$ ($\varepsilon = 0.675$), with the magnetic field applied at strengths of 1 $T$, 3 $T$, and 8 $T$. It is immediately apparent that the microrotation profile is disrupted within the stenosis, mirroring the velocity's flattened parabolic profile. An increase in hematocrit reduces microrotation both inside and outside the stenosis by nearly 30%. Additionally, applying the external magnetic field without accounting for the MMR effect does not lead to any significant change in microrotation. However, when the MMR effect is included, microrotation decreases substantially in both the stenotic region and downstream, with a near 99.9% reduction at a magnetic field strength of 8 $T$ for both hematocrit values. These findings confirm that the internal rotation of the erythrocytes is nearly "frozen," as they are oriented parallel to the applied magnetic field.

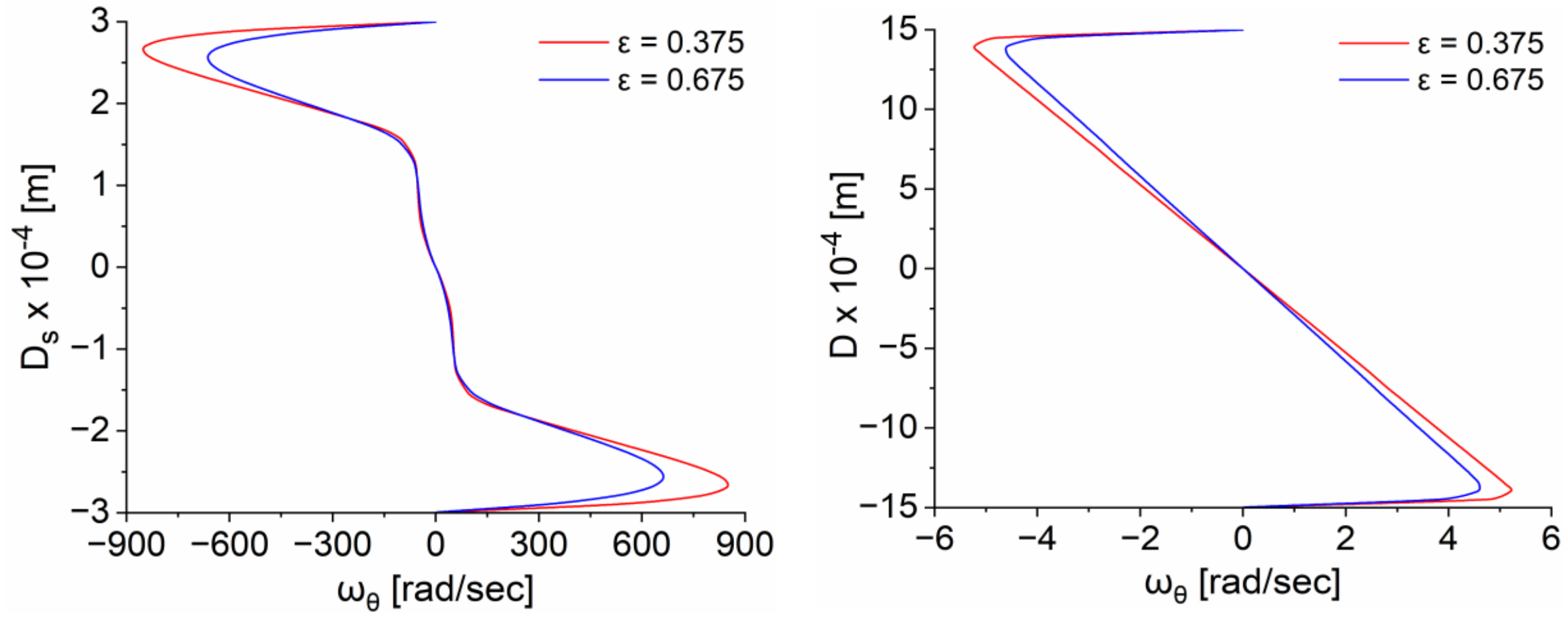

*(a)*

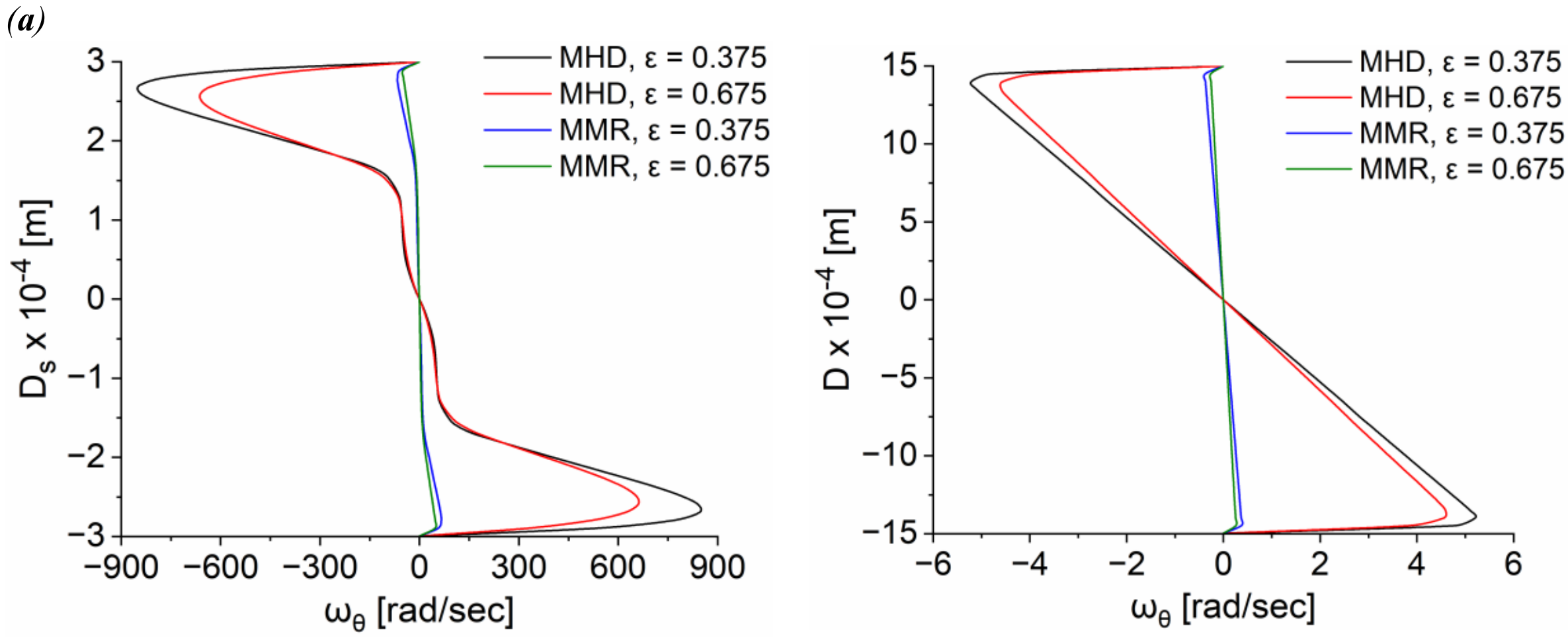

*(b)*

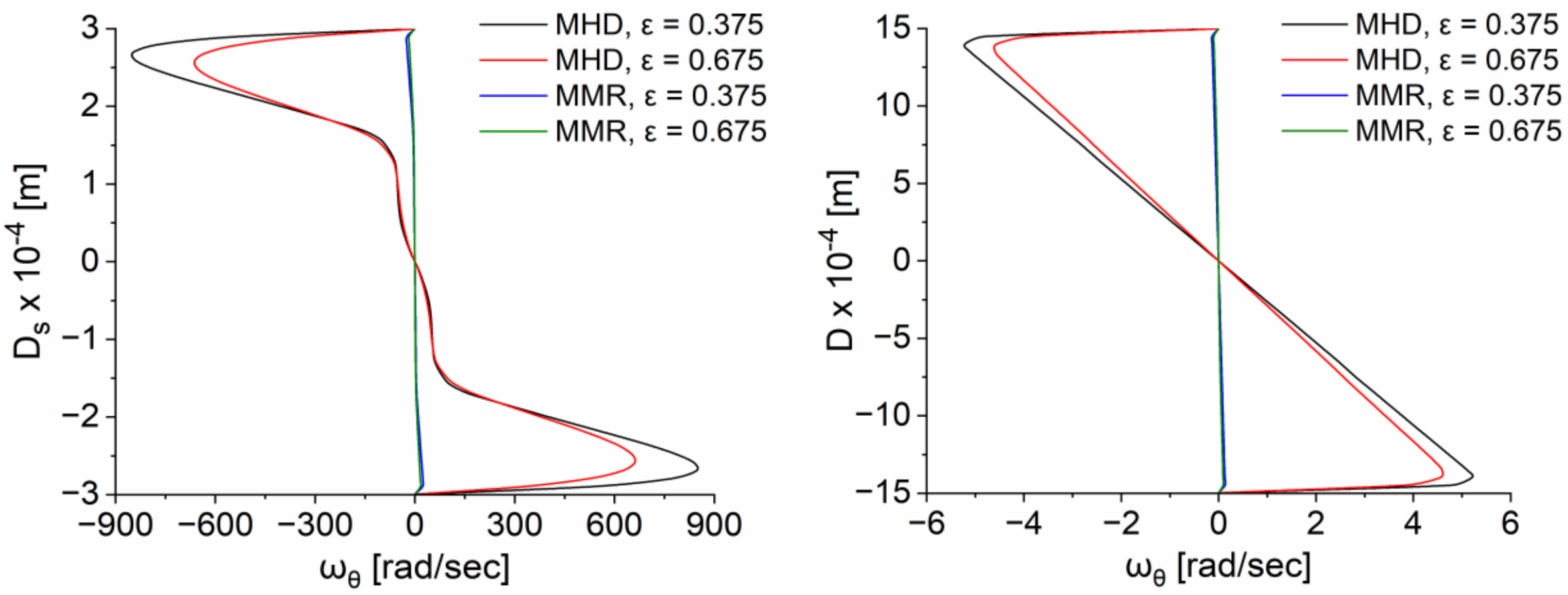

*(c)*

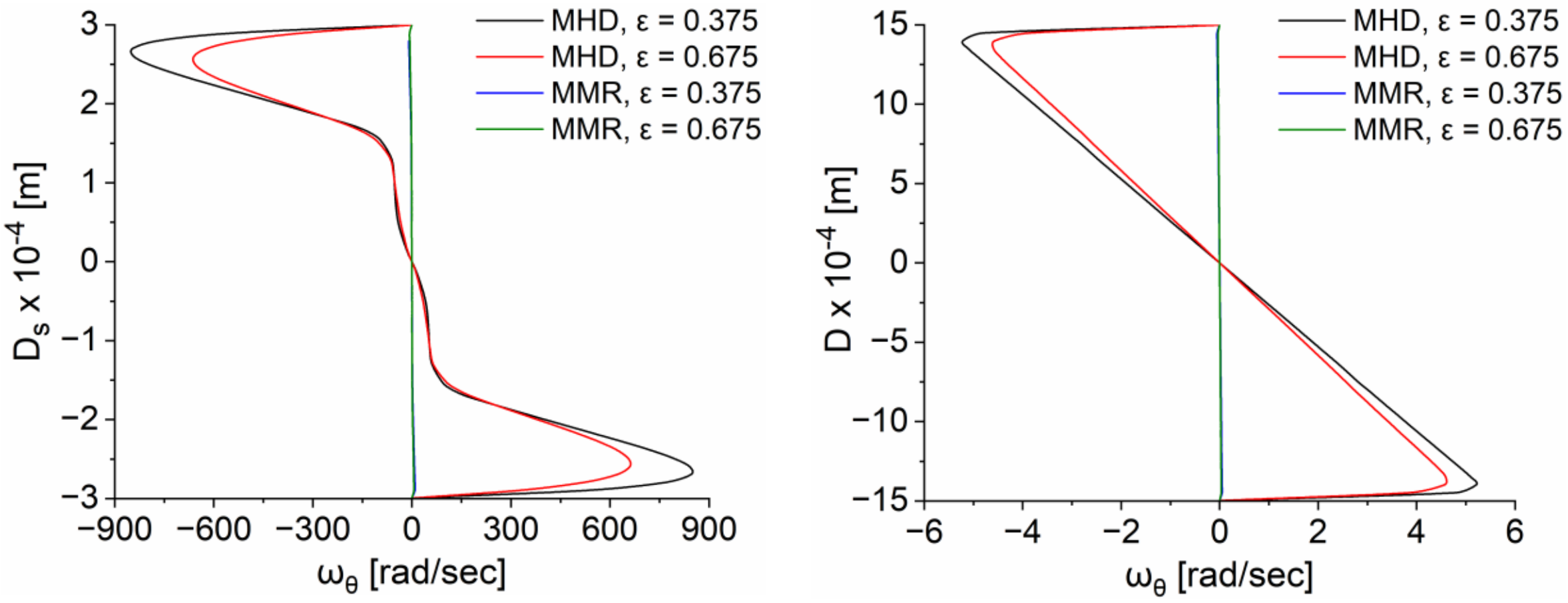

*(d)*

FIG. 29. Microrotation profiles at the center of the stenotic region (left) and downstream the latter (right) with an 80% stenosis for (a) the Newtonian and micropolar blood flow without an applied magnetic field and (b) the micropolar blood flow by ignoring and considering the MMR term with a magnetic field of 1 $T$ , (c) 3 $T$ and (d) 8 $T$ . The hematocrit is varied at $\varphi =$ 25% and $\varphi = 45\%$ ($\varepsilon = 0.375$ and $\varepsilon = 0.675$, respectively).

Figure 30 shows the wall shear stress $\tau_w$ distribution for the 80% stenosis. Because of the much stronger geometric constriction, the $\tau_w$ peak at the throat is substantially higher and more sharply localized than in the 50% case, while the downstream region exhibits a long section of very low $\tau_w$ associated with the near-stagnant recirculation documented earlier in the velocity and microrotation fields. This pattern—very high $\tau_w$ at the throat and low WSS in the post-stenotic separation region—matches the trends reported in pulsatile and turbulent stenosis simulations, where the throat is exposed to damaging levels of shear and the distal segment to low, oscillatory $\tau_w$ [53-56].

As for the 50% stenosis, the micropolar model yields higher peak $\tau_w$ than the Newtonian model at fixed flow rate, again reflecting the additional microstructural resistance and the no-spin wall condition. This is consistent with the behavior found in Hogan and Henriksen for $\delta =$ 0 [16] and with our earlier 2D channel results, where micropolarity enhances near-wall shear stresses relative to the Newtonian case [15]. The effect of MMR is now more visible: increasing the magnetic field strength leads to a progressive reduction and slight broadening of the $\tau_w$ peak, especially at higher hematocrit values. The shear stress inside the stenosis remains elevated, but the MMR term partially damps the extreme values, which is physically plausible because the magnetic torque tends to suppress microrotation and flatten the core velocity profile within the already narrow throat.

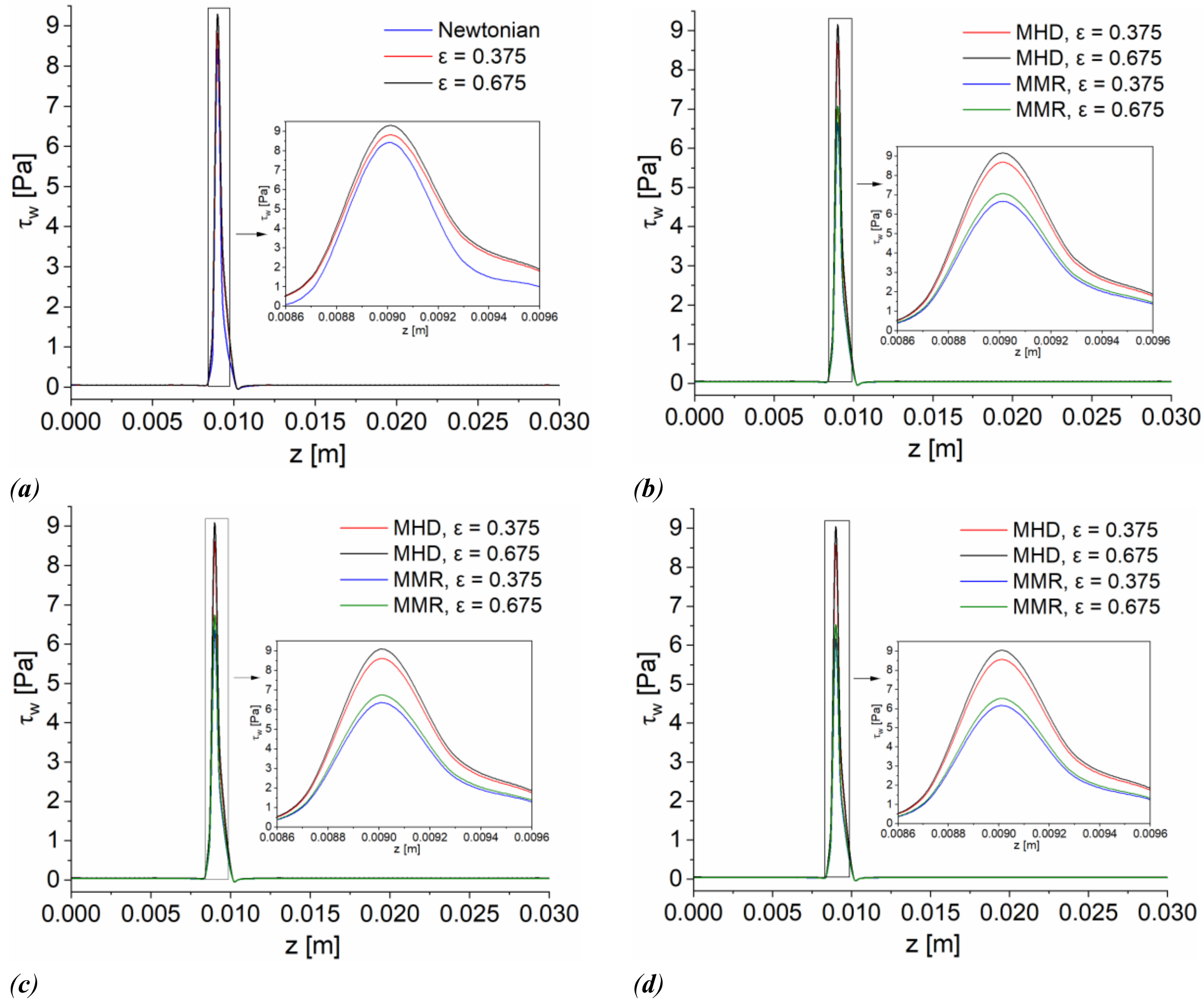

FIG. 30. Wall shear stress $\tau_w$ for the 80% stenosis for (a) the Newtonian and micropolar blood flow without an applied magnetic field and (b) the micropolar blood flow by ignoring and considering the MMR term with a magnetic field of 1 $T$ , (c) 3 $T$ and (d) 8 $T$ . The hematocrit is varied at $\varphi = 25\%$ and $\varphi = 45\%$ ($\varepsilon = 0.375$ and $\varepsilon = 0.675$, respectively). The rectangle region indicates the area magnified in the inset, which highlights the differences in the peak wall shear stress near the stenosis throat.

The pressure distributions for the 80% stenosis are shown in Figure 31. In all cases, a very sharp pressure drop occurs across the stenotic section, followed by an almost flat pressure downstream, in line with the low velocities observed past the stenosis (see Figure 26). This behavior agrees qualitatively with previous numerical and experimental studies of severe stenoses [53, 54, 57], which report large pressure losses concentrated within the stenosis and only weak pressure variation downstream of the stenosis.

For the Newtonian, micropolar, and MHD cases (without MMR), the centerline pressure curves are nearly indistinguishable due to the severity of the stenosis; the flow is already so strongly constricted that further changes in rheology or conductivity only produce very small variations that are not visible on the scale of the figure. For clarity, we therefore do not plot all hematocrit and magnetic field combinations for these models, as we did in all previous results.

In contrast, when MMR is included, a noticeable additional pressure drop appears across the throat. This is consistent with the strong braking of the microrotation and velocity fields discussed in Section III.B.

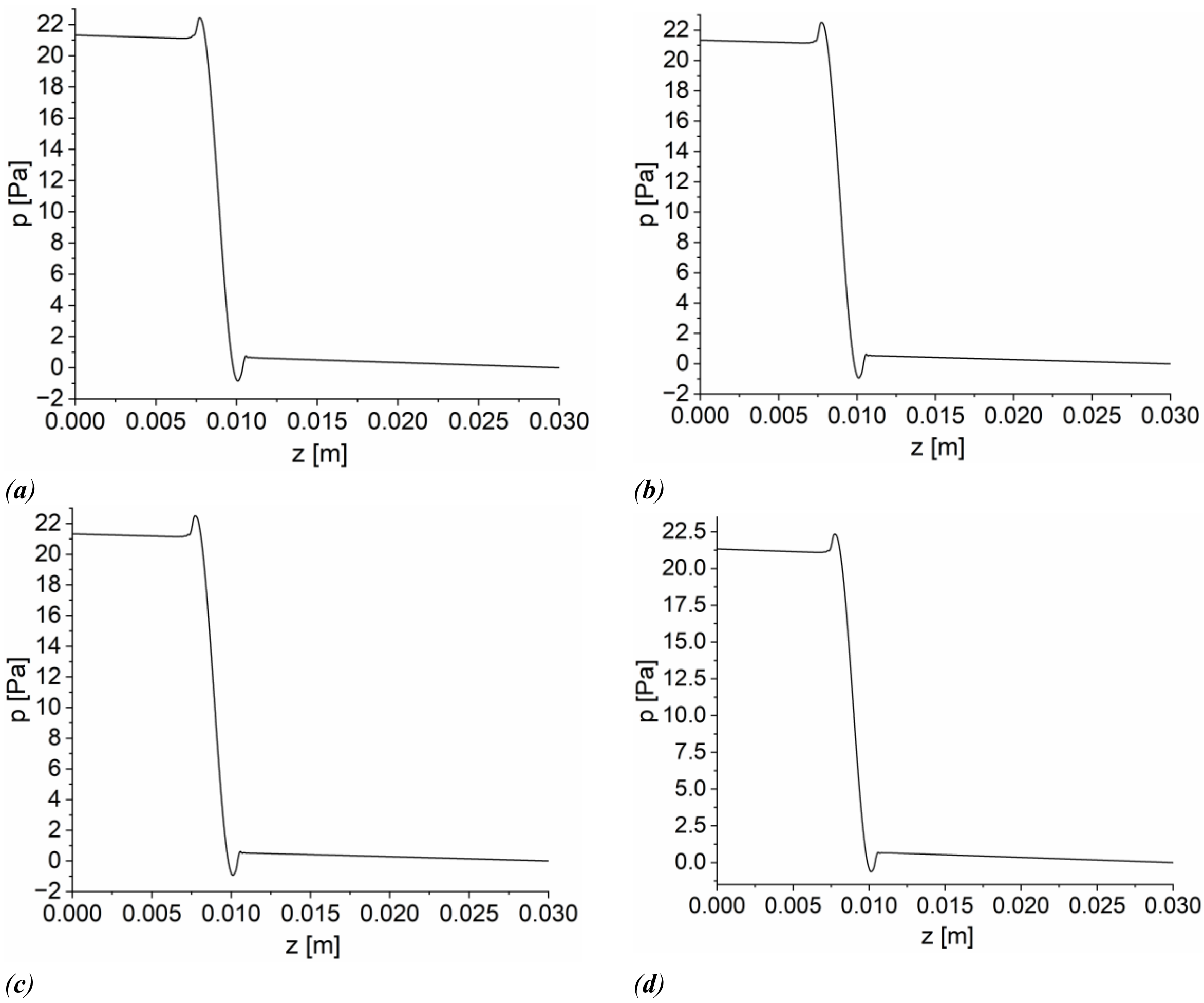

FIG. 31. Pressure drop for the 80% stenosis for (a) the Newtonian, and (b) micropolar blood flow without an applied magnetic field, and the micropolar blood flow by (c) ignoring and (d) considering the MMR term with a magnetic field of 1 $T$. The hematocrit is kept constant at $\varphi = 45\%$ ($\varepsilon = 0.675$).

## IV. CONCLUSIONS

This paper concerns the numerical study of a 3D micropolar MHD blood flow through stenosis by acknowledging and ignoring the effects of micromagnetorotation (MMR). Two stenotic regions are considered, commonly found in cardiovascular diseases: one at 50% and the other at 80%. Key flow variables are derived, including streamlines, vorticity and microrotation contours, and velocity and microrotation profiles both inside and outside the stenotic areas, as well as wall shear stress and pressure drop. For comparative analysis, these variables are assessed across four scenarios: Newtonian blood flow, micropolar blood flow, MHD micropolar blood flow without the MMR effect, and MHD micropolar blood flow with the MMR effect. Two hematocrit levels are considered, one at $\varphi = 25\%$ and one at $\varphi = 45\%$. The applied magnetic field is varied across three values:, 1 $T$, 3 $T$, and 8 $T$. The values selected for the hematocrit and the applied magnetic field were obtained from other numerical and experimental studies related to various biomedical applications.

For the computational analysis, two new transient solvers, epotMicropolarFoam and epotMMRFoam, were developed for the first time using the open-source OpenFOAM library. These solvers can simulate any MHD micropolar flow with magnetic particles (such as blood) with or without the MMR effect, respectively. Their development was based on existing transient OpenFOAM solvers, including icoFoam, epotFoam, and micropolarFoam, for Newtonian, MHD, and micropolar flows, respectively. Both epotMicropolarFoam and epotMMRFoam apply the low-magnetic-Reynolds number approximation, neglecting the magnetic induction equation, and instead using an electric potential formulation. The solvers were validated against the analytical results of an MHD micropolar Poiseuille blood flow from the paper by Aslani et al., using different values for the hematocrit and the intensity of the applied magnetic field. The results showed excellent agreement between the numerical and analytical results for all considered flow cases, with an error not exceeding 2% for both velocity and microrotation.

Considering the micropolar blood flow through stenosis without an applied magnetic field, the analysis shows that the internal rotation of erythrocytes does not produce noticeable changes for the 50% stenosis, even at high hematocrit values, due to the relatively small size of the artery, which minimizes any micropolar effects on the flow. However, these micropolar effects become more apparent in the 80% stenosis, showing a significant reduction in velocity, vorticity, and microrotation. Furthermore, the streamlines, vorticity, and microrotation contours are altered, with any vortices or disturbances downstream of the stenosis nearly braked, especially as hematocrit increases. This happens because the high degree of stenosis significantly reduces the artery's diameter, making micropolar phenomena more evident. Moreover, both wall shear stress and pressure drop are elevated, confirming that micropolarity can significantly alter the mechanical loading and hemodynamic burden in stenosed vessels.

When the magnetic field is applied to the stenosis without considering the MMR effect, it does not cause any significant change in blood flow for any variable, regardless of the magnetic field intensity. This result was anticipated, as the Lorentz force has a minimal impact on the stenosis due to blood’s relatively low electrical conductivity and the size of the artery. However, when the MMR term is included, the blood flow is significantly affected in all cases, with considerable reductions in velocity, vorticity, and microrotation. Velocity and vorticity can decrease by up to 30%, while microrotation can decrease by up to 99.9%. Wall shear stress and pressure drop are significantly altered, with substantially higher peak wall shear stress, and higher overall pressure drop, especially under strong magnetic fields. Physically, this result means that the erythrocytes are polarized in the direction of the externally applied magnetic field, with no internal rotation allowed, leading to a significant reduction in vorticity and velocity. Additionally, any vortices or disturbances downstream from the stenosis are severely dampened. These findings further validate the damping effect of micromagnetorotation observed in previous studies. It should be noted that this flow damping caused by MMR is enhanced as the hematocrit and the degree of stenosis increase.

In conclusion, the effect of micromagnetorotation, which has so far been ignored in MHD micropolar blood flows, appears to play an important role in influencing blood flow, both within the stenotic region and downstream of it. This term should not be overlooked due to its serious implications (e.g., endothelium stress, pathological progression) on blood flow configurations, such as stenoses, because of the additional damping in the blood flow.

## ACKNOWLEDGMENTS

This research was funded by the Action “Flagship actions in interdisciplinary scientific fields with a special focus on the productive fabric,” implemented through the National Recovery and Resilience Fund Greece 2.0 and funded by the European Union–NextGenerationEU (Project ID: TAEDR-0535983).

The authors report no conflict of interest.